\documentclass{osa-article}

\journal{oe}

\usepackage{amsmath}
\usepackage{caption}
\usepackage{subcaption}
\usepackage{hyperref}

\newcommand{\myvec}[1]{\mathbf{#1}}
\newcommand{\uvec}[1]{\hat{\mathbf{#1}}}
\providecommand{\abs}[1]{\lvert#1\rvert}

\DeclareMathOperator\erf{erf}

\begin{document}

\title{Three-dimensional, time-dependent analysis of high- and low-$Q$ free-electron laser oscillators}

\author{Peter J.M. van der Slot,\authormark{1,*,$\dagger$} and Henry P. Freund\authormark{2,3,$\dagger$}}

\address{\authormark{1}Laser Physics and Nonlinear Optics, Mesa$^{+}$ Institute for Nanotechnology, Department of Science and Technology, University of Twente, Enschede, the Netherlands\\
\authormark{2}Department of Electrical and Computer Engineering, University of New Mexico, Albuquerque, New Mexico, USA\\
\authormark{3}NOVA Physical Science and Simulations, Vienna, Virginia, USA}
\noindent\authormark{$\dagger$}These authors contributed equally to this work.\\
\authormark{*}Corresponding author: \email{p.j.m.vanderslot@utwente.nl}



\newcommand{\orcidauthorA}{0000-0001-7473-1752} 
\newcommand{\orcidauthorB}{0000-0002-2223-8760} 

\abstract{Free-electron lasers (FELs) have been designed to operate over virtually the entire electromagnetic spectrum from microwaves through x-rays and in a variety of configurations including amplifiers and oscillators. Oscillators can operate in both the low and high gain regime and are typically used to improve the spatial and temporal coherence of the light generated. We will discuss various FEL oscillators ranging from systems with high-quality resonators combined with low-gain undulators to systems with a low-quality resonator combined with a high-gain undulator line. The FEL gain code MINERVA and wavefront propagation code OPC are used to model the FEL interaction within the undulator and the propagation in the remainder of the oscillator, respectively. We will not only include experimental data for the various systems for comparison when available, but also present for selected cases how the two codes can be used to study the effect of mirror aberrations and thermal mirror deformation on FEL performance.}

\section{Introduction}
\begin{figure}[t] 
\centering
\includegraphics[width=10cm]{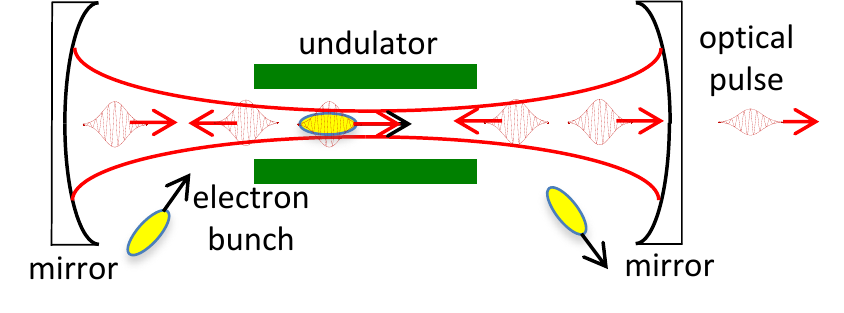}
\caption{\label{fig:fel_osc}Schematic representation of a typical layout for an FEL oscillator. The electron beam consisting of electron pulses is guided through the undulator and around the mirrors. The optical pulse that is amplified within the undulator is recirculating within the resonator and extracted through one of the mirrors \cite{freund_2018}. }
\end{figure}

Free-electron laser (FEL) oscillators (FELO) have been part of the overall research activity since the beginnings of the field. An FEL oscillator consists of an undulator placed within an optical resonator, also known as an optical cavity, that typically consists of two spherical mirrors separated by a distance $L_{\textrm{cav}}$. Note that the undulator is not necessarily centered within the optical resonator. Further, an electron beam is directed into the cavity, through the undulator and out of the cavity as is schematically shown in  Figure~\ref{fig:fel_osc} which is taken from Reference \cite{freund_2018}. The optical pulse generated by the electrons streaming through the undulator is extracted from the cavity either by a partially reflective (\textit{i.e.}, transmissive) mirror or through a hole in the mirror. Observe that multiple optical pulses may be contained within the resonator depending upon the repetition rate of the electron bunches and the roundtrip time for the optical pulses to circulate through the resonator.

The first FEL oscillator driven by a relativistic electron beam was realized in 1977 \cite{deacon_1977}, shortly after coherent amplification of radiation of the same wavelength was experimentally demonstrated \cite{elias_1976}. Coherent undulator radiation was first demonstrated using the so-called Ubitron, which was configured both as amplifier and oscillator, and which is nowadays considered to be the first (non-relativistic) FEL \cite{phillips_1988}. In the early days of FEL development, the technology available to generate high-quality, high-brightness electron beams, that are necessary for efficient FEL operation, was relatively primitive by today’s standards. At that time, the limited available gain, which in part may be due to space restrictions, favoured oscillator configurations to bring the laser into saturation and provide fully coherent output as mirrors were readily available all the way from the visible to well into the microwave spectral range. Although we limit ourselves here to infrared wavelengths or shorter, exciting FEL research also takes place at longer wavelengths. 

The unique properties of FEL oscillators, in particular to fill the spectral gap for coherent light sources in the visible to microwave spectral range, resulted quickly in the establishment of several facilities for applied research. The characteristics of user facilities using FEL oscillators depends in part on the accelerator technology used, which include electrostatic acceleration \cite{elias_1986, ramian_1992, gover_2004, gover_2005}, energy recovery linear accelerators (linacs) \cite{minehara_2000, neil_2006, thomas_2007, thompson_2015, vinokurov_2016, shevchenko_2019}, non-recovery linacs \cite{smith_1990, vanderWiel_1993, tomimasu_1996, ortega_1996, ortega_2002, schwettman_1996, horiike_2002, tanaka_2004, vanderzande_2013, schollkopf_2015, zen_2016,  klopf_2018, li_2018, schollkopf_2019, zhao_2020}, microtrons \cite{gallerano_1999, jeong_2004} and synchrotrons \cite{hara_1996, litvinenko_1998, hosaka_2006, wu_2006, wu_2007}. Research performed at these user facilities includes amongst others biological and material science \cite{edwards_2007}. A more complete overview of the various FEL oscillators together with the main system parameters can be found in Ref.~\cite{neyman_2017}.

As mentioned, the optical properties of FEL oscillators are in part determined by the electron accelerator used. For example, electrostatic accelerators are limited in the maximum electron beam energy they can produce and, consequently, the spectral range is limited to microwave and THz frequencies \cite{neyman_2017}. The much higher electron beam energies of storage rings allows the FEL oscillators to operate in the visible down to the deep UV wavelengths \cite{neyman_2017}, while RF linacs can in principle produce any beam energy from a few MeV up to tens of GeV, allowing FELs to cover a large fraction of the electromagnetic spectrum from microwaves all the way down to hard x-rays \cite{neyman_2017}. 

Furthermore, electrostatic accelerators are capable of long to continuous wave electron pulses, that can result in single longitudinal mode oscillation with a relative linewidth of about 10$^{-8}$ over a time of 5~$\mu$s, which was set by the voltage droop present in the accelerator \cite{elias_1986} at the time of measurement. On the other hand, the short duration of and spread in electron energy for the electron pulses produced by RF linacs typically leads to a relative linewidth of the order of 10$^{-4}$ to 10$^{-2}$ \cite{newnam_1985, bamford_1989, oshea_1993, prazeres_1993, vandermeer_1993, oshea_1994, nguyen_1995, xie_1995, berryman_1996, minehara_1999, neil_2000a, lumpkin_1990, jeong_2001, thompson_2015, wu_2018, schollkopf_2019}, unless additional spectral filtering is used. For example, using the Vernier effect of a double cavity resonator to reduce the number of intra-cavity longitudinal modes and subsequent external filtering, an RF-linac-based FEL oscillator was able to provide a single longitudinal mode \cite{oepts_1993}. FEL oscillators based on storage rings provide a lower relative linewidth within the range $10^{-5}$ to $5 \times 10^{-4}$ \cite{billardon_1986, kulipanov_1990, yamada_1992, yamazaki_1993, takano_1993a, couprie_1993, litvinenko_1998, trovo_2002, nolle_2000, sei_2009} due to the higher electron beam quality of storage rings.   

Finally, the average and peak power are to a large extend determined by the accelerator technology. Superconducting energy-recovery linacs are the primary drivers for high average power FEL oscillators \cite{benson_1999, neil_2000, nishimori_2001, hajima_2003a, antokhin_2004, benson_2007, shevchenko_2020}; however, other types of FEL oscillators have used various techniques to increase the output pulse energy including but not limited to optical klystrons \cite{vinokurov_1977, deacon_1984, yamazaki_1991, yamada_1992, couprie_1993, yamazaki_1993, takano_1993, yamazaki_1995, berryman_1996, litvinenko_1998, wu_2005, wu_2006, sei_2009}, cavity dumping \cite{cutolo_1989, benson_1990, bhowmik_1990, takahashi_2009, rana_2021},  pulse stacking in an external cavity \cite{burghoorn_1992,smith_1997, faatz_1994}, electron beam energy ramping \cite{marks_2017,xu_2019} or dynamic cavity desynchronization \cite{jaroszynski_1990, bakker_1993, knippels_1994, song_2000, sumitomo_2019, zen_2020}. Tapering of the undulator has also been used to improve the peak power in the pulse \cite{edighoffer_1984,curtin_1990, jaroszynski_1995, benson_2001, asgekar_2012}, however, tapering of the undulator in FEL oscillators leads to complicated dynamics and strong dependence on system parameters \cite{saldin_1993, christodoulou_2002, dattoli_2012, ottaviani_2016}.

Wavelength tuning of FEL oscillators is typically obtained by selecting a set of electron beam energies and varying the undulator gap. Operation at other wavelength ranges is possible through harmonic generation \cite{newnam_1985, bamford_1990, prazeres_1990, curtin_1990, prazeres_1991, prazeres_1993, yamazaki_1993, berryman_1996, jaroszynski_1996, hayakawa_2002a, yamada_1998, neil_2001, hajima_2001, trovo_2002, litvinenko_2003, hayakawa_2003, deninno_2004, sei_2010, kubarev_2011}. The simultaneous generation of multiple wavelengths may be beneficial for certain applications. The fixed relation between the fundamental and its harmonics can be overcome using different undulators in a single optical cavity \cite{jaroszynski_1994, ortega_1996, prazeres_1998, wu_2015, yan_2016}, two undulators driven by a single electron beam at a single beam energy but each having its own optical resonator \cite{zako_1999} or having different beam energies \cite{swent_1991, smith_1998} to allow operation at two independently chosen wavelengths.   

FEL oscillators share many characteristics with amplifier systems, however, there are also differences. For example, slippage between the optical and electron pulse in combination with cavity desynchronization will effect the FEL oscillator dynamics. Here, a cavity is considered synchronized when it is tuned to a length where the roundtrip time for the optical pulse matches an integer multiple of the time separating two subsequent electron pulses. Maxima in the gain and extraction efficiency are found for different cavity desynchronizations. In particular, dynamic cavity desynchronization \cite{jaroszynski_1990} can be used to first set a cavity length corresponding to maximum gain for a quick initial growth of the optical pulse energy and then switched to a cavity length for maximum extraction efficiency that is obtained at small cavity desynchronization. The latter is typically also associated with the generation of short optical pulses \cite{bakker_1993, knippels_1995, ortega_1996, knippels_1997, calderon_2000, hajima_2003, mcneil_2011, zen_2020, hajima_2021}. For short pulse FEL oscillators, \textit{i.e}, where the root-mean-square (rms) width of the electron bunch is smaller than the total slippage distance, the oscillator is operating in the superradiant regime \cite{hahn_1993, bakker_1994, piovella_1995, jaroszynski_1997, hajima_2001a, hajima_2003, nishimori_2006, duris_2018, hajima_2021}. An analytic theory for this regime predicts a scaling of the optical pulse energy $E_L$ with the bunch charge $q$ as $E_L \propto q^{3/2}$ and a scaling of the temporal width $\tau_L$ of the optical pulse as $\tau_L \propto q^{-1/2}$ \cite{piovella_1995} which has been experimentally confirmed \cite{jaroszynski_1997}, although higher extraction efficiencies have also been observed for a perfectly synchronized cavity length where the analytic theory brakes down \cite{hajima_2003, zen_2020}. Furthermore, short pulse FEL oscillators operating in the large-slippage regime also show a stable oscillation in the macropulse power, known as limit-cycle oscillations, for certain cavity lengths detuned from perfect synchronization, which has been observed both experimentally \cite{jaroszynski_1993, bakker_1994, knippels_1996, christodoulou_2002, deninno_2003, kiessling_2018} and in numeric simulations \cite{colson_1982, hahn_1993, blau_1995, piovella_1995, blau_2015}. 

Techniques well known from  laser physics have also been applied to FEL oscillators either experimentally or in a conceptual study, such as injection seeding \cite{amir_1991, oepts_1992, szarmes1996, takahashi_2007, evain_2012, hajima_2017} and mode locking \cite{oepts_1992, szarmes1996, shvets_1997, mcneil_2011}. On the other hand, the unique nature of the gain process in FELs allows for micro structuring of the gain medium \cite{boehmer_1981, boscolo_1982, schnitzer_1985} to enhance the gain, which is not readily available for ordinary lasers. One important application of this technique is to transfer coherence from a low-frequency optical beam to a higher-frequency optical beam. This process is known as high-gain harmonic generation (HGHG) \cite{yu_1991}. This technique has been successfully used to generate fully coherent light at soft x-ray frequencies \cite{allaria_2012}. Alternative methods currently investigated, in particular to reach even shorter wavelengths, rely on Bragg reflections from atomic layers using appropriated crystals \cite{kim_2008}.

The drive to higher gain per unit length and to shorter wavelengths pushed  accelerator development; in particular, research into photo-cathode based electron sources~\cite{sheffield_1988, dowell_1993} that typically produce higher quality electron beams and contributed to today's state-of-the-art accelerator technology. The availability of electron beams with improved beam quality has driven FEL development over the years, especially the capability to generate shorter wavelengths. Due to a lack of suitable mirrors, initial focus was on FELs relying on self-amplified spontaneous emission (SASE), with a  proof-of-principle experiment at 530 nm~\cite{arnold_2001}, followed with demonstrations at shorter wavelengths~\cite{rossbach_2019}. To overcome poor temporal coherence, HGHG is used to provide completely coherent output pulses down to 4-nm wavelength~\cite{allaria_2015}, while for even smaller wavelengths FEL oscillators using the above-mentioned Bragg mirrors are considered~\cite{kim_2008}. 

Along with increasingly sophisticated FEL oscillator designs and experiments,  analytic theories \cite{saldin_1993, saldin_1993a, piovella_1995, dattoli_1995, dattoli_2007, curcio_2018} have been developed for initial assessment of  oscillator performance.  Furthermore, the more complete simulation codes have also become adept at treating increasingly complicated designs and include both steady-state and time-dependent simulations in one- and three-dimensions.

In the remainder of this paper we will focus on simulation of a few different FEL oscillators. To model these oscillators, we have selected the FEL gain code MINERVA, which models the light amplification within the undulator line, and the optical wavefront propagation code OPC, which models the light propagation in the remainder of the resonator. The mathematical formulations of these codes are presented in Section \ref{section:num_form}. Subsequently, we present some considerations on resonators relevant for FEL oscillators in Section \ref{section:opt_res}. We continue with a discussion on a low-gain/high-$Q$ oscillators, taking the Infrared Demo and Infrared Upgrade experiments performed at the Thomas Jefferson National National Accelerator Facility (JLab) as  examples. Both are discussed in Section~\ref{section:lg_hq}. We compare this with a high-gain/low-$Q$ oscillator, also known as a regenerative amplifier (RAFEL), also operating in the infrared and investigate the performance of such a system in the soft x-ray spectral range. These systems are described in Section~\ref{section:hg_lq}. We conclude with a summary and discussion in Section~\ref{section:sum_concl}.

\section{Numerical formulation} \label{section:num_form} 
Simulations of FEL oscillators have appeared in the literature by including optical propagation algorithms to existing FEL simulation codes as well as self-contained FEL oscillator codes \cite{colson_1983, colson_1983a, mcVey_1989, parazzoli_1999, dattoli_1999, dattoli_2004, dattoli_2007a, dattoli_2009, dattoli_2012, blau_2015}. In this paper, we describe the use of an existing FEL simulation code to treat the interaction within the undulators and to link that to a code specifically designed to propagate the optical field through various resonator configurations. Such optics codes are sufficiently general to be able to treat a variety of resonator designs, and many may have been originally created to deal with other types of lasers. In this simulation environment, the FEL code hands off the optical field at the resonator exit to the optics code which then propagates the field around the resonator and back to the undulator entrance, after which it is handed off to the FEL code for another pass through the undulator.

Among the earliest applications of this method was to simulations of the IR-Demo \cite{benson_1999,neil_2000} as described in Section~\ref{section:irdemo}, and the 10-kW Upgrade experiments at JLab \cite{vanderslot_2009} which employed the MEDUSA FEL code \cite{freund_2000} and OPC \cite{karssenberg_2006} and successfully reproduced many of the essential features of the experiment. OPC has also been interfaced with the GENESIS FEL code \cite{reiche_1999} to study VUV FEL oscillators \cite{mcneil_2007}. In this paper, we discuss the linkage of the MINERVA simulation code with OPC.

As will be described below, while MINERVA represents the optical field as a superposition of Gaussian modes with two independent polarization directions, OPC propagates the field on a grid. This necessitates the mapping of the Gaussian modes in MINERVA at the undulator exit to the grid supported by OPC, and the decomposition of the optical field in OPC at the undulator entrance back into Gaussian modes for both polarization components.

\subsection{The MINERVA simulation code} \label{section:minerva}
MINERVA is a time-dependent, three-dimensional simulation code that models the interaction between electrons and a co-propagating optical field through an undulator line which may include strong-focusing quadrupoles and/or dipole chicanes \cite{freund_2017,freund_2018,freund_submitted, minerva_2021}. 
The optical fields are described using a superposition of Gauss-Hermite modes using two independent polarization directions taken to be in the $x$- and $y$-directions with the $z$-direction taken along the axis of the undulator line. The Gauss-Hermite modes constitute a complete set and together with the two independent polarizations are able to describe the generation of arbitrary polarizations that might arise for any given undulator geometry. The vector potential representing the optical field in terms of the Gauss-Hermite modes is given by

\begin{equation}\label{eq:field_decomposition}
\begin{split}
       \delta\myvec{A}(\myvec{x},t) =  &\hat{\myvec{e}}_x \sum_{\substack{l,n=0\\h=1}}^{\infty} e_{l,n,h}^{(x)}(\myvec{x},t) \left(\delta A_{l,n,h}^{(1,x)}(z,t) \sin(\varphi_{h}^{(x)}(\myvec{x},t)) + \delta A_{l,n,h}^{(2,x)}(z,t) \cos( \varphi_{h}^{(x)}(\myvec{x},t)) \right) + \\
       &\hat{\myvec{e}}_y \sum_{\substack{l,n=0\\h=1}}^{\infty} e_{l,n,h}^{(y)}(\myvec{x},t)\left(\delta A_{l,n,h}^{(1,y)}(z,t) \sin(\varphi_{h}^{(y)}(\myvec{x},t)) + \delta A_{l,n,h}^{(2,y)}(z,t) \cos( \varphi_{h}^{(y)}(\myvec{x},t)) \right),
\end{split}
\end{equation}

\noindent where $(l,n)$ are the transverse mode numbers, $h$ is the harmonic number and $\delta A_{l,n,h}$ are the mode amplitudes. Furthermore, $\hat{\myvec{e}}_j$ is a unit vector in the $j$-direction, $j=x,y$ and

\begin{equation}\label{eq:mode_eigenfunc}
    e_{l,n,h}^{(j)}(\myvec{x},t)=\frac{w^{(j)}_{0,h}}{w^{(j)}_{h}(z-z_0,t)}e^{-r^2/(w_{h}^{(j)}(z-z_0,t))^2} H_{l}(\zeta_x^{(j)}(x,z,t)) H_{n}(\zeta_y^{(j)}(y,z,t))
\end{equation}

\noindent are the transverse eigenfunctions for polarization direction $j$ $r=\sqrt{x^2+y^2}$, $H_l$ and $H_n$ are the Hermite polynomials of order $l$ and $n$, respectively, $\zeta^{(j)}_x(x,z,t)=\sqrt{2}x/w_h^{(j)}(z-z_0,t)$,  $\zeta^{(j)}_y(y,z,t)=\sqrt{2}y/w_h^{(j)}(z-z_0,t)$ and $w^{(j)}_{0,h}$ and $w^{(j)}(z-z_0,t)$ are the beam radii in the waist at $z=z_0$ and at location $z$ along the optical axis. Note, $e_{l,n,h}^{(j)}$ and $w^{(j)}$ 
are dependent on time through the source dependent expansion (SDE) \cite{sprangle_1987}, as different parts of the optical pulse will experience different gain. Finally, the optical phases $\varphi^{(j)}_h(\myvec{x},t)$ are given by

\begin{equation}\label{eq:mode_phase}
   \varphi^{(j)}_h(\myvec{x},t)=h(k_0 z-\omega_0 t)+\alpha_h^{(j)}(z,t)\left( \frac{r}{w^{(j)}_{h}(z-z_0,t)} \right)^2, 
\end{equation}

\noindent where $k_0 =\omega_0/c$ is the vacuum wavenumber belonging to the carrier frequency $\omega_0$, $c$ is the speed of light \textit{in vacuo} and $\alpha_h^{(j)}(z,t)$ is the curvature of the wavefront for harmonic $h$. In this formulation we assume that the mode amplitudes, curvature of the phase fronts and radii of the modes are slowly varying functions of $z$ and $t$. The total optical power is given by 

\begin{equation}\label{eq:modes_power}
    P(z,t) = \sum_{\substack{l,n=0\\h=1}}^{\infty}P_{l,n,h}(z,t)=\frac{m_e^2 c^5}{8e^2}k_0^2\sum_{\substack{l,n=0\\h=1}}^{\infty}2^{l+n-1}l!n!\sum_{j=x,y}(w_{0,h}^{(j)})^2\left((\delta a_{l,n,h}^{(1,j)}(z,t))^2+(\delta a_{l,n,h}^{(2,j)}(z,t))^2 \right),
\end{equation}

\noindent where $\delta a_{l,n,h}^{(i,j)}(z,t)=e\delta A_{l,n,h}^{(i,j)}(z,t)/m_e c^2$ are the components of the normalized mode amplitudes, $i=1,2$, $j=x,y$ and $m_e$ is the electron rest mass.

The optical fields are driven by accelerated electrons as they co-propagate through the magnetic fields that are placed along the transport line. Of main interest here is the amplification of the optical field by the electrons within the undulators that can be placed in arbitrary configuration along the electron beam transport line. Within the slowly varying phase and amplitude approximation, the evolution of the normalized mode amplitudes $\delta a_{l,n,h}^{(i,j)}(z,t)$ is given by

\begin{equation}\label{eq:mode_evolution}
    \frac{d}{dz} \begin{pmatrix} \delta a_{l,n,h}^{(1,j)}(z,t) \\ \delta a_{l,n,h}^{(2,j)}(z,t)\end{pmatrix} + K_{l,n,h}^{(j)}(z)\begin{pmatrix} \delta a_{l,n,h}^{(2,j)}(z,t) \\ -\delta a_{l,n,h}^{(1,j)}(z,t)\end{pmatrix} = \begin{pmatrix} S_{l,n,h}^{(1,j)}(z,t) \\ S_{l,n,h}^{(2,j)}(z,t)\end{pmatrix},
\end{equation}

\noindent where $d/dz$ is the convective derivative and $K_{l,n,h}^{(j)}(z,t)$ is given by

\begin{equation}
    K_{l,n,h}^{(j)}(z,t)=(1+l+n)\left[ \frac{\alpha_h^{(j)}(z,t)}{w_h^{(j)}(z,t)}\frac{dw_h^{(j)}(z,t)}{dz} -\frac{1}{2}\frac{d\alpha_h^{(j)}(z,t)}{dz} - \frac{1+(\alpha_h^{(j)}(z,t))^2}{k_0 (w_h^{(j)}(z,t))^2}\right].
\end{equation}

\noindent The source terms $S_{l,n,h}^{(i,j)}(z,t)$ in Equation~\ref{eq:mode_evolution} are given by

\begin{equation}\label{eq:mode_source_terms}
    \begin{pmatrix} S_{l,n,h}^{(1,j)}(z,t) \\ S_{l,n,h}^{(2,j)}(z,t)\end{pmatrix} = \frac{1}{\pi}\frac{\omega_b^2}{k_0 c^2 }\frac{1}{2^{l+n-1}}\frac{1}{l!n!} \frac{1}{(w_{0,h}^{(j)})^2}\left< e_{l,n,h}^{(j)}(\myvec{x}) \frac{v_j(z,t)}{\abs{v_z(z,t)}}\begin{pmatrix} -\cos(\varphi_h^{(j)}(z,t)) \\ \sin(\varphi_h^{(j)}(z,t)) \end{pmatrix} \right>,
\end{equation}

\noindent where $\omega_b$ is the nominal plasma frequency, $v_j$ is the component of the electron velocity in direction $j$, $v_z$ is the longitudinal velocity component and $\left<(\dots )\right>$ is an average over the electron distribution given by

\begin{equation}\label{eq:mode_source_particle_average}
\begin{split}
    &\left<(\dots )\right>= \\ &\int_{0}^{2\pi}\frac{d\psi_0}{2\pi}\int_{1}^{\infty}\frac{d\gamma_0}{\sqrt{\pi/2}\Delta\gamma}\frac{e^{-(\gamma_0-\gamma_{\textrm{avg}})^2/ 2\Delta\gamma^2}}{1+\erf(\gamma_{\textrm{avg}}/\sqrt{2}\Delta\gamma)}\iint \frac{dx_0dy_0}{2\pi\sigma_r^2}\iint\frac{dp_{x0}dp_{y0}}{2\pi\sigma_p^2}e^{-r^2/2\sigma_r^2}e^{-p_{\perp 0}^2/2\sigma_p^2} (\dots).
\end{split}
\end{equation}

\noindent In Equation~\ref{eq:mode_source_particle_average}, $\gamma_{\textrm{avg}}$ and $\Delta\gamma$ are the relativistic factors corresponding to the initial average electron energy and energy spread, and $\sigma_r$ and $\sigma_p$ describe the width of the initial distribution function in transverse and momentum phase space, respectively.

To minimize the number of modes required to accurately describe the amplified optical field within the undulators, the so-called source dependent expansion \cite{sprangle_1987} is used. Within this approximation, the spot size and curvature of the eigenmodes for each of the polarization directions $j=x,y$ are allowed to evolve according to

\begin{equation} \label{eq:modes_sde_w}
    \frac{dw_h^{(j)}(z,t)}{dz}=\frac{2\alpha_h^{(j)}(z,t)}{hk_0 w_h^{(j)}(z,t)}-w_h^{(j)}(z,t)Y_h^{(j)}(z,t)
\end{equation}

\noindent and 

\begin{equation}\label{eq:modes_sde_alpha}
    \frac{1}{2}\frac{d\alpha_h^{(j)}(z,t)}{dz}=\frac{1+(\alpha_h^{(j)}(z,t))^2}{h k_0 (w_h^{(j)}(z,t))^2} -X_h^{(j)}(z,t) -\alpha_h^{(j)}(z,t)Y_h^{(j)}(z,t),
\end{equation}

\noindent where

\begin{equation}\label{eq:modes_sde_x}
\begin{split}
    &X_h^{(j)}(z,t)= \\
    &\frac{2}{(\delta a_{0,0,h}^{(j)}(z,t))^2}\left[\left(S_{2,0,h}^{(1,j)}(z,t) +S_{0,2,h}^{(1,j)}(z,t)\right)\delta a_{0,0,h}^{(1,j)}(z,t) - \left( S_{2,0,h}^{(2,j)}(z,t) +S_{0,2,h}^{(2,j)}(z,t) \right)\delta a_{0,0,h}^{(2,j)}(z,t)\right]
\end{split}
\end{equation}

\noindent and 

\begin{equation}\label{eq:modes_sde_y}
\begin{split}
    &Y_h^{(j)}(z,t)=\\
    &-\frac{2}{(\delta a_{0,0,h}^{(j)}(z,t))^2}\left[\left(S_{2,0,h}^{(1,j)}(z,t) +S_{0,2,h}^{(1,j)}(z,t)\right)\delta a_{0,0,h}^{(1,j)}(z,t) + \left( S_{2,0,h}^{(2,j)}(z,t) +S_{0,2,h}^{(2,j)}(z,t) \right)\delta a_{0,0,h}^{(2,j)}(z,t)\right].
\end{split}
\end{equation}

\noindent Furthermore, in Equations~\ref{eq:modes_sde_x} and \ref{eq:modes_sde_y}, $(a_{0,0,h}^{(j)}(z,t))^2=(a_{0,0,h}^{(1,j)}(z,t))^2+(a_{0,0,h}^{(2,j)}(z,t))^2$. Note that SDE (equations \ref{eq:modes_sde_w} to \ref{eq:modes_sde_y}) recovers vacuum diffraction when no electron beam is present [$X_h^{(j)}(z,t)=Y_h^{(j)}(z,t) \equiv 0$].

The source terms in Equation~\ref{eq:mode_evolution} depend on the evolution of the electron coordinates and velocities that are given by the Newton-Lorentz equations

\begin{equation}\label{eq:elec_dxdt}
    \frac{dx}{dz}=\frac{v_x}{v_z},
\end{equation}

\begin{equation}\label{eq:elec_dydt}
    \frac{dy}{dz}=\frac{v_y}{v_z},
\end{equation}

\begin{equation}\label{eq:elec_NL}
    \frac{d\myvec{p}}{dz}=-\frac{e}{v_z}\left[ \delta\myvec{E} +\frac{\myvec{v}}{c}\times\left(\myvec{B} + \delta\myvec{B} \right)\right],
\end{equation}

\noindent and the evolution of the ponderomotive phase $\psi$ given by

\begin{equation}\label{eq:elec_phase}
    \frac{d\psi}{dz}=k+k_u - \frac{\omega}{v_z}.
\end{equation}

\noindent Using the Coulomb gauge, the optical fields $\delta\myvec{E}=-\frac{1}{c}\frac{\partial\delta\myvec{A}}{\partial t}$ and $\delta\myvec{B}= \nabla\times\delta\myvec{A}$ are derived from the optical vector potential given by Equation~\ref{eq:field_decomposition}, while $\myvec{B}$ is the static magnetic field produced by the magnetic elements along the electron transport line. The model presented here ignores space-charge effects, however, this can be easily incorporated \cite{freund_2018}.

The dynamical equations for the particles and fields are integrated simultaneously using a 4th order Runge-Kutta algorithm. Hence, the number of equations in the simulation is $N_{\mathrm{equations}} = N_{\mathrm{slices}}[6 N_{\mathrm{particles}} + 4(N_{\mathrm{modes}} + N_{\mathrm{harmonics}})]$, where $N_{\mathrm{slices}}$ is the number of slices in the simulation, and $N_{\mathrm{particles}}$ is the number of particles in each slice, $N_{\mathrm{modes}}$ is the total number of modes in the fundamental and all the harmonics, and $N_{\mathrm{harmonics}}$ is the number of harmonics. The Runge-Kutta algorithm allows the step size to change so optimized step sizes can be used in each magnetic element or in the drift spaces while this imposes no limitation on the placement of components along the electron beam path.

The formulation self-consistently tracks the generation of the optical field with arbitrary polarizations depending on the undulator configuration. The polarization state of the output light, therefore, can be determined by calculation of the Stokes parameters \cite{born_1999}. Measurements of the polarization in FELs have been characterized by the Stokes parameters \cite{lee_2008, allaria_2014}.

Magnetic field elements, such as undulators, quadrupoles and dipoles, can be placed in arbitrary sequences to specify a variety of different transport lines, and the gap lengths between different undulators can be varied as well. Field configurations can be set up for single or multiple undulator segments and can contain quadrupoles placed between the segments or be superimposed on the lattice to create a FODO lattice. Magnetic elements can also be dipole chicanes to model optical klystron, high-gain harmonic generation (HGHG) or phase shifters. See Section~\ref{section:b-models} for analytical models for the various static magnetic field elements.

\subsection{Analytic models for the static magnetic fields} \label{section:b-models}
The various types of undulators are modeled using three-dimensional analytical representations. Two models are available for the planar undulator, representing a flat-pole face and a parabolic-pole face with weak two-plane focusing, respectively. The flat-pole-face undulator with the magnetic field oscillating in the $y$-plane is described by 

\begin{equation}\label{eq:mag_fpf}
    \myvec{B}_u(\myvec{x})=B_u(z)\left[\left(\sin(k_u z) - \frac{\cos(k_u z)}{k_u B_u(z)} \frac{dB_u}{dz} \right)\uvec{e}_y \cosh(k_u y) + \uvec{e}_z \sinh(k_u y)\cos(k_u z)\right],
\end{equation}

\noindent where $B_u$ is the amplitude of the undulator field, $k_u=2\pi/\lambda_u$, $\lambda_u$ is the undulator period and $B_u(z)$ describes the entrance and exits taper at the ends of an undulator segment. The field described in Equation~\ref{eq:mag_fpf} is both curl- and divergence free when the amplitude is constant.
The analytic entrance and exit taper function $B_u(z)$ is given by

\begin{equation}\label{eq:mag_taper}
    B_u(z)=\left\{ \begin{matrix} B_{u0} \sin^2\left(\frac{k_u z}{4N_{tr}} \right) & 0\leq z \leq N_{tr}\lambda_u\\ 
    B_{u0} & N_{tr}\lambda_u \leq z \leq L_{tr} \\
    B_{u0} \cos^2\left(\frac{k_u(L_{tr}-z)}{4N_{tr}}\right) & L_{tr}\leq z \leq L_u \end{matrix} \right.,
\end{equation}

\noindent where $B_{u0}$ is the undulator amplitude in the homogeneous part of the undulator, $L_{w}$ is the length of the undulator segment, $N_{tr}$ is the number of undulator periods in the transition region and $L_{tr}=L_{w}-N_{tr}\lambda_u$ is location of the start of the exit taper. The field in the taper regions has zero divergence, and the $z$-component of the curl also vanishes. The transverse components of the curl do not vanish, but are of the order of $1/(k_u B_u(z)) dB_u/dz$, which is usually small.   

A parabolic-pole-face undulator with the magnetic field orientation in the $y$-plane and weak focusing in the $x$- and $y$-planes is described by the analytic model

\begin{equation}\label{eq:mag_ppf}
    \myvec{B}_u(\myvec{x})=B_u(z)\left[\left(\cos(k_u z) - \frac{\sin(k_u z)}{k_u B_u(z)} \frac{dB_u}{dz} \right)\uvec{e}_\perp(x,y) -\sqrt{2} \uvec{e}_z \cosh\left(\frac{k_u x}{\sqrt{2}}\right)\sinh\left(\frac{k_u y}{\sqrt{2}}\right)\sin(k_u z) \right],
\end{equation}

\noindent where

\begin{equation}
    \uvec{e}_\perp(x,y)=\uvec{e}_x\sinh\left(\frac{k_u x}{\sqrt{2}}\right)\sinh\left(\frac{k_u y}{\sqrt{2}}\right) + \uvec{e}_y\cosh\left(\frac{k_u x}{\sqrt{2}}\right)\cosh\left(\frac{k_u y}{\sqrt{2}}\right)
\end{equation}

\noindent and the taper function $B_u(z)$ is again given by Equation~\ref{eq:mag_taper}. As in the case of the flat-pole-face model, Equation~\ref{eq:mag_ppf} is divergence free and the $z$-component of the curl also vanishes. Both, the flat-pole-face and parabolic-pole-face ideally produce linearly polarized light.

Circular polarized light is produced by helical undulators that are described by 

\begin{equation}\label{eq:mag_helical}
\begin{split}
    \myvec{B}_u(\myvec{x}) = & 2B_u(z)\left[\left(\cos(\chi) - \frac{\sin(\chi)}{k_u B_u(z)} \frac{dB_u}{dz} \right)I_{1}^{'}(k_u r)\uvec{e}_r - \right.\\
    &\left. \left(\sin(\chi) - \frac{\cos(\chi)}{k_u B_u(z)} \frac{dB_u}{dz} \right)\frac{I_{1}(k_u r)}{k_u r}\uvec{e}_{\theta} +I_1(k_u r) \sin(\chi)\uvec{e}_z \right]
\end{split}
\end{equation}

\noindent in cylindrical coordinates $r, \theta, z$, where $\chi=k_u z -\theta$, $I_1$ is the regular Bessel function of the first kind, the prime ($'$) indicates the derivative of the function with respect to its argument and $B_u(z)$ is again given by Equation~\ref{eq:mag_taper}. 

Finally, a magnetic field with varying degree of ellipticity is produced by an APPLE-II undulator which can be approximated by a superposition of two flat-pole-face undulators that are oriented perpendicularly to each other and one of which can be displaced with respect to the other along the axis of symmetry. As such, the field is represented by

\begin{equation}\label{eq:mag_apple2}
\begin{split}
    \myvec{B}_u(\myvec{x}) = B_u(z) & \Bigl[ 
    \left(\sin(k_u z+\phi) - \frac{\cos(k_u z+\phi)}{k_u B_u(z)} \frac{dB_u}{dz} \right)\cosh(k_u x)\uvec{e}_x  + \\
    & \left(\sin(k_u z) - \frac{\cos(k_u z)}{k_u B_u(z)} \frac{dB_u}{dz} \right)\cosh(k_u y)\uvec{e}_y + \\
    & \left(\sinh(k_u x)\cos(k_u z+\phi)+ \sinh(k_u y)\cos(k_u z)\right)\uvec{e}_z \Bigr],
\end{split}
\end{equation}

\noindent where $\phi$, $0\leq\phi\leq\pi$, represents the phase shift between the two linear undulators and as before, $B_u(z)$ is given by Equation~\ref{eq:mag_taper}. This model is valid near the axis the axis of symmetry. The ellipticity, $u_e$ of the APPLE-II undulator is given by 

\begin{equation}\label{eq:mag_apple2_elip}
    u_e = \frac{1-\abs{\cos(\phi)}}{1+\abs{\cos(\phi)}}.
\end{equation}

\noindent We note that the APPLE-II undulator configured as linear undulator has $u_e=0$ while $u_e=1$ for a helical configuration.

The remaining static magnetic field models are used to describe quadrupole magnets, using 

\begin{equation}\label{eq:mag_quad}
    \myvec{B}_Q(z)=B_Q(z)\left(y\uvec{e}_x+x\uvec{e}_y\right),
\end{equation}

\noindent where $B_Q(z)$ is the constant field gradient over some range $z_1\leq z\leq z_2$, and dipole magnets using a field model that is described by a constant field oriented perpendicularly to the axis of symmetry over some range $z_3\leq z\leq z_4$. Both, the field models for quadrupole and dipole magnets, have hard-edge field transitions and are curl- and divergence-free over the range where these are defined.

\subsection{Performance of the MINERVA FEL code} \label{section:c-validation}
To illustrate the performance of the MINERVA FEL code, we briefly compare the predictions of this code with experimental data from two SASE FEL experiments. We consider the “Sorgente Pulsata ed Amplificata di Radiazione Coerente” (SPARC) experiment, a SASE FEL located at ENEA Frascati \cite{giannessi_2011}, and the Linac Coherent Light Source (LCLS), which is a SASE FEL at the Stanford Linear Accelerator Center \cite{emma_2010}.

\subsubsection{The SPARC SASE FEL}
The experimental parameters of SPARC \cite{giannessi_2011} are as follows. The electron beam energy was 151.9 MeV, with a bunch charge of 450 pC, and a bunch width of 12.67 ps. The peak current was approximately 53 A for a parabolic temporal bunch profile. The $x$- and $y$-emittances were 2.5 mm-mrad and 2.9 mm-mrad respectively, and the rms energy spread was 0.02 percent. There were six undulators each of which was 77 periods in length with a period of 2.8 cm and an amplitude of 7.88 kG. Each undulator was modeled using Equation~\ref{eq:mag_fpf} with one period for the entrance up-taper and another for the exit down-taper. Further, in the simulation eight undulators were used to show saturation of the system. The gap between the undulators was 0.4 m in length and the quadrupoles (0.053 m in length with a field gradient of 0.9 kG/cm) forming a strong focusing lattice were located 0.105 m downstream from the exit of the previous undulator. Note that the quadrupole orientations were fixed and did not alternate. The electron beam was matched into the undulator/focusing lattice. The resonance occurred at a wavelength of 491.5 nm. In the experiment, the pulse energies were measured in the gaps between the undulator segments.

\begin{figure}[t] 
\centering
\includegraphics[width=8cm]{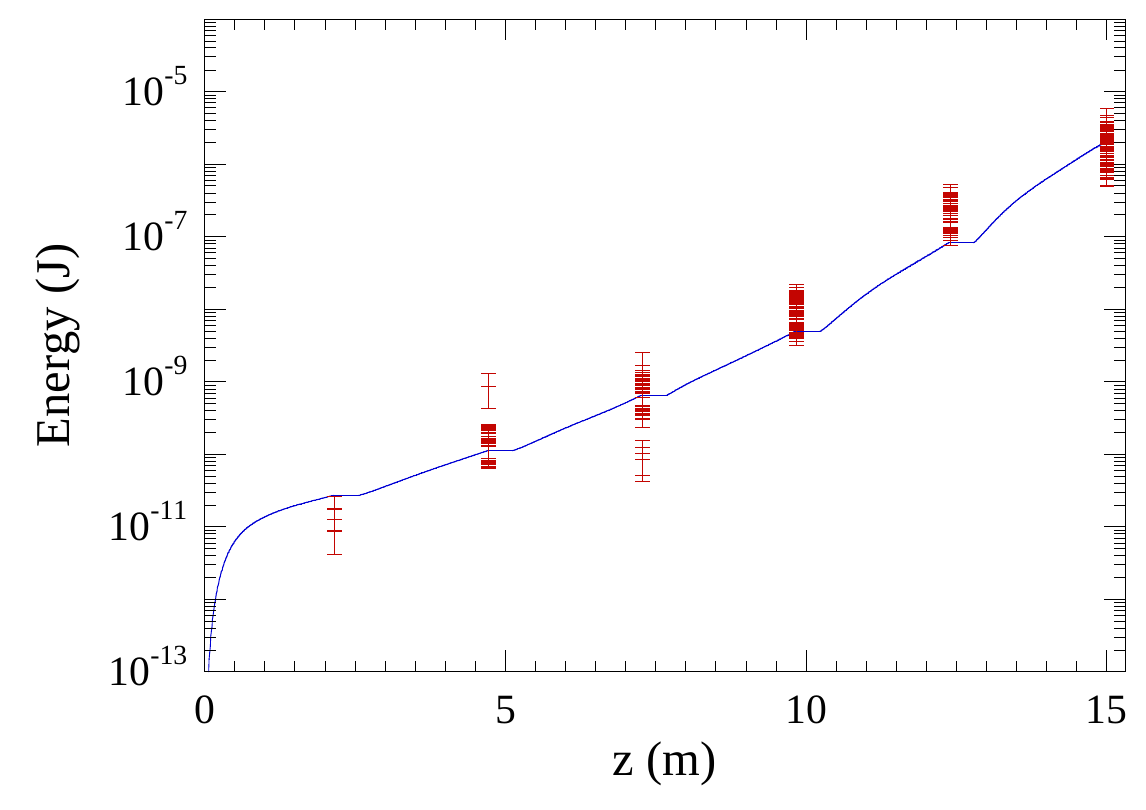}
\caption{\label{fig:sparc_E_z}Simulated (blue line) and measured (red crosses) pulse energy as a function of the propagation distance $z$ for the SPARC experiment. The different points corresponds to individual measurements and the error bar indicates the error in each of the measurements. The simulation result is an average over 20 noise realizations.}
\end{figure} 
\begin{figure}[t] 
\centering
\includegraphics[width=8cm]{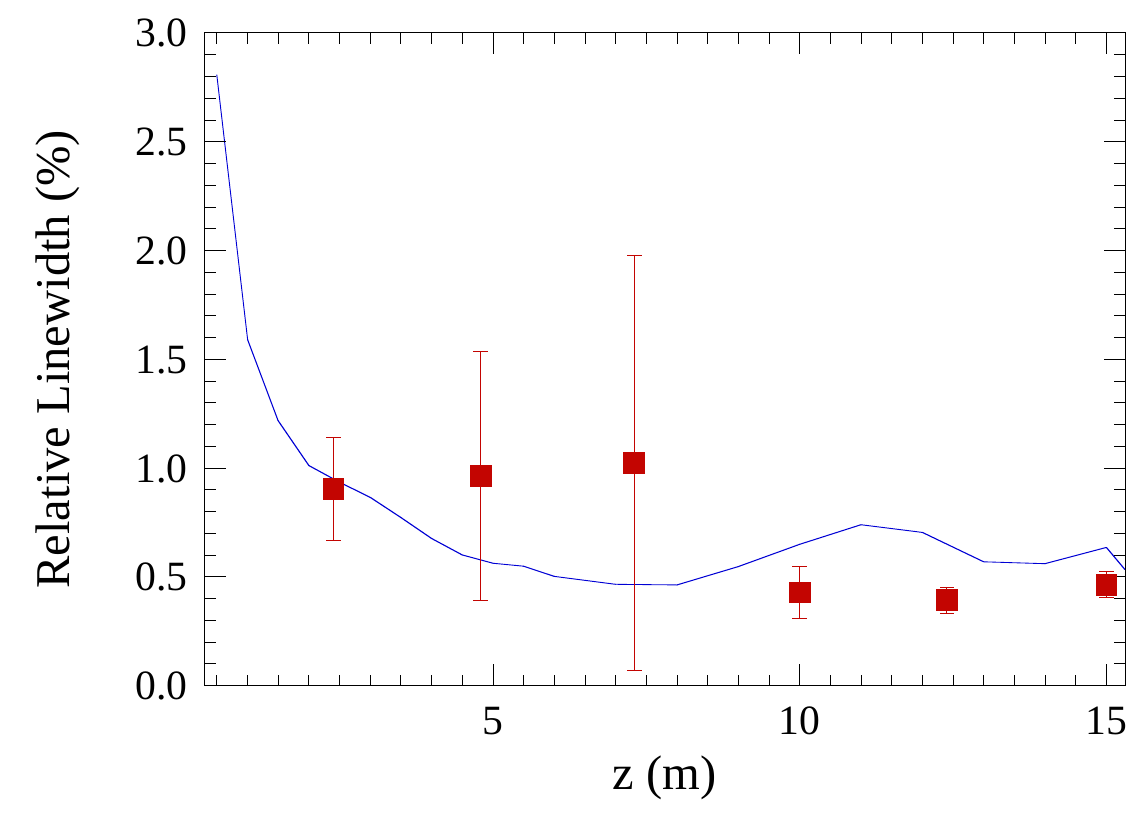}
\caption{\label{fig:sparc_df}Simulated (blue line) and measured (red data points) relative linewidth as a function of the propagation distance $z$ for the SPARC experiment. The error bar indicates the error in the measurement. The simulation result is an average over 20 noise realizations.}
\end{figure} 

A comparison of the pulse energy as found in the simulation (blue line) and from the experiment (red markers) is shown in Figure~\ref{fig:sparc_E_z}, where the simulation result is averaged over 20 simulation runs with different noise seeds. This yields convergence to better than 5 percent. Energy conservation in the simulation is maintained to within better than one part in $10^4$. Each marker represents a single measurement that is repeated several times, while the error bar indicates the error in the measurement.
The experimental data is courtesy of L. Giannessi. Agreement between the simulation and the measured performance is excellent. 

A comparison between the evolution of the relative linewidth as determined from simulation (blue line) and by measurement (red markers, data courtesy of L. Giannessi) is shown in Figure~\ref{fig:sparc_df}. Agreement between the simulation and the measured linewidth is within about 35 percent after 15 m. As shown, the predicted linewidths are in substantial agreement with the measurements.

\subsubsection{The LCLS SASE FEL}
To illustrate the performance of MINERVA at much shorter wavelengths, we also briefly compare with experimental data from LCLS \cite{emma_2010}, which is a SASE FEL user facility that became operational in 2009 operating at a 1.5 \AA\   wavelength. 

To operate at 1.5 \AA, LCLS employs a 13.64 GeV/250 pC electron beam with a flat-top temporal pulse shape of 83 fs duration. The normalized emittance ($x$ and $y$) is 0.4 mm-mrad and the rms energy spread is 0.01 percent. The undulator line consisted of 33 segments with a period of 3.0 cm and a length of 113 periods. In the simulation each segment is modeled using Equation~\ref{eq:mag_fpf} with one period each in entry and exit tapers. A mild down-taper in field amplitude of -0.0016 kG/segment starting with the first segment (with an amplitude of 12.4947 kG and $K_{\textrm{rms}}$ = 2.4748) and continuing from segment to segment was used. This is the so-called gain taper. The electron beam was matched into a FODO lattice consisting of 32 quadrupoles each having a field gradient of 4.054 kG/cm and a length of 7.4 cm. Each quadrupole was placed a distance of 3.96 cm downstream from the end of the preceding undulator segment, and the gap lengths between the undulators followed a repetitive sequence of short (0.48 m)-short (0.48 m)-long (0.908 m).

\begin{figure}[t] 
\centering
\includegraphics[width=8cm]{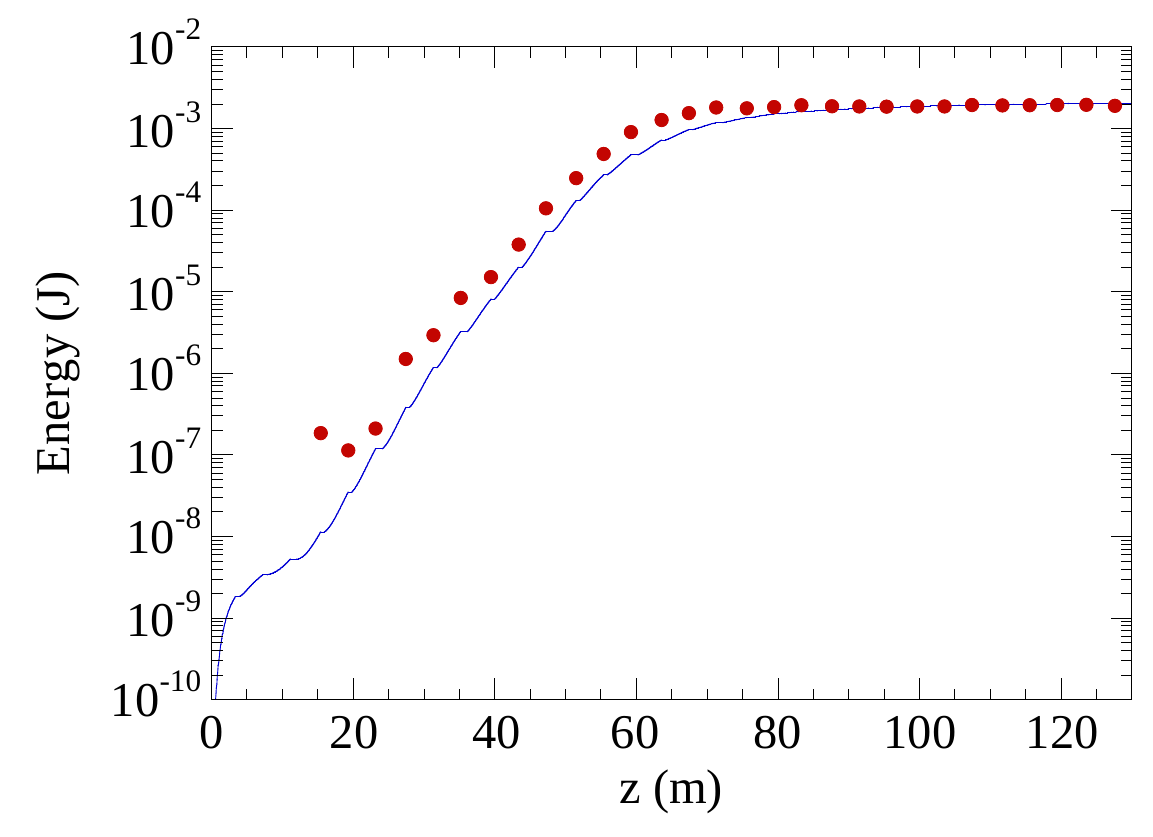}
\caption{\label{fig:lcls_E_z}Simulated (blue line) and measured (red circles) optical pulse energy as a function of the propagation distance $z$ for the LCLS experiment. The simulation result is an average over 25 noise realizations.}
\end{figure} 
The LCLS produces pulses of about 1.89 mJ at the end of the undulator line \cite{emma_2010}, and saturation is found after about 65 – 75 m along the undulator line. A comparison between the measured pulse energies (red circles), as obtained by giving the electrons a kick to disrupt the FEL process, and the simulation (blue) is shown in Figure~\ref{fig:lcls_E_z}. The experimental data is courtesy of P. Emma and H.-D. Nuhn at SLAC, and the simulation results represent an average over an ensemble of 25 runs performed with different noise seeds. As shown in Figure~\ref{fig:lcls_E_z}, the simulations are in good agreement with the measurements in the exponential growth region with close agreement for the gain length. The simulation exhibits saturation at the same distance as the experiment in the range of 65 – 75 m at a pulse energy of 1.5 mJ. After saturation, in view of the gain taper, the pulse energy grows more slowly to about 2.02 mJ at the end of the undulator line, which is approximately 8 percent higher than the observed pulse energy.

The agreement between simulation and experiment for the pulse energy is poorer during the early stages of the interaction. This may be due to a variety of reasons. On the experimental side, as the pulse energy grows by 5 to 6 orders of magnitude from the initial shot noise to saturation, it is difficult to calibrate the detectors for the low pulse energies at the early stages of the interaction. Also, while kicking the electrons provides a fast measurement method, it is accompanied by a larger background signal and the possibility of restarting the FEL process downstream the undulator line when the kick is performed at the beginning of the undulator line \cite{emma_2010}. On the simulation side, there may be some inaccuracies in the shot noise algorithm or the finite number of optical modes used that underestimates the initial noise level.

\subsection{The optical propagation code OPC}
When modeling FEL oscillators, the electron-light interaction within the undulator segments needs to be modeled as well as the propagation of the light through the remainder of the oscillator. The light propagators internal to FEL gain codes, including that in MINERVA, are typically not very efficient when considering free-space propagation over large distances and interaction with various optical elements. 

The wavefront propagator\cite{karssenberg_2006, opc_2021} is based upon the scalar paraxial Helmoltz equation \cite{siegman_1986}

\begin{equation}\label{eq:opc_helmholtz}
    \frac{\partial^2 u}{\partial x^2} +\frac{\partial^2 u}{\partial y^2} -2 i k_0 \frac{\partial u}{\partial z} = 0,
\end{equation}

\noindent where $u$ is the complex scalar wave amplitude that describes the transverse profile of a time-harmonic optical beam, $k_0=2\pi/\lambda_0$, and $\lambda_0$ is the free-space wavelength. By applying a two-dimensional spatial Fourier transform, Equation~\ref{eq:opc_helmholtz} transforms into

\begin{equation}\label{eq:opc_fourier_helmholtz}
    (k_x^2 + k_y^2)\Tilde{u} -2 i k_0 \frac{\partial \Tilde{u}}{\partial z} = 0,
\end{equation}

\noindent where

\begin{equation}\label{eq:opc_fourier_def}
    \Tilde{u}(k_x,k_y,z) \equiv \int\limits_{-\infty}^{\infty}\int\limits_{-\infty}^{\infty}dx dy\,u(x,y,z)e^{-i(k_x x+ k_y y)}
\end{equation}

\noindent is the two-dimensional Fourier transform of $u(x,y,z)$. Given a known profile $u_0(x,y)=u(x,y,0)$ at $z=0$, \textit{e.g.}, at the exit of the undulator line, the analytic solution in reciprocal space to Equation~\ref{eq:opc_fourier_helmholtz} is given by 

\begin{equation}\label{eq:opc_reciprocal_sol}
    \Tilde{u}(k_x,k_y,z)=\Tilde{u}_0(k_x,k_y)e^{iz\lambda_0(k_x^2 + k_y^2)/\pi}.
\end{equation}

\noindent The solution in normal space at location $z$ is found by the inverse spatial Fourier transform of Equation~\ref{eq:opc_reciprocal_sol},

\begin{equation}\label{eq:opc_real_sol}
    u(x,y,z) \equiv \frac{1}{4\pi^2}\int\limits_{-\infty}^{\infty}\int\limits_{-\infty}^{\infty}dk_x dk_y\, \Tilde{u}(k_x,k_y,z)e^{i(k_x x +k_y y)}).
\end{equation}

\noindent This wavefront propagation is known as spectral propagation and is very suitable for propagation over short and large distances. However, when propagating over long distances, care has to be taken that the complete optical field remains contained within the transverse domain of the two-dimensional spatial Fourier transform (Equation~\ref{eq:opc_fourier_def}), \textit{e.g.}, as a result of a fast Fourier transform implementation, to avoid artificial reflections from the edges of this domain.

By substituting Equation~\ref{eq:opc_reciprocal_sol} into Equation~\ref{eq:opc_real_sol}, using the initial profile $u_0(x,y)$ and Equation~\ref{eq:opc_fourier_def}, changing the order of integration and performing the integration over $k_x$ and $k_y$ we obtain the Fresnel diffraction integral \cite{haus_1984}

\begin{equation}\label{eq:opc_fresnel}
    u(x,y,z)=-\frac{i}{\lambda_0 z}\int\limits_{-\infty}^{\infty}\int\limits_{-\infty}^{\infty}d\xi d\eta\, u_0(\xi,\eta)e^{i \pi[(x-\xi)^2+(y-\eta)^2]/\lambda_0 z}.
\end{equation}

\noindent Note that the Fresnel diffraction integral can be derived using various methods, including the free-space Green's function \cite{goodman_1996} or by applying the paraxial approximation to the spherical wavelets in Huygens' integral \cite{siegman_1986}. It can also be shown that propagation through a cascaded set of paraxial optical components described by an overall ray-optical ABCD matrix can be described by a single Huygens' integral given by \cite{siegman_1986}

\begin{equation}\label{eq:opc_huygens_ABCD}
    u(x,y,z)=e^{-ik_0 z}\int\limits_{-\infty}^{\infty}\int\limits_{-\infty}^{\infty}d\xi d\eta\, K(x,y,\xi,\eta)u_0(\xi,\eta),
\end{equation}

\noindent where Huygens' kernel $K$ is given by

\begin{equation}\label{eq:opc_huygens_kernel}
    K(x,y,\xi,\eta)\equiv \frac{i}{\lambda_0\sqrt{B_x B_y}}e^{-i\frac{\pi}{B_x\lambda_0}(A_x\xi^2-2\xi x+D_x x^2)}
    e^{-i\frac{\pi}{B_y\lambda_0}(A_y\eta^2-2\eta y+D_y y^2)},
\end{equation}

\noindent and $A_i, B_i, C_i$ and $D_i$ are the components of the system ABCD matrix for $i=x,y$. The ABCD matrix can describe real or complex orthogonal paraxial optical systems that may contain astigmatism \cite{siegman_1986}. Note, in the presence of apertures, the optical field must be propagated from aperture to aperture since apertures cannot be included in the ABCD matrix.

An efficient calculation of the integrals appearing in the spectral propagation method and Fresnel's diffraction integrals relies on fast Fourier transforms. By applying a transformation to Equation~\ref{eq:opc_huygens_ABCD} it can be put into a form that also allows the use of fast Fourier transforms to efficiently evaluate the integrals. We define $M_x=a_2/a_1$, where $a_1$ and $a_2$ are measures for the size of the optical field, such that the optical field becomes negligible for $\abs{x}\geq a_{1,2}$ at the input and output plane, respectively. Similarly, $M_y=a_3/a_4$ where $a_3$ and $a_4$ are measures for the size of the optical field, such that the optical field becomes negligible for $\abs{y}\geq a_{3,4}$ in the input and output plane, respectively.  Using the transformation $x'=a_1 x, \xi'=a_2\xi, y'=a_3 y, \eta' = a_4 \eta,$

\begin{equation}
    v_0(\xi',\eta')=\sqrt{a_1 a_3} u_0(\xi,\eta)e^{-i\pi\frac{(A_x-M_x)\xi^2}{B_x\lambda_0}}e^{-i\pi\frac{(A_y-M_y)\eta^2}{B_y\lambda_0}},
\end{equation}

\noindent and

\begin{equation}
    v(x',y')=\sqrt{a_2 a_4} u(x,y)e^{i\pi\frac{(D_x-M_x^{-1})x^2}{B_x\lambda_0}}e^{i\pi\frac{(D_y-M_y^{-1})y^2}{B_y\lambda_0}}.
\end{equation}

\noindent Equation~\ref{eq:opc_huygens_ABCD} can be transformed into

\begin{equation} \label{eq:opc_fresnel_modified}
    v(x',y')=i\sqrt{N_{c,x}N_{c,y}}\int\limits_{-1}^{1}\int\limits_{-1}^{1}d\xi' d\eta' \, K(x',y',\xi',\eta')v_0(\xi',\eta')
\end{equation}

\noindent with the kernel $K$ given by

\begin{equation}
    K(x',y',\xi',\eta')=e^{i\pi N_{c,x}(x'-\xi')^2}e^{i\pi N_{c,y}(y'-\eta')^2}.
\end{equation}

\noindent In Equation~\ref{eq:opc_fresnel_modified}, $N_{c,x}$ and $N_{c,y}$ are equivalent collimated Fresnel numbers \cite{siegman_1986} defined as 

\begin{equation}\label{eq:opc_fresnel_numbers}
    N_{c,x} = \frac{M_x a_1^2}{B_x\lambda_0},\; N_{c,y} = \frac{M_y a_3^2}{B_y\lambda_0}.
\end{equation}

\noindent We shall refer to Equation~\ref{eq:opc_fresnel_modified} as the modified Fresnel diffraction integral and this integral can also be efficiently evaluated using fast Fourier transforms with the added benefit that independent magnification factors in the $x$- and $y$-directions can be applied in going from the input to the output plane. This means that the output plane can grow or shrink in size with the optical beam when it expands or contracts in propagating from the input to the output plane that are defined by the system ABCD matrix used in the propagation.

\subsubsection{optical elements}
To allow modeling of FEL oscillators, the optical field needs to interact with various types of optical elements, such as lenses, mirrors or diaphragms. Diaphragms are implemented as hard edge apertures that reduce the optical field to zero outside the aperture. Various apodization functions are available that can be applied to the edge of the aperture. Lenses and mirrors are implemented as elements that apply a phase shift to the optical field. This is done by multiplying the optical field by $e^{-iq(x,y)}$ where $q(x,y)$ is the local phase shift in the transverse plane. For example, a thin lens is modeled as $q(x,y)=k_0(x^2 +y^2)/2f$ with $f$ the focal strength of the lens. Mirrors are modeled as thin lenses using $f=R/2$ where $R$ is the radius of curvature of the mirror. More complicated optical elements as thick lenses or a combination of lenses can be implemented by determining the overall ABCD matrix and using the modified Fresnel diffraction integral (Equation~\ref{eq:opc_fresnel_modified}). OPC also allows for more complicated phase masks that can be created using Zernike polynomials \cite{born_1999}. These can be used to implement not only various type of mirror and lens aberrations, but also mirror distortion resulting from thermal loading \cite{vanderslot_2007, vanderslot_2008} (see Section~\ref{section:mirrors} for more detail). In the latter case OPC can set a scaling factor for the aberrations depending on the (average) optical power loading of the mirror.

Typically, it is assumed that the FEL gain bandwidth is sufficiently small that dispersive effects within the optical elements can be neglected. However, this is not the case when crystal Bragg mirrors are used, \textit{e.g.}, to model oscillator configurations at x-ray wavelengths. For such mirrors, the angle of reflection and mirror loss strongly depend on the type of crystal used, its orientation and the x-ray photon energy \cite{shvydko_2004}. To properly model the reflection of Bragg mirrors, the incident optical field needs to be Fourier transformed into the frequency domain to handle the different photon energies in the optical field and a two-dimensional spatial Fourier transform of the incident field is needed to deal with the various angles of incidence on the mirror. 

Several optical elements can be combined to form a more complex optical component, \textit{e.g.}, by combining a mirror with a diaphragm element, extraction of radiation from a resonator through a hole in one of the mirrors can be modeled. Another example is to use an external finite element program to simulate mirror surface distortion due to thermal loading of the mirror. This surface distortion can then be converted to a phase mask, the amplitude of which can be dynamically scaled at each roundtrip to model mirror distortion during the start-up of the oscillator. 

Finally, the collection of wavefront propagators together with the optical components and interfaces with several FEL gain codes is known as the optical propagation code (OPC). As OPC is specifically designed to work several FEL gain codes to model FEL oscillators, some of the optical components allows forking of the optical propagation path. For example, at the mirror used for coupling the light out of the resonator, the tracking of the light within the resonator can temporarily be suspended to propagate the light outside the resonator to some diagnostic point where several optical properties of the light can be obtained. Afterwards, the propagation of the light within the resonator can be continued. This allows simultaneous monitoring of the light properties at some external diagnostic point as well as the build-up of the light within the resonator.

\subsubsection{Modeling mirrors} \label{section:mirrors}
The default optical components used by OPC are ideal thin lenses and spherical mirrors. However, by using amplitude and phase masks, additional optical components can be added, in particular more realistic lenses and mirrors. Here, we limit ourselves to phase masks that are generated using the circle polynomials $R^{|m|}_n$of Zernike \cite{born_1999}. These polynomials are used to generate a phase difference $d\theta$ defined on a transverse plane that is applied to the optical field as

\begin{equation}
    d\theta=D_{mn} R^{|m|}_n(\rho)\times \left\{\begin{matrix}\cos{(m\phi)} & m\le0\\ \sin{(m\phi)}&m<0\end{matrix}\right.,
\end{equation}

\noindent where $\rho$ is a scaled radial distance, $\rho=\sqrt{x^2+y^2}/\rho_c$ with $\rho_c$ some characteristic length scale, $\phi$ is the angle $\tan{^{-1}(y/x)}$ and $D_{mn}$ is the amplitude of the polynomial. The indices $m,n$ describe the type of aberration, for example $m=0$ and $n=4$ corresponds to spherical aberration and $m=1$ and $n=3$ to coma \cite{born_1999}. Therefore, adding a phase mask and apply it at the location of a lens of mirror adds the associated aberrations to create a more realistic model of the component.

It is also possible to use these Zernike polynomials to generate new optical components. For example adding the polynomials $m,n=2,2$ with $m,n=0,2$ with appropriate amplitudes generates a cylindrical lens of a certain focal strength. This can be used to correctly model the performance of a spherical mirror with a non-zero angle of incidence, \textit{e.g.}, when used in a ring resonator, to model for the different foci in tangential and sagittal direction. 

Another application is to use the Zernike polynomials to model mirror surface distortions, in particular due to thermal loading of the mirror. Here, we assume that round mirror is uniformly cooled at its circumference and that the heat loading is axi-symmetric. A finite element program can be used to calculate the steady state deviation $\delta z(r/r_0)$, where $r_0$ is some characteristic length, that results from the heat generation in the mirror due to some absorbed power $P_{\textrm{abs}}$. A fit using $R_n^0(r/r_0)$ Zernike polynomials with $n$ taken even is then used to determine the amplitudes to be used in generating the phase mask. To convert the mirror distortion $\delta z$ into a localized change in phase for the optical field we set

\begin{equation} \label{eq:phase_heating}
    d\theta (r/r_0) = - \frac{4\pi\delta z(r/r_0)}{\lambda},
\end{equation}

\noindent where the minus sign is required to comply with the phase advance used in OPC and the phase change is proportional to twice the mirror displacement. It is reasonable to assume that a stable transverse field distribution is established after starting from noise when the intracavity optical starts to heat the mirror. Therefore, the phase distortion due to mirror heating can be obtained by scaling Equation \ref{eq:phase_heating} with a factor $P_i(t)\tau/(P_{\textrm{abs}}T_{\textrm{rep}})$ with $P_{\textrm{abs}}$ the absorbed power used to calculate $\delta z$, $P_i$ the instantaneous absorbed power, $\tau$ the duration of the optical pulse and $T_{\textrm{rep}}$ the repetition time for the optical pulses.

\section{Some considerations for optical resonators used in free-electron lasers} \label{section:opt_res}
Most optical resonators used in laser oscillators consists of two spherical mirrors separated by a distance $L_{\textrm{cav}}$, as schematically shown in Figure~\ref{fig:fel_osc}. When the radius of curvature $R_c$ of the mirrors is equal, the resonator is called concentric when $L_{\textrm{cav}}=2R_c$ and confocal when $L_{\textrm{cav}}=R_c$. Other configurations with unequal radii of curvature of different distances between the mirrors are known as generalized resonators.

\begin{figure}[t] 
\centering
\includegraphics[width=10cm]{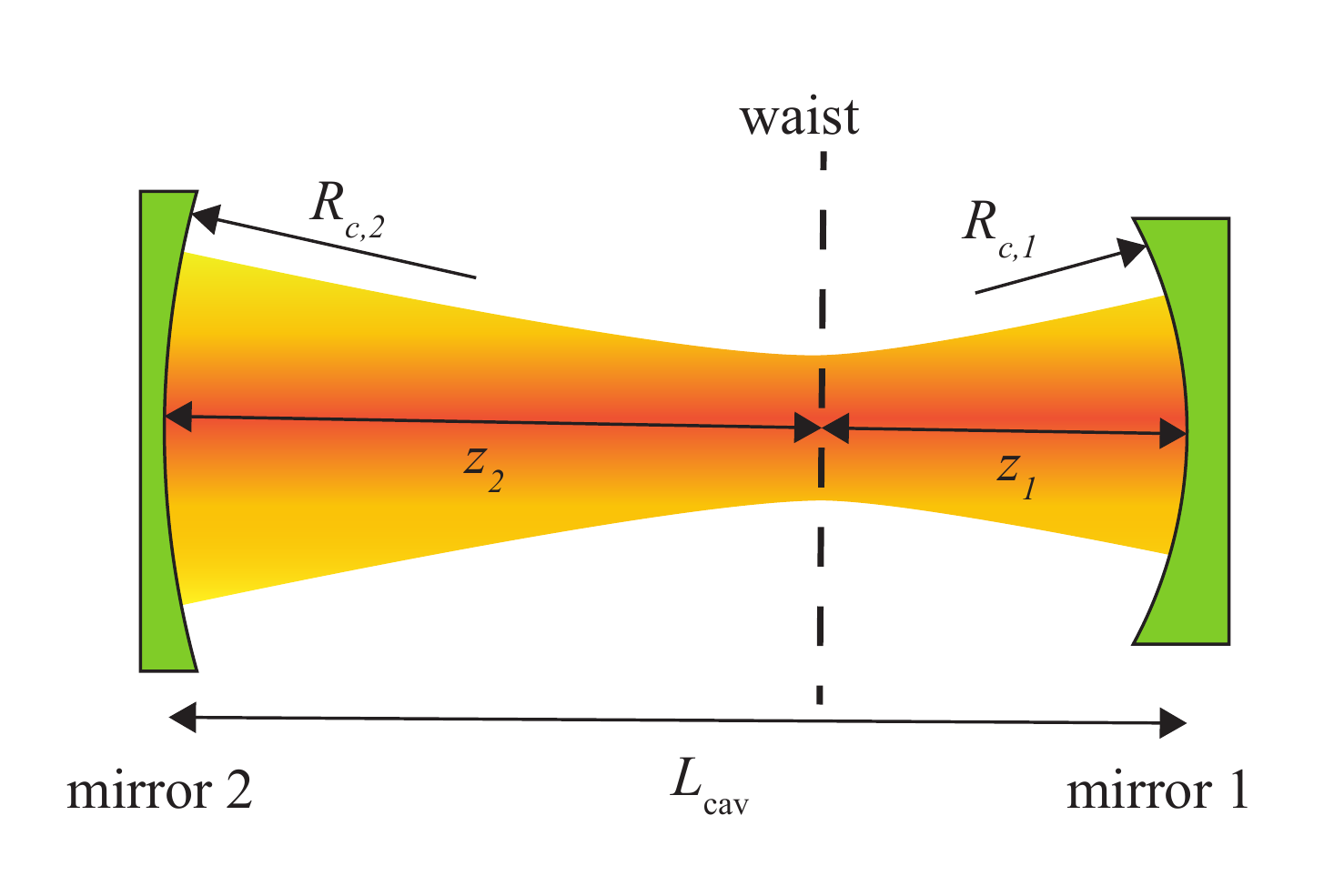}
\caption{\label{fig:gen_res}Schematic representation of a generalized two-mirror resonator, defining some of its parameters.}
\end{figure}  

Most of the low-gain FEL oscillators employ stable concentric resonators to ensure that the optical pulse train does not “walk” out of the cavity. Consider a generalized resonator with a cavity length $L_{\textrm{cav}}$ as shown in Figure~\ref{fig:gen_res} which shows the radii of curvature, $R_{c,i} (i=1,2)$, for the two mirrors as well as the distances $z_i$ from the mode waist to each of the mirrors. The transverse optical modes of the cold resonator, \textit{i.e.}, without a gain medium (electron beam) present, can be described as either Gauss-Hermite  \textit{cf.} Equations~\ref{eq:field_decomposition}-\ref{eq:mode_phase} or Gauss-Laguerre modes. 

Defining 

\begin{equation}
   g_i=1-\frac{L_{\textrm{cav}}}{R_{c,i}}
\end{equation}

\noindent then the Rayleigh range $z_R$ for the Gaussian eigenmodes in the cavity is given by

\begin{equation}
    z_{R}=  L_{\textrm{cav}} \sqrt{\frac{g_1g_2(1-g_1g_2)}{g_1+g_2-2g_1g_2}}
\end{equation}

\noindent Since the Rayleigh range must be positive, we must have $0\leq g_1g_2 \leq 1$, 
which is the condition for resonator stability \cite{siegman_1986}.
This means that Gaussian beams can only be eigenmodes of stable resonators.
The stability condition implies that

\begin{equation}\label{eq:stability}
    L_{\textrm{cav}}<R_1+R_2.
\end{equation}

\noindent We also note that the waist of the Gaussian mode is at a distance

\begin{equation}
    z_1 = L_{\textrm{cav}} \frac{g_2(1-g_1)}{g_1+g_2-2g_1g_2}
\end{equation}

\noindent from mirror 1 and at a distance $z_2=L_{\textrm{cav}}-z_1$ from mirror 2. Note that the sign convention used for the ray-optic ABCD formalism \cite{siegman_1986} applies here as well, meaning that $R_c>0$ for a concave mirror and $R_c<0$ for a convex mirror and that the distances $z_i$ have a sign.

The Gauss-Hermite transverse modes as described by Equations~\ref{eq:field_decomposition}-\ref{eq:mode_phase} are eigenmodes of an empty generalized resonator having a particular Rayleigh range and location of the waist defined by the resonator. Whenever the resonator contains internal apertures clipping the optical field, partial waveguiding, or if the light is extracted via a hole, a single transverse Gaussian mode can no longer be an eigenmode of the resonator. The Fox-Li method \cite{fox_1961} can be used to find the eigenmodes of such resonators \cite{vanderslot_2013}.

The resonant frequencies, or longitudinal modes, of the resonator are those frequencies for which the total roundtrip phase is an integer multiple of $2\pi$. As a result, the resonant frequencies differ for the different transverse optical modes \cite{siegman_1986}. The frequency spacing between two subsequent resonances of the same transverse mode is known as the free-spectral range $\Delta\nu_{\textrm{FSR}}$. FEL resonators typically have a $\Delta\nu_{\textrm{FSR}}$ that is much smaller than the gain bandwidth due to the large mirror separation used. Consequently, most FEL oscillators are operating on multiple longitudinal modes. 

As shown in Figure~\ref{fig:fel_osc}, the resonator may contain more than one optical pulse if the roundtrip time in the resonator is larger than the time between subsequent electron bunches. The resonator design must ensure synchronism between the optical pulses and the electron bunches and optimal performance is typically found near the synchronous or zero-detuning cavity length where the incoming electron bunch coincides with the returning optical pulse train. This zero-detuning cavity length, $L_0$, is given by

\begin{equation}\label{eq:oscillator rep rate}
    \frac{M}{f_{\textrm{rep}}}=\frac{2L_{\textrm{0}}-L_u}{c}+\frac{L_u}{v_g},
\end{equation}

\noindent where $M$ is the number of simultaneous electron bunches in the optical cavity, $f_{\textrm{rep}}$ is the repetition rate of the electron bunches, $L_u$ is the undulator length and $v_g$ is the group velocity of the light within the undulator. Solving for $L_0$ gives 

\begin{equation}\label{eq:zero_detuning_length}
    L_{\textrm{0}}=\frac{cM}{2f_{\textrm{rep}}}-\frac{L_u}{2}\frac{c}{v_g}\left(1-\frac{v_g}{c}\right),
\end{equation}

\noindent which reduces to the well-known expression 

\begin{equation}\label{eq:zero_detuning_low_gain}
    L_0=\frac{cM}{2f_{\textrm{rep}}},
\end{equation}

\noindent for low gain oscillators where $v_g \approx c$. As will be discussed in the section~\ref{section:hg_lq} on high-gain/low-$Q$ oscillators, exponential gain in the undulator results in a reduction in the group velocity which has the effect of shortening the zero-detuning cavity length compared to that for low-gain/high-$Q$ oscillators (Equation~\ref{eq:zero_detuning_low_gain}).

In general, the dynamics in an oscillator are set by the interplay between the instantaneous gain and loss, where in a free-electron laser the gain is provided by the electrons streaming through the undulator and loss is due to out-coupling of the radiation and other loss mechanisms that may be present within the resonator. The total loss of the resonator determines the quality factor $Q$ of the resonator \cite{siegman_1986}. To obtain laser oscillation, the roundtrip small-signal gain has to be higher than the loss per roundtrip. The so-called threshold gain is the roundtrip small-signal gain needed to just balance the loss per roundtrip. When the light builds up inside the resonator, the gain is reduced and a stationary state is obtained when the saturated gain equals the loss per roundtrip. Note that this is the same condition as for threshold, \textit{i.e}, the saturated gain is equal to the threshold gain.  If we denote the output power on the $n^{\textrm{th}}$ pass as $P_n$, then the power after the $(n+1)^{\textrm{th}}$ pass is $P_{n+1} = (1 - L)(1 + G)P_n$. Equilibrium is characterized by $P_{n+1} = P_n$; at which point $G = L/(1 - L)$. 

In FELs where the bunch charge/peak current is relatively small and the gain cannot reach the exponential regime in a single pass through the undulator, the resonator $Q$ must be high enough that the oscillator can lase. In such cases, most of the energy extracted from the electrons remains circulating in the resonator and a small fraction of the power is out-coupled. This implies that the loss, which is often dominated by the fraction of radiation that is coupled out of the resonator, but also includes losses at the mirrors or within the resonator, can impose difficulties if mirror absorption is large enough to result in excessive heating or degradation of the reflectivity. In such cases, the mirrors must be cooled or otherwise protected. For example, storage ring FELs have a low small-signal gain and thus require high-quality mirrors that are susceptible to the higher-harmonics generated. Short-wavelength x-ray FEL oscillators also have a low small-signal gain and the available x-ray mirror technology is pushed to its limits to produce mirrors with sufficiently low loss. High average power operation may require cryogenic cooling of the mirrors to prevent mirror distortion due to thermal loading.

The efficiency of a low-gain/high-$Q$ oscillator is predicted to be $\eta$ = 1/2.4$N_u$~\cite{freund_2018}, where $N_u$ denotes the number of uniform periods in the undulator; hence, high efficiency requires relatively short undulators. Because of this, oscillator design is a balance between having a sufficiently long undulator for the gain to exceed the losses, set by the quality $Q$ of the resonator, but not so long that the efficiency is negatively impacted.

\section{Low-Gain/High-\texorpdfstring{$Q$}{Q} Oscillators} \label{section:lg_hq}
In this section, we describe simulations of two low-gain/high-\texorpdfstring{$Q$}{Q} infrared FEL oscillator experiments conducted at JLab: the IR-Demo \cite{benson_1999,neil_2000,neil_2000a} and the 10-kW Upgrade \cite{neil_2006}. Both of these experiments were conceived as demonstrations for high average power infrared FELs based on energy recovery linacs. 

\subsection{The IR-Demo Experiment at JLab} \label{section:irdemo}
The IR Demo experiment \cite{benson_1999,neil_2000,neil_2000a} was based on a superconducting energy recovery linac at JLab that produced 0.4 (± 0.1) ps rms electron bunches with energies of about 38 MeV bunch charges of 60 pC (60 A peak current) at a repetition rate of 18.7 MHz corresponding to an average current of 1. 2 mA. The transverse emittance of the bunches was 7.5 (±1.5) mm-mrad (rms) and the rms energy spread was about 0.25 percent. The experiment uses a 42-period flat-pole-face planar undulator with a period of 2.7 cm and an on-axis amplitude of 5.56 kG yielding an undulator strength parameter of $K_{\textrm{rms}}$ = 0.99. Therefore, the simulation model uses Equation~\ref{eq:mag_fpf} to model the undulator with one period entry and exit tapers. The resonant wavelength was 4.8 $\mu$m and a concentric resonator was used with a optical cavity length of about 8 m \cite{benson_1998}. The radii of curvature of the mirrors was 4.045 m with a cold cavity Rayleigh range of 40 cm. Transmissive out-coupling through the downstream mirror was used and had a transmittance of approximately 10 percent. The outer radius of the mirrors was 2.54 cm. OPC uses the modified Fresnel propagator (Equation~\ref{eq:opc_fresnel_modified}) to handle the divergence and convergence of the optical beam inside the resonator using a fixed number of grid points.        

\begin{figure}[t] 
\centering
\includegraphics[width=8cm]{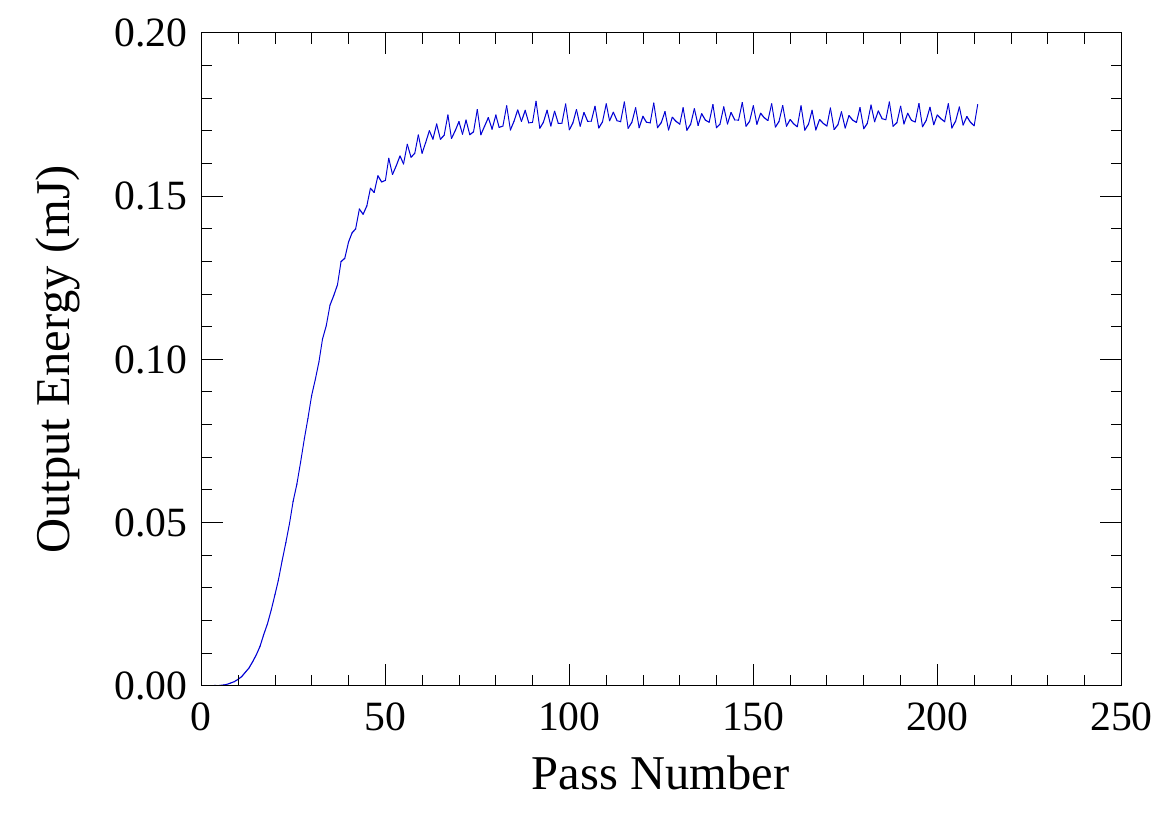}
\caption{\label{fig:ir_demo_E_n}Simulated pulse energy as a function of roundtrip number $n$ at a near optimal cavity detuning of $L_{\textrm{cav}}=8.01049$~m for the IR Demo experiment. Remaining parameters as described in the text.}
\end{figure} 

Simulations of this experiment were conducted with MEDUSA/OPC. MEDUSA \cite{freund_2000} and MINERVA both employ Gaussian representations for the optical field and integrate the three-dimensional Lorentz force equations for the electron trajectories without performing an average over the wiggle-motion. However, MEDUSA is the more primitive code and MINERVA contains many additions and incorporates superior algorithms not present in MEDUSA. Nevertheless, in cases where the two codes have been compared, their predictions for FEL performance are in agreement to within about 10 percent.

The evolution of the output energy found in simulation versus pass is shown in Figure~\ref{fig:ir_demo_E_n} for a cavity length near the peak of the detuning curve of 8.01049 m. Observe that saturation is found after about 70 passes at a pulse energy of 0.17 mJ corresponding to an average power of about 318 W at a repetition rate of 18.7 MHz. This represents an average efficiency of about 0.70 percent which is close to the theoretically predicted efficiency of about 1.0 percent for an undulator with 40 uniform periods

\begin{figure}[tb] 
\centering
\includegraphics[width=10cm]{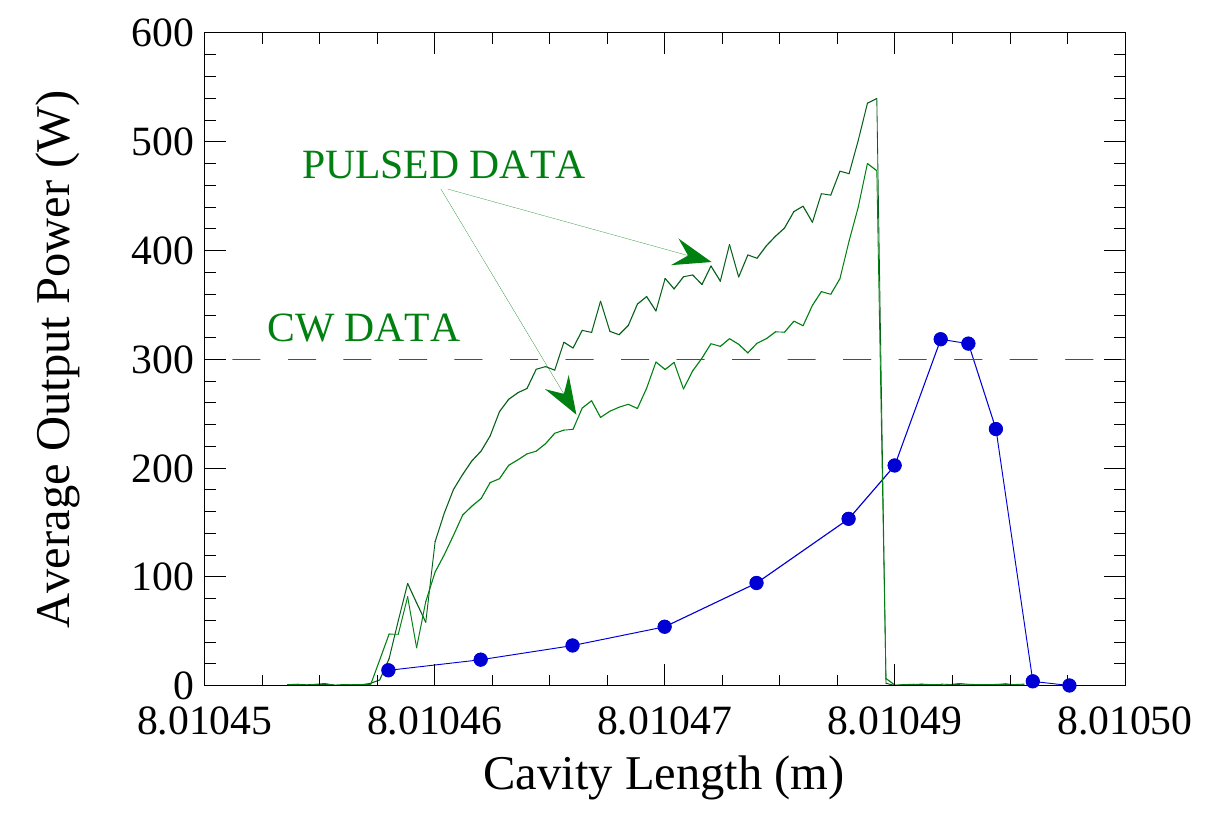}
\caption{\label{fig:ir_demo_cav_tune}Tuning curve for the IR Demo experiment including measurements from CW and pulsed runs (green) and from simulation (blue). Other parameters as described in the text.}
\end{figure}

The average power found in simulation near the peak in the detuning curve is consistent with the observation of 311 W during CW operation of the IR Demo \cite{benson_1999,neil_2000,neil_2000a}. This is also shown in Figure~\ref{fig:ir_demo_cav_tune} which shows the detuning curves found in simulation (blue) and during pulsed operation of the IR Demo (green). The average power during CW operation is indicated by the dashed line. Observe that the zero-detuning point in the simulation is shifted by about 5 $\mu$m from that observed in the experiment. This may be due to an uncertainty in the cavity length measurement by up to 10 $\mu$m. The full width of the detuning curve is found to be within about 30 – 35 $\mu$m.

Higher bunch charges were used in the pulsed runs than in the CW runs in the IR Demo. Lower bunch charges were used in the CW runs in order to reduce distortion due to mirror heating. Indeed, mirror distortion made it difficult to obtain a detuning curve during CW operation. Mirror heating was not a serious problem in the pulsed runs. When the gain to loss ratio is small, as in the CW runs and in the simulation, it is expected that the tuning curve for a low-gain/high-Q oscillator will display a peak at the zero detuning point followed by a rapid decline as the cavity length decreases, and this is what is found in simulation. However, when the gain to loss ratio increases, then the detuning curve will exhibit a shoulder as indicated in the detuning curves for the pulsed runs. This shoulder will be much more prominent in the RAFEL simulations discussed below.

\subsection{The 10-kW Upgrade Experiment at JLab} \label{section:irupgrade}
This experiment represented an upgrade to the original IR-Demo experiment \cite{neil_2000} and numerous elements represent upgrades. For example, the accelerating modules were upgraded to achieve higher energy, bunch charge, and average current. Effort was also made to reduce/mitigate the beam breakup instability in view of the higher average current. The original undulator was replaced to achieve high gain and resonance at a shorter wavelength. In order to handle the higher power, the resonator was lengthened to reduce the mirror loading and cryogenic, edge-cooling was used for the mirrors.
As a result, the kinetic energy was increased to 115 MeV, while the energy spread of 0.3 percent remained approximately the same. The charge of the electron bunch was almost doubled to 115 pC with a pulse length of 390 fs. The normalized emittance of 9 mm-mrad in the wiggle plane and 7 mm-mrad in the plane orthogonal to the wiggle plane was similar as in the IR demo experiment, and the maximum repetition rate of 74.85 MHz for the electron bunches remained unchanged. Note, that although the IR-Demo experiment was capable of running at 74.85 MHz, the results reported in Section~\ref{section:irdemo} corresponded to operation at 17.85 MHz.  To operate at somewhat shorter wavelengths, the planar undulator, which is modeled in the simulation using Equation~\ref{eq:mag_ppf} with one period up- and down-taper, had a longer period of 5.5 cm, a total of 30 periods, and a peak on-axis magnetic field of 3.75 kG. The electron beam was focused into the undulator with the focus at the center of the undulator. The concentric resonator was also updated \cite{shinn_2003} and had a length of about 32 m with a cold-cavity Rayleigh length of 0.75 m. The total loss in the resonator was 21 percent with about 18 percent out-coupled per pass from the downstream mirror. For these settings, the wavelength was 1.6 $\mu$m.

In simulating this experiment, the number of particles in MINERVA was 5832 per slice, while the separation between slices was 5.4 fs. The number of optical modes was dynamically adjusted each roundtrip to accommodate the evolution of the optical field inside the resonator. OPC uses again the modified Fresnel propagator (Equation~\ref{eq:opc_fresnel_modified}) to handle the divergence and convergence of the optical beam inside the resonator.

\begin{figure}[t] 
\centering
\includegraphics[width=10cm]{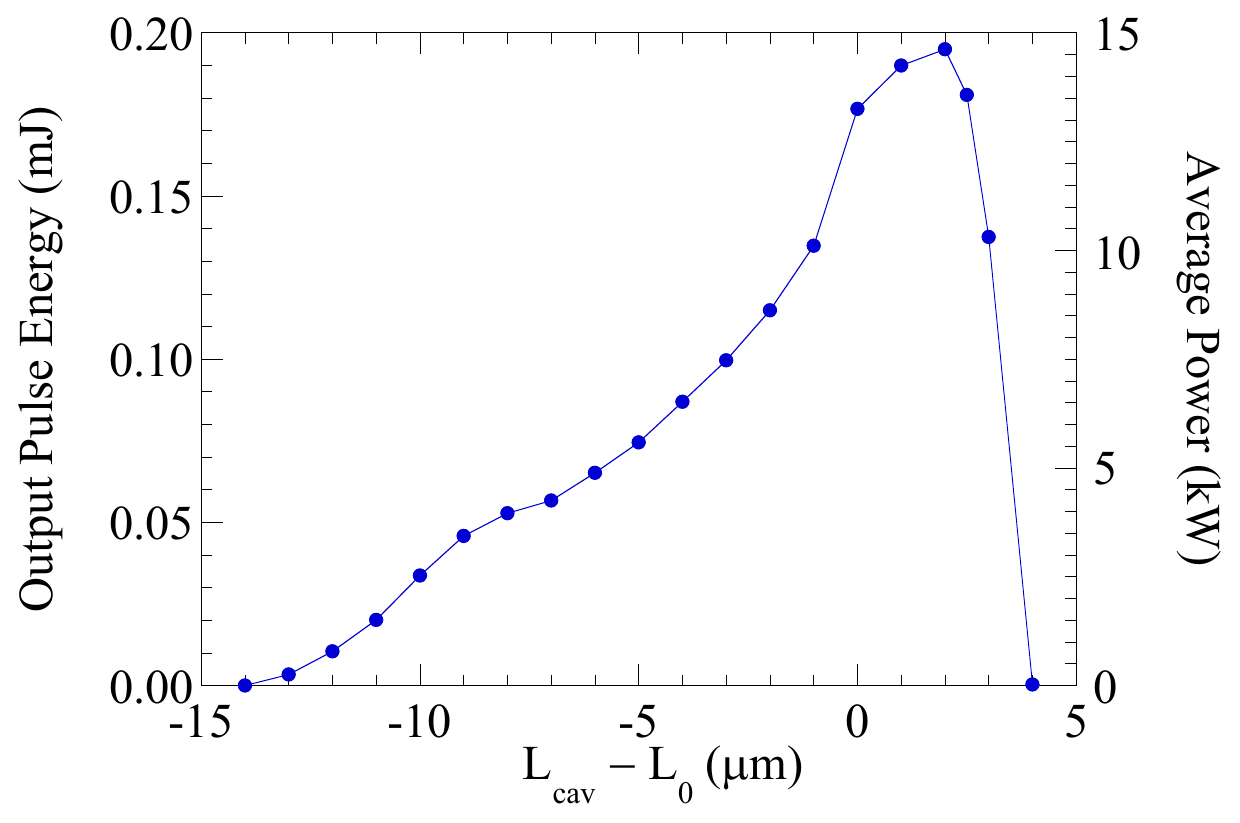}
\caption{\label{fig:ir_upgrade_cav_tune}Tuning curve for the IR upgrade experiment. Simulation parameters as described in the text.}
\end{figure} 

The length of the optical cavity must be selected so that the returning optical pulse is in synchronism with the electron bunches. The roundtrip time for the optical pulses in the cavity is $\tau_r = 2L_{\textrm{cav}}/c$ and the separation between electron bunches is $\tau_{\textrm{sep}} = 1/f_{\textrm{rep}}$, where $L_{\textrm{cav}}$ is the cavity length and $f_{\textrm{rep}}$ is the electron bunch repetition rate. Perfect synchronism (referred to as zero-detuning) is obtained when $\tau_r = M\tau_{\textrm{sep}}$, which leads to Equation~\ref{eq:zero_detuning_low_gain}. Here $M$ is the number of optical pulses in the cavity. In this case there were 16 optical pulses in the cavity and the zero-detuning length is $L_0$ = 32.041946079 m. The cavity detuning curve obtained from simulations is shown in Figure~\ref{fig:ir_upgrade_cav_tune} as a function of the difference between the cavity length $L_{\textrm{cav}}$ and the zero-detuning length. With a maximum pulse energy of 0.194 mJ and a repetition rate of 74.85 MHz, we find that the maximum output power of 14.52 kW occurs for a positive detuning of 2 $\mu$m and is close to the measured value of 14.3 ± 0.72 kW \cite{benson_2007}. As a result, the predicted extraction efficiency is about 1.4 percent, which is close to the theoretical value of  $\eta$ = 1.7 percent. We remark that the previous simulation of this experiment with MEDUSA/OPC \cite{vanderslot_2009} yielded an average output power of 12.3 kW, and the present formulation is in better agreement with the experiment than in the earlier simulation. As in the previous simulation \cite{vanderslot_2009}, the roughly triangular shape of the detuning curve is also in agreement with the experimental observation.

\begin{figure}[t] 
\centering
\includegraphics[width=10cm]{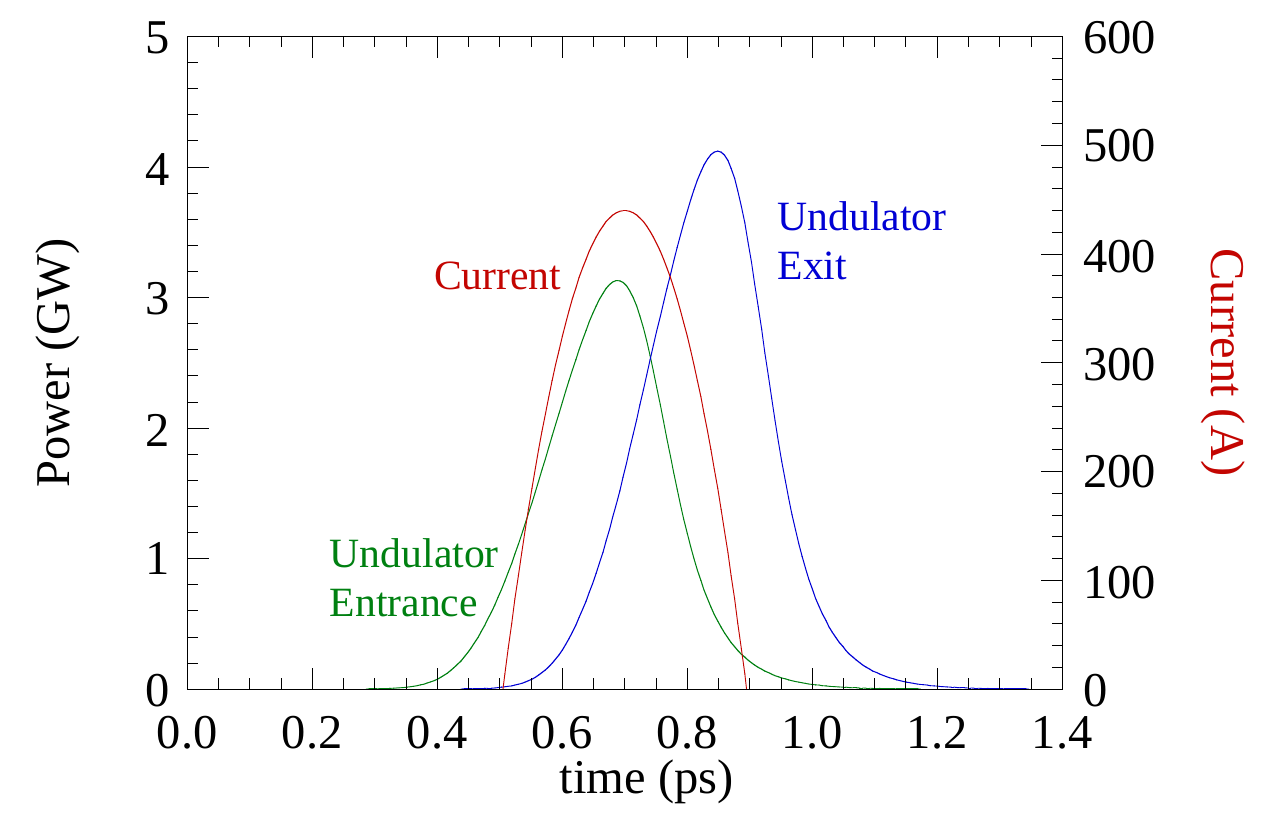}
\caption{\label{fig:ir_upgrade_p_t}Temporal profiles of the power in the optical pulse after 100 passes at the undulator entrance (green) and exit (blue) as well as the current in the electron bunch (right axis, red) for the IR upgrade experiment. The cavity length is $L_{\textrm{cav}}= 32.041946079$~m and other parameters as described in the text.}
\end{figure} 

The temporal profiles of the optical pulse at the undulator entrance and exit as well as that of the electron bunch current are shown in Figure~\ref{fig:ir_upgrade_p_t} for the zero-detuning cavity length after pass 100 which corresponds to a stable, saturated steady-state. Observe that the electron bunch is centered in the time window, which has a duration of 1.4 ps. That this is at zero-detuning is indicated by the fact that the incoming optical pulse at the undulator entrance is in close synchronism with the electron bunch. It is also evident that the center of the optical pulse advances by about 0.16 ps as it propagates through the undulator, and this is in good agreement with the theoretical slippage estimate of $N_u\lambda/c$ for a low-gain FEL, where $N_u$ is the number of periods in the undulator. Finally, it should be remarked that this is in the steady-state regime where the losses in the resonator and the out-coupling are compensated for by the gain in the undulator.

\section{High-Gain/Low-\texorpdfstring{$Q$}{Q} Oscillators} \label{section:hg_lq}
An alternate approach to oscillator design is to use a high-gain undulator where the radiation grows exponentially on a single pass through the undulator. Because the gain is high, the permissible resonator loss can be relatively high; hence, a large fraction of the power can be coupled out of the resonator. This type of oscillator has been referred to as a Regenerative Amplifier FEL (RAFEL), and the concept has been experimentally demonstrated at the Los Alamos National Laboratory \cite{nguyen_1999}. Hence, a RAFEL may also be thought of as a low-$Q$ oscillator and has advantages both for (1) high power designs since the mirror loading can be kept below mirror deformation or damage thresholds, and (2) for VUV and x-ray oscillators.

RAFELs differ from low-gain oscillators in a number of ways. One difference is that, as shown in Madey’s theorem \cite{madey_1979}, a low-gain oscillator exhibits no gain directly on the resonance. In contrast, the growth rate in the exponential gain regime has a peak on-resonance, and this is reflected in the wavelengths excited in a RAFEL. A second difference is the overall efficiency. The saturation efficiency, $\eta$, of a low-gain oscillator is $\eta$ = 1/2.4$N_u$, where $N_u$ is the number of periods in the undulator. Since the radiation exponentiates in each pass through the undulator in the RAFEL, the efficiency is given by that found in the high-gain Compton regime where $\eta$ = $\rho$ where $\rho$ is the Pierce parameter. A third difference is in the linewidth, which scales inversely with $N_u$ in a low-gain oscillator but which is given by the linewidth of the exponential interaction in the high-gain Compton regime. A fourth difference is in the longitudinal and transverse mode structure, which is determined largely by the resonator properties in a low-gain oscillator. In a RAFEL, by contrast, the exponential gain leads to saturation in a very small number of passes through the resonator and the mode structure is largely governed by the interaction in the undulator. A fifth difference is in the effect of slippage. Slippage in a low-gain oscillator scales with $N_u$. However, the high-gain in a RAFEL results in a reduction in the group velocity such that slippage scales with $N_u$/3. However, one point of similarity that the RAFEL shares with low-gain oscillators, is the presence of limit-cycle oscillations.

In this section, we discuss simulations of an infrared RAFEL with the intention of illustrating many of the general properties of a RAFEL and how it compares both to low-gain/high-$Q$ oscillators and SASE FELs \cite{freund_2013} and simulations of an x-ray RAFEL concept making use of hole out-coupling \cite{freund_2019}. We note that the infrared RAFEL simulations were performed with MEDUSA/OPC while the x-ray RAFEL simulations were performed with MINERVA/OPC.

\subsection{An infrared RAFEL}

\begin{table}[t]
\caption{\label{tab:ir_rafel}Electron beam, undulator and resonator parameters for the IR RAFEL.} 
\centering
\begin{tabular}{llll}
\hline
\textbf{Electron beam}	&  & & \\
\hline
\hspace{2 mm}Energy		& 55 & MeV & \\
\hspace{2 mm}Charge		& 800 & pC & \\
\hspace{2 mm}Bunch duration & 1.2 & ps & parabolic \\
\hspace{2 mm}Repetition rate & 87.5 & MHz & \\
\hspace{2 mm}Normalized emittance & 15 & mm-mrad & \\
\hspace{2 mm}Energy spread & 0.25\% & & \\
\hspace{2 mm}Matched beam radius & 392 & $\mu$m & \\
\hline
\textbf{Undulator}	&  & & two-plane focusing \\
\hline
\hspace{2 mm}Period & 2.4 & cm & \\
\hspace{2 mm}Magnitude & 6.5 - 7.0 &kG & \\
\hspace{2 mm}$K_{\textrm{rms}}$ & 1.03 - 1.11 & & \\
\hspace{2 mm}Length & $100\lambda_u$ & & 98 uniform \\
\hline
\textbf{Resonator} & & & Concentric \\
\hspace{2 mm}Wavelength & 2.2 & $\mu$m & \\
\hspace{2 mm}Length & 6.852 & m & \\
\hspace{2 mm}Radii of curvature & 3.5 & m & \\
\hspace{2 mm}Rayleigh range & 0.5 & m & \\
\hspace{2 mm}Hole radius & 5.0 & mm & \\
\hspace{2 mm}Out-coupling & 97\% & & \\
\hline
\end{tabular}
\end{table}

The electron beam, undulator, and resonator parameters are summarized in Table~\ref{tab:ir_rafel}. Observe that the temporal profile of the electron bunch is parabolic with a full width of 1.2 ps, and that the undulator is a two-plane-focusing (i.e., parabolic pole face) design. Consequently, Equation~\ref{eq:mag_fpf} is used to model the undulator with the first and last periods tapered up and down to model the injection and ejection of the beam. Since there is exponential growth and, hence, optical guiding of the radiation, we use a matched beam in the undulator. The resonator is concentric with the power coupled out through a 5.0 mm hole in the downstream mirror, which provides for a typical average out-coupling of about 97 percent. Given the repetition rate, $f_{\textrm{rep}}$, the nominal zero-detuning cavity length, $L_0$, is 6.85239904 m when $M$ = 4 and assuming $v_g=c$ (see Equation~\ref{eq:zero_detuning_low_gain}). The temporal window is an important numerical consideration, and must be chosen to be large enough to accommodate the maximum cavity detuning length that is consistent with pass-to-pass amplification so that the optical pulse remains within the time window for all the usable choices of cavity length. In practice for this example, we choose a temporal window of 4.0 ps and include 182 temporal slices, which corresponds to the inclusion of one temporal slice every three wavelengths. OPC uses the modified Fresnel propagator (Equation~\ref{eq:opc_fresnel_modified}) to handle the divergence and convergence of the optical beam inside the resonator using a grid with a fixed number of grid points.

\begin{figure}[t] 
\centering
\includegraphics[width=8cm]{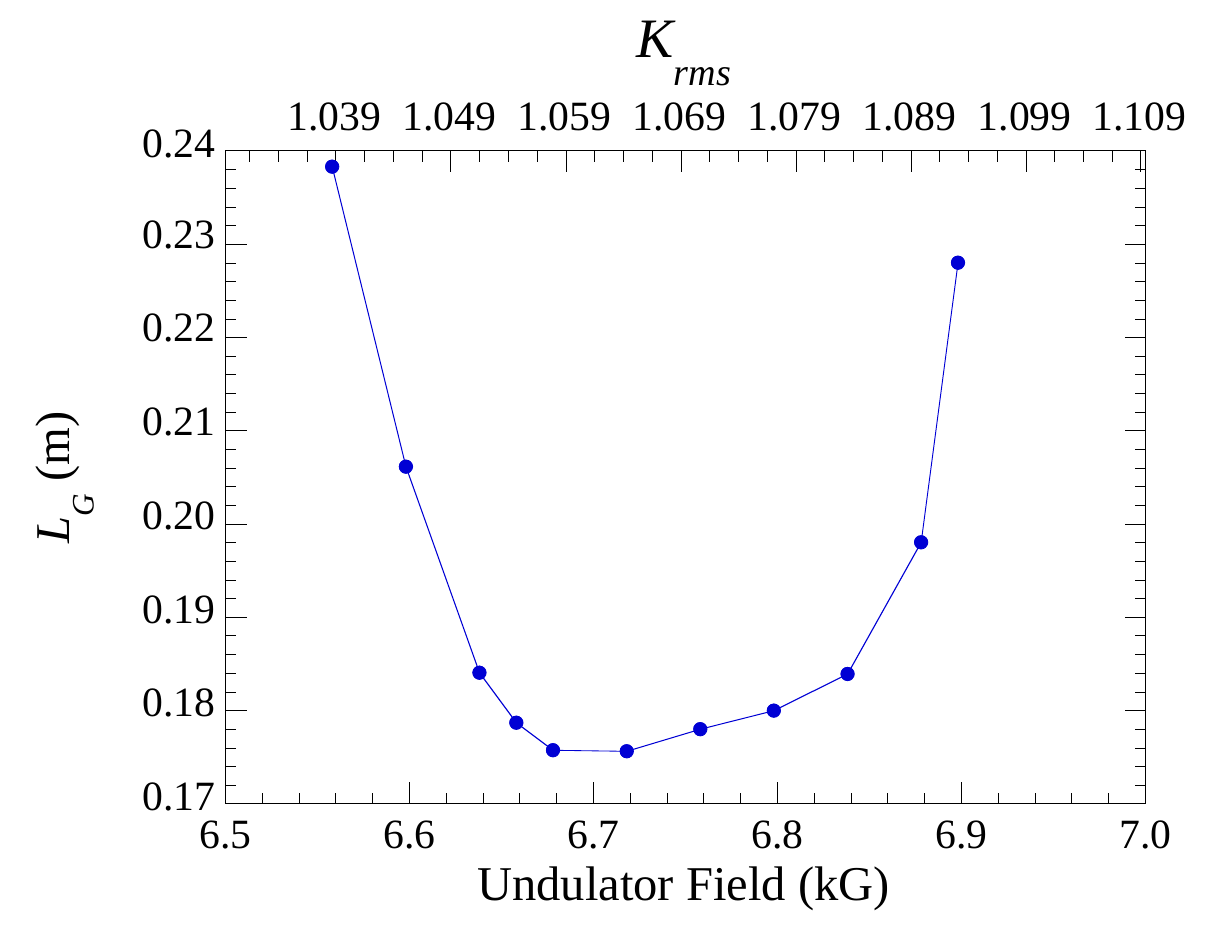}
\caption{\label{fig:ir_rafel_gl_bw}Gain length $L_G$ versus on-axis undulator field amplitude $B_u$ for the IR RAFEL. Other parameters as in Table~\ref{tab:ir_rafel}.}
\end{figure} 

We first consider the single-pass gain because that will affect the performance of the RAFEL. Since the undulator is long enough to achieve exponential growth, we show the gain length, $L_G$, found in simulation versus the undulator field strength in Figure~\ref{fig:ir_rafel_gl_bw}. The optimal (\textit{i.e.}, minimal) gain length of 0.176 m occurs for an on-axis undulator field strength of about 6.7 kG which corresponds to an rms undulator strength parameter of $K_{\textrm{rms}}$ = 1.06. This is in good agreement with the prediction based on the parameterization of the interaction by Ming Xie \cite{mingxie_1995}.

It should be noted that the well-known resonance condition $\lambda = \lambda_u(1 + K_{\textrm{rms}}^{2})/2\gamma^2$ predicts an undulator field of about 6.81 kG ($K_{\textrm{rms}} = 1.08$) at a 2.2 $\mu$m wavelength. This shift in the resonance is due to three-dimensional effects and is in disagreement with the shift in the resonance associated with low-gain oscillators.
\begin{figure}[t] 
\centering
\includegraphics[width=8cm]{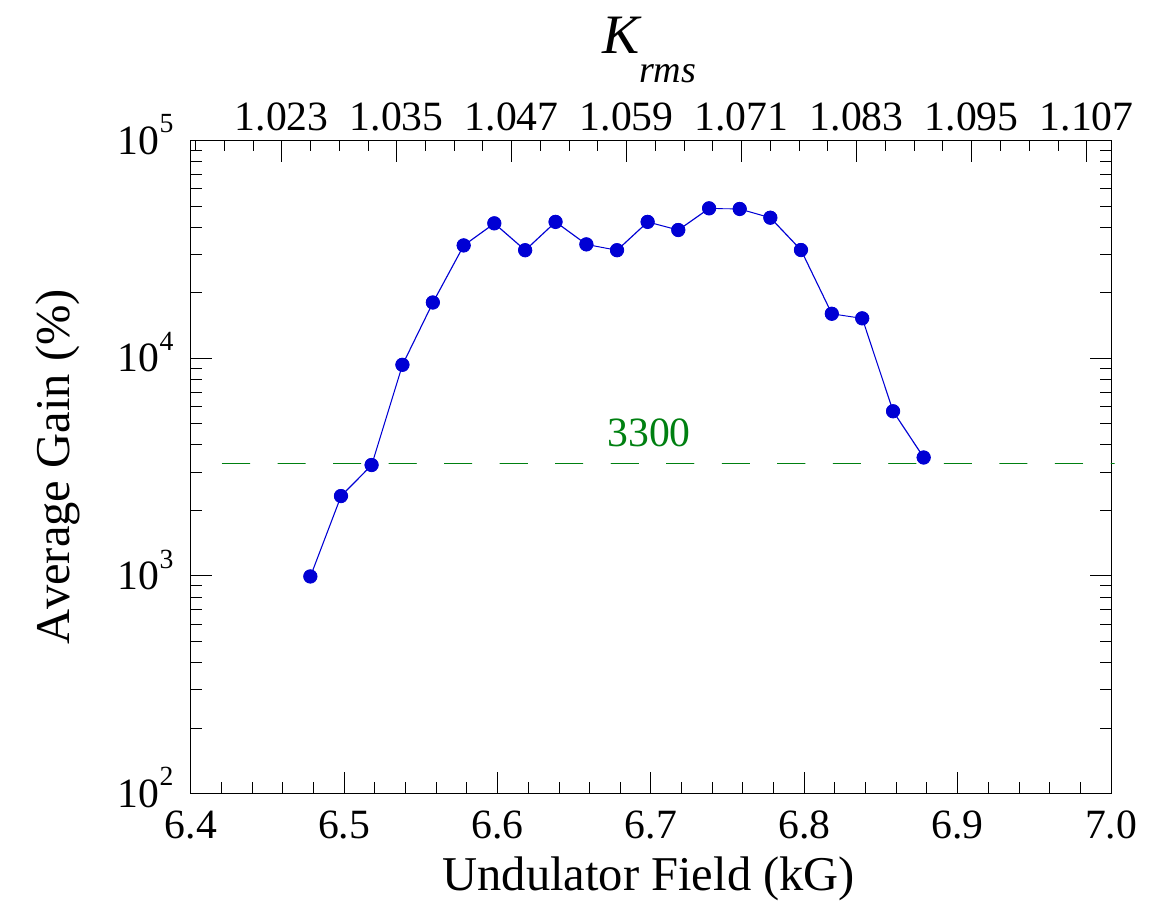}
\caption{\label{fig:ir_rafel_gav}Average gain produced by the IR RAFEL over the first 10 passes versus the undulator field $B_u$ for a cavity length of  $L_{\textrm{cav}}=8.62538144$~m ($\Delta L_{\textrm{cav}}=-8\lambda$. Other parameters as in Table~\ref{tab:ir_rafel}.}
\end{figure} 
The gain length has implications over the permissible range of $K_{\textrm{rms}}$ for which the RAFEL will operate (\textit{i.e.}, over which there is pass-to-pass amplification). Since the RAFEL will saturate when the gain balances the loss, and the loss for the resonator is about 97 percent, this implies that the RAFEL will operate as long as the single-pass gain exceeds about 3200 – 3300 percent. In order to identify this range more closely, we perform multi-pass simulations and take the average gain over the first 10 passes. We take an average because there are fluctuations in the gain on pass-to-pass basis (\textit{i.e.}, limit-cycle like oscillations), which will be discussed in more detail below. The average gain is shown as a function of the on-axis undulator field under the assumption of a cavity length of 8.65238 in Figure~\ref{fig:ir_rafel_gav}. This represents a cavity detuning with respect to the zero-detuning length of $\Delta L_{\textrm{cav}} = -8\lambda$. It is clear from Figure~\ref{fig:ir_rafel_gav} that the gain is relatively constant over the range of about 6.65 – 6.85 kG ($K_{\textrm{rms}} =$ 1.054 – 1.085) and falls off rapidly as the field diverges outside this range, which is consistent with the behavior of the gain length shown in Figure~\ref{fig:ir_rafel_gl_bw}. The cutoff for a gain of about 3300 percent occurs for field levels of about 6.518 kG ($K_{\textrm{rms}} = 1.03$) at the low end and 6.878 kG ($K_{\textrm{rms}} = 1.09$) at the high end, and we do not expect the RAFEL to function outside of this range of undulator fields.

\begin{figure}[t] 
\centering
\includegraphics[width=8cm]{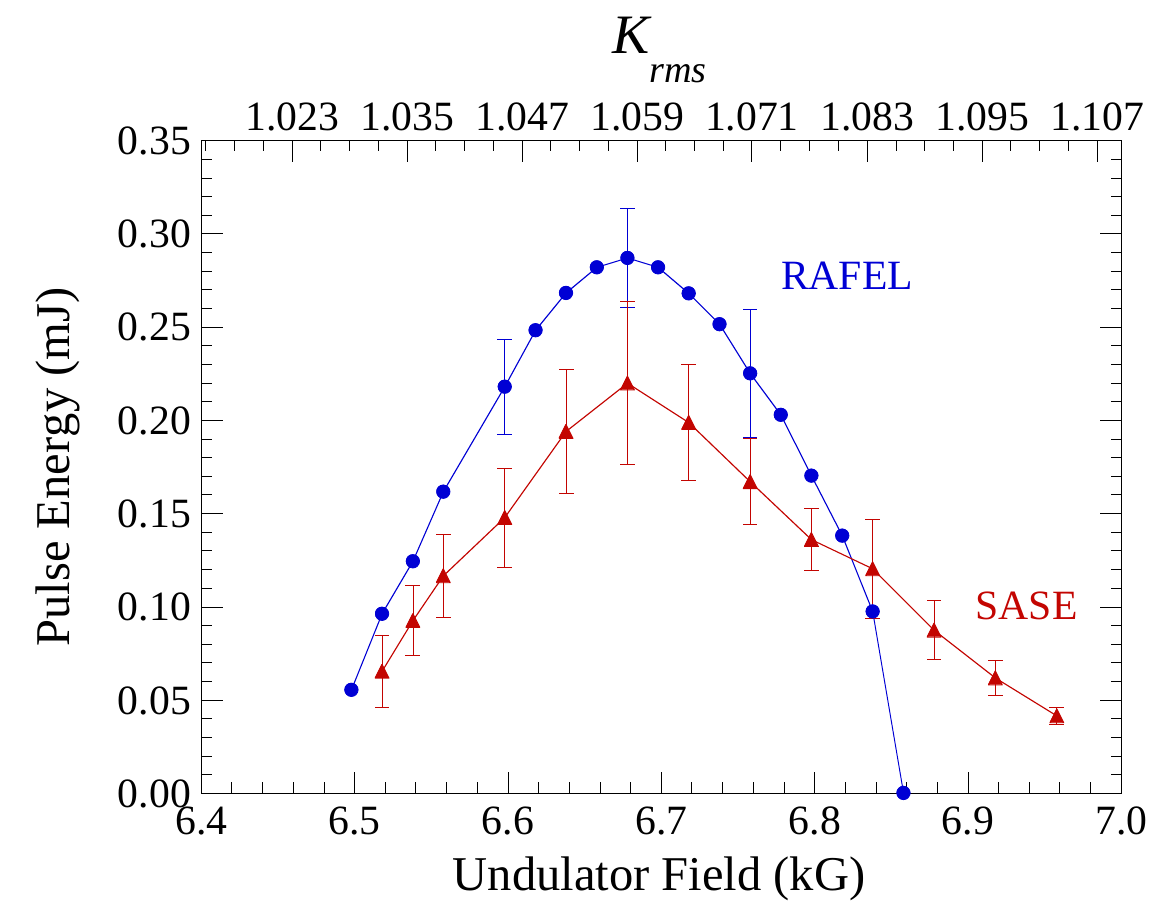}
\caption{\label{fig:ir_rafel_E_Bw}Output pulse energy for the IR RAFEL (blue circles) and an equivalent SASE FEL (red triangles) as a function of the undulator field strength $B_u$. The cavity detuning for the IR RAFEL is $\Delta L_{\textrm{cav}}=-8\lambda$, while the other simulation parameters are given in Table~\ref{tab:ir_rafel}. The error bars indicate energy fluctuations due to limit-cycle like oscillations for the RAFEL and pulse-to-pulse rms energy fluctuations in case of SASE.}
\end{figure}

In order to demonstrate how the saturation efficiency of a RAFEL differs from that of a low-gain oscillator, it is instructive to compare the RAFEL with an equivalent SASE FEL. The performance of the RAFEL is shown in Figure~\ref{fig:ir_rafel_E_Bw}, where we plot the average pulse energy (blue circles) in the steady-state as a function of the on-axis undulator field for the same cavity detuning ($\Delta L_{\textrm{cav}} = -8\lambda$) as used for Figure~\ref{fig:ir_rafel_gav}. The error bars characterize the limit-cycle oscillations. The RAFEL reaches its’ peak pulse energy for an undulator field of 6.678 kG ($K_{\textrm{rms}} = 1.058$) and falls to zero outside the range predicted in Figure~\ref{fig:ir_rafel_gl_bw}. We also plot the equivalent SASE saturated pulse energy (red triangles). In order to deal with the statistical fluctuations inherent in the SASE output, a large number of runs were made with different shot noise distributions, and found the average (red triangles) and standard deviations (error bars). Note that the SASE results represent the pulse energies over whatever length of undulator is required to reach saturation. Observe that (1) the RAFEL configuration saturates with a higher pulse energy than the SASE configuration, (2) the fluctuations in the RAFEL in the steady-state regime are comparable in magnitude to the statistical fluctuations found in SASE, however, the fluctuations of the RAFEL are deterministic in nature, and (3) the FWHM of the tuning range in $K_{\textrm{rms}}$ is comparable for both the RAFEL and SASE configurations.

The RAFEL saturates with about a 0.28 mJ pulse energy, which corresponds to an extraction efficiency of about 0.64 percent. This compares well with the empirical formula \cite{mingxie_1995} that predicts a saturation efficiency of about 0.76 percent. In contrast, the saturation efficiency of a low-gain oscillator is predicted to be $\eta = 1/(2.4 N_u) = 0.43$\%.

The spectral linewidth of the RAFEL also differs from that of a low-gain oscillator. The full width of the spectrum for a typical low-gain oscillator is given by $\Delta\omega/\omega = 1/N_u = 0.01$ for the example under consideration. This can be translated into a tuning range over the undulator field as follows

\begin{equation}\label{eq:DBw/Bw}
    \left|\frac{\Delta B_u}{B_u}\right|=\frac{1+K_{\textrm{rms}}^{2}}{K_{\textrm{rms}}^{2}}\left|\frac{\Delta \omega}{\omega}\right|.
\end{equation}

\noindent This implies a full width tuning range of $\Delta B_u = 0.063$ kG ($\Delta K_{\textrm{rms}} = 0.001$), which is much narrower than what we find in simulation. The relative SASE linewidth is given by ($\Delta\omega/\omega)_{\textrm{rms}} = \rho$ \cite{huang_2007} where $\rho$ denotes the Pierce parameter. Here, $\rho = 0.0097$ and ($\Delta\omega/\omega)_{\textrm{rms}} = 0.0097$. Converting this to a tuning range in the undulator field and going from the rms width to a FWHM tuning range, we obtain ($\Delta B_u/B_u)_{\textrm{FWHM}} = 0.022$, which compares well with the simulation results that give ($\Delta B_u/B_u)_{\textrm{FWHM}} = 0.019$. Hence, the RAFEL behaves more like a SASE FEL than a typical low-gain oscillator in regards to the spectral linewidth.

Another way in which the RAFEL differs from a low-gain oscillator is in the cavity detuning. In a low-gain FEL oscillator, the zero-detuning length is obtained by assuming that the group velocity $v_g$ equals the speed of light \textit{in vacuo} $c$ throughout the resonator. However, $v_g$ is reduced in a RAFEL by the interaction in the undulator, and results in smaller synchronous cavity length. This makes the cavity detuning dependent on the gain of the FEL. The change in group velocity also affects slippage. In a low-gain oscillator, the group velocity reduction is small and the slippage is one wavelength per undulator period; hence, the slippage distance is $l_{\textrm{slip}} = N_u\lambda$. However, the slippage per undulator period is reduced in a high-gain RAFEL, or in any FEL where there is exponential growth because the gain medium reduces both the phase and group velocities. The reduced phase velocity results in the optical guiding of the radiation, while the reduced group velocity results in less slippage. It has been shown that $l_{\textrm{slip}} = N_u\lambda/3$ at the resonant wavelength \cite{saldin_2000}. For the example under consideration, this yields $l_{\textrm{slip}} = 72$ $\mu$m which is much less than the slippage length of 220 $\mu$m if the RAFEL behaved as a low-gain oscillator.

In order to estimate the effect of this on the detuning length, we note that the zero-detuning length is found by equating the roundtrip time of the radiation through the cavity with the spacing between electron bunches (1/$f_{\textrm{rep}}$), \textit{cf.} Equation~\ref{eq:zero_detuning_length}. As a result, in the high gain regime where $v_g = c/(1 + 1/3\gamma_{\parallel}^2$), the difference in synchronous cavity length $\Delta L_{\textrm{0}}$ for a high-gain RAFEL and a low-gain FEL oscillator is given by Equation~\ref{eq:zero_detuning_length} as

\begin{equation}\label{eq:Delta L0}
    \Delta L_{\textrm{0}}=\frac{L_u}{6\gamma^2}(1+K_{\textrm{rms}}^{2})=-\frac{N_u\lambda}{3}.
\end{equation}

\noindent This is comparable to what is found in simulation.
\begin{figure}[t] 
\centering
\includegraphics[width=8cm]{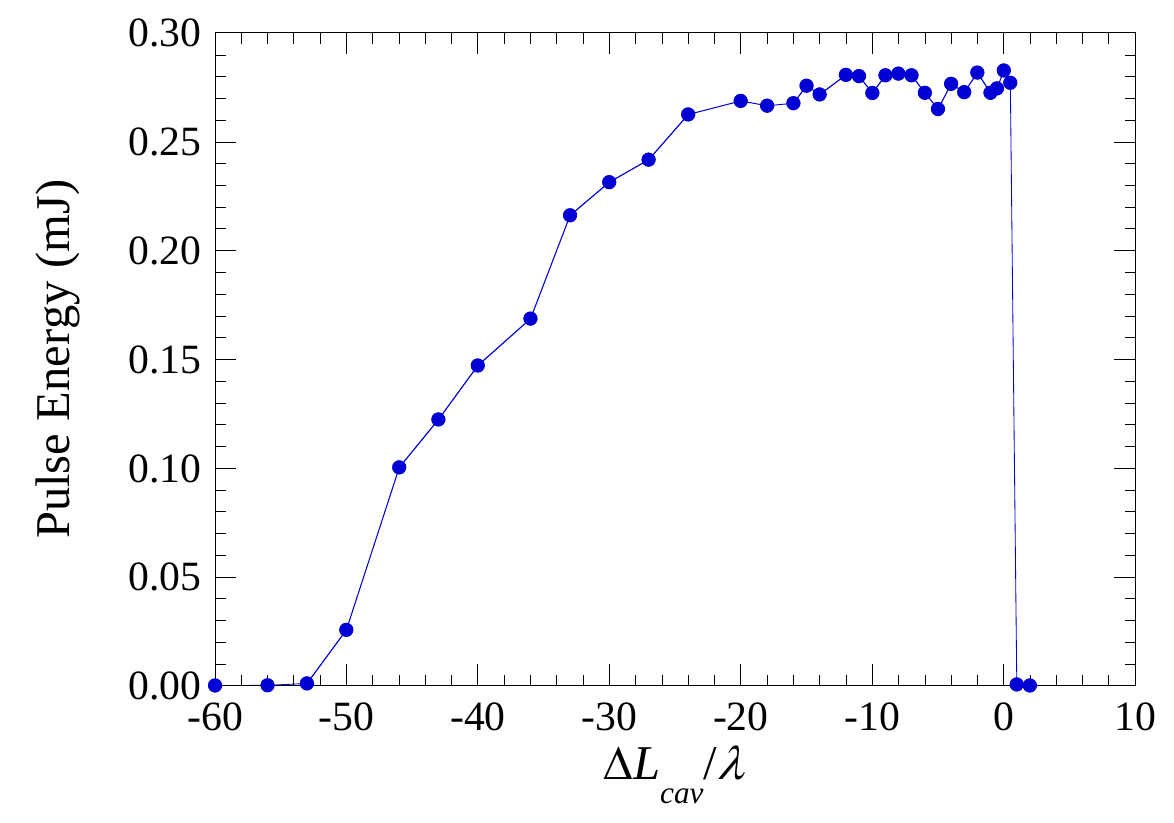}
\caption{\label{fig:ir_rafel_E_cav_tune}Cavity detuning curve of the IR RAFEL for an on-axis undulator field of $B_u=6.658$~kG ($K_{\textrm{rms}}$ = 1.055), while the other parameters are given in Table~\ref{tab:ir_rafel}.}
\end{figure} 
The detuning curve found in simulation is shown in Figure~\ref{fig:ir_rafel_E_cav_tune}, where we plot the output pulse energy versus cavity detuning for a undulator field of 6.658 kG ($K_{\textrm{rms}}$ = 1.055). Here we define the cavity detuning relative to the nominal zero detuning length (Equation~\ref{eq:zero_detuning_low_gain} with $M=1$) so that $\Delta L_{\textrm{cav}} = L_{\textrm{cav}} - L_0$. As shown in Figure~\ref{fig:ir_rafel_E_cav_tune}, we find a full width detuning range of about 50 $\mu$m - 110 $\mu$m and a FWHM detuning range of about 40 $\mu$m which are in reasonable agreement with the estimate based on the one-dimensional analysis of slippage.

\begin{figure}[t] 
\centering
\includegraphics[width=8cm]{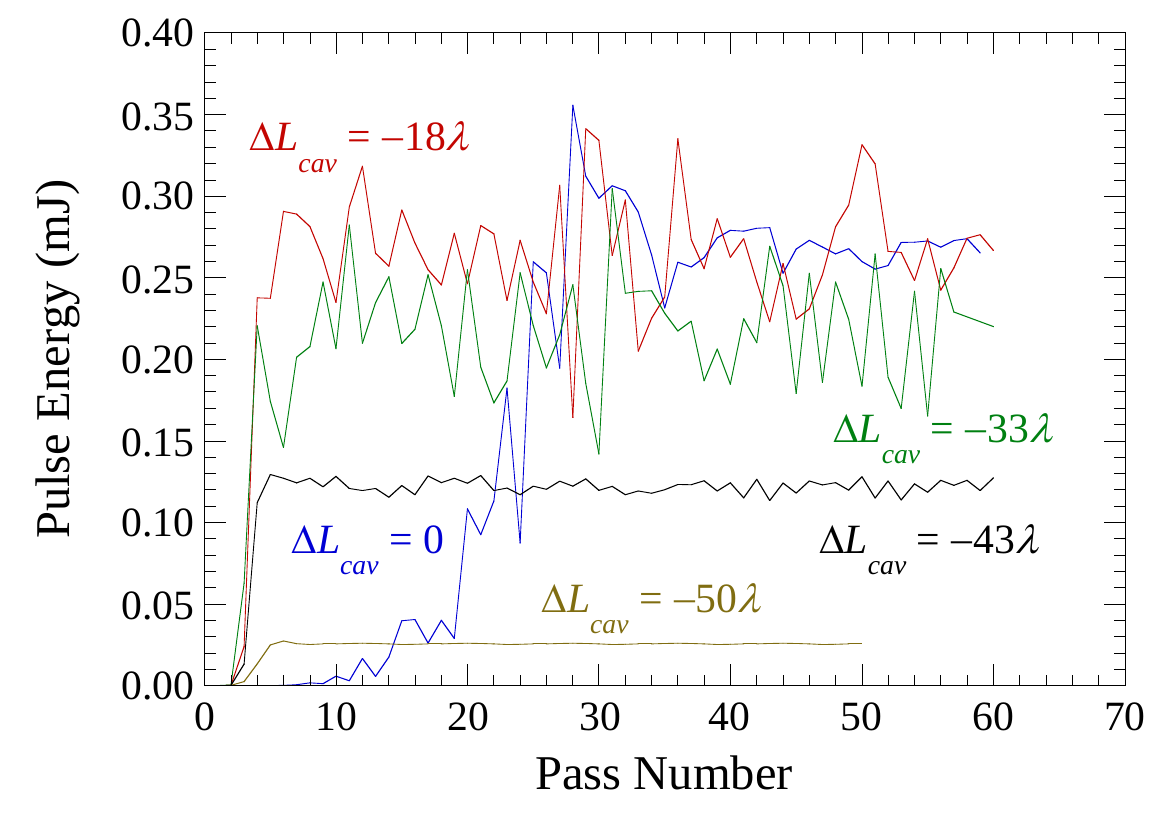}
\caption{\label{fig:ir_rafel_E_n}Temporal evolution of the pulse energy for cavity detunings of  $\Delta L_{\textrm{cav}}=0, -18\lambda, -33\lambda, -43\lambda$ and $-50\lambda$. The on-axis undulator field is $B_u=6.658$~kG ($K_{\textrm{rms}}$ = 1.055), while the other parameters are given in Table~\ref{tab:ir_rafel}.}
\end{figure}

The temporal evolution of the pulse energy is shown for an undulator field of 6.658 kG ($K_{\textrm{rms}}$ = 1.055) in Figure~\ref{fig:ir_rafel_E_n}, where we plot the pulse energy versus pass number through the undulator for the choice of several cavity detunings that samples the complete detuning curve. It is clear that significant fluctuations are found over a large range of detunings and that both the magnitude and period of the fluctuations decrease as the magnitude of the detuning increases, although the magnitude of the fluctuations decreases as well near the zero-detuning length.

\begin{figure}[t] 
\centering
\includegraphics[width=8cm]{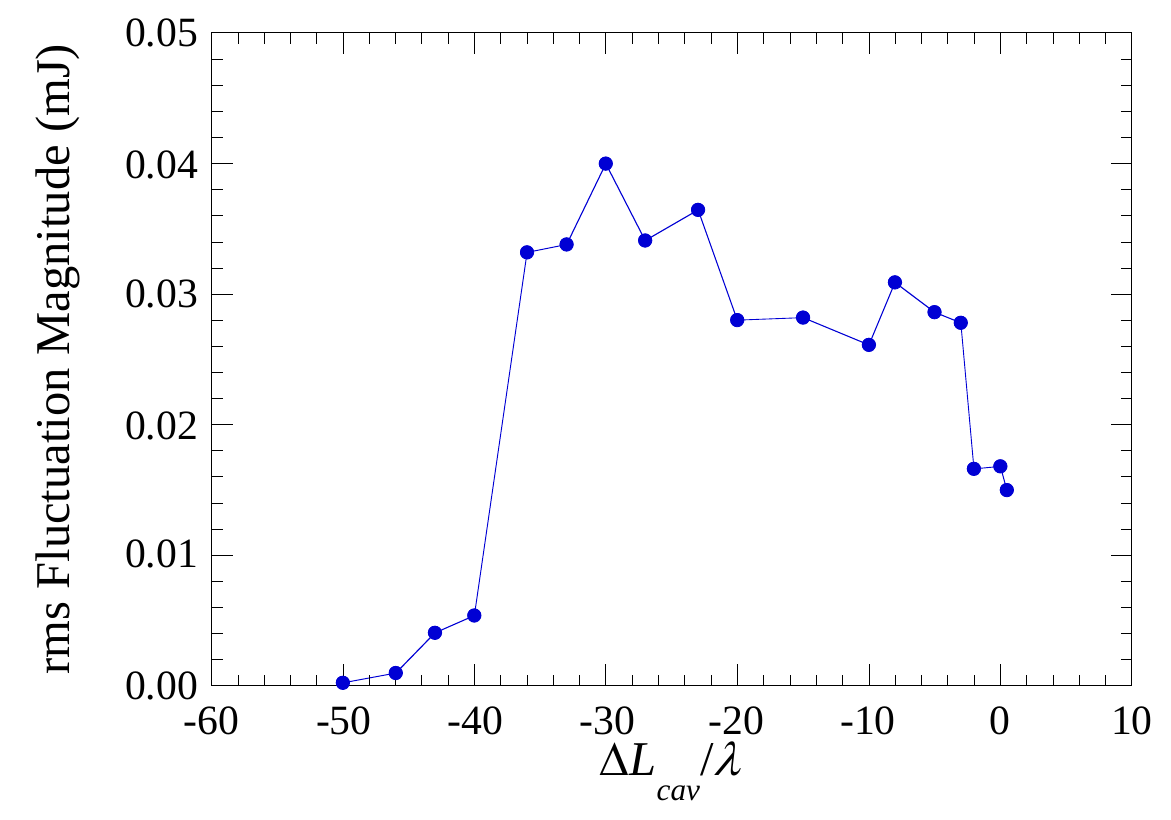}
\caption{\label{fig:ir_rafel_fluc}Standard deviation of the optical pulse energy as a function of cavity detuning $\Delta L_{\textrm{cav}}/\lambda$ for a undulator field strength of $B_u=6.658$~kG ($K_{\textrm{rms}}$ = 1.055), while the other parameters are given in Table~\ref{tab:ir_rafel}.}
\end{figure}

The fluctuations seen in simulation can be rapid and irregular. There are two possible explanations for this. One is that due to the high gain and high out-coupling, small changes in the mode structure from pass to pass can result in relatively large changes in the gain and, hence, the pulse energy. These “small” changes can include variations in the transverse mode structure (both in terms of the modal decomposition and spot size), and the temporal pulse shape. The second explanation, related to the first, is that since we have employed hole out-coupling, these relatively small changes in the transverse mode structure at the mirror can give rise to large differences in the out-coupling of the optical mode. It is not surprising, therefore, that the magnitude of the fluctuations vary depending on the cavity detuning. In Figure~\ref{fig:ir_rafel_fluc}, we show the variation in the rms magnitude of the fluctuations in the out-coupled pulse energy as a function of the cavity detuning. It is clear from Figure~\ref{fig:ir_rafel_fluc} that the fluctuation level is relatively constant at about the 0.03 mJ level over most of the detuning range, but with rapid declines at the ends of the detuning range. Also, the oscillation period is of the order of a few passes through the resonator.

Fluctuations/oscillations have been observed in low-gain oscillators and are referred to as limit-cycle oscillations. The observation of limit-cycle behavior in the low-gain FELIX FEL oscillator corresponds to an oscillation period of \cite{jaroszynski_1993}

\begin{equation}\label{eq:delta_tau_1}
    \Delta \tau=-\tau_{\textrm{slip}}\frac{L_{\textrm{cav}}}{\Delta L_{\textrm{cav}}},
\end{equation}

\noindent where $\tau_{\textrm{slip}} = l_{\textrm{slip}}/c$ is the slippage time. For the case of FELIX, $L_{\textrm{cav}}$ = 6 m, $\lambda$ = 40 $\mu$m, and $N_u$ = 38, and the cavity detuning ranges over about 160 $\mu$m. As a result, $\tau_{\textrm{slip}}$ = 5.1 ps and $\tau_{r}$ (= 2$L_{\textrm{cav}}/c$) = 40 ns is the nominal roundtrip time; hence, this implies that the limit cycle oscillation occurs over a period of about 3 $\mu$s or 75 passes for a cavity detuning of -100 $\mu$m. 

In contrast, if we apply the slippage time for the high gain RAFEL under consideration

\begin{equation}\label{eq:delta_tau_2}
    \Delta \tau=-\frac{N_u\lambda}{3\Delta L_{\textrm{cav}}}\frac{L_{\textrm{cav}}}{c}=-\frac{\tau_{r}}{2}\frac{N_u\lambda}{3\Delta L_{\textrm{cav}}}.
\end{equation}

\noindent As such, we expect the oscillation period to occur on the scale of a small number of passes for the indicated cavity detuning range. This is indeed what is observed in Figure~\ref{fig:ir_rafel_E_n}. For example, the oscillations occur approximately every 2 – 4 passes for $\Delta L_{\textrm{cav}}/\lambda$ = -8, which is consistent with Equation~\ref{eq:delta_tau_2}. However, there is not a great deal of variation with detuning possible when the oscillations occur on such a fast time scale, and we must take Equations~\ref{eq:delta_tau_1} and \ref{eq:delta_tau_2} as approximate measures of the oscillation period. Still, the observed oscillation period is well described by the formula for the oscillation period for limit-cycle oscillations found in in low-gain oscillators when the appropriate slippage is taken into account. For a high-gain RAFEL, the much lower slippage results in very short oscillation periods.

The limit-cycle like oscillations in the RAFEL are correlated with the fluctuations/oscillations in the transverse mode structure. The transverse mode structure in a low-gain oscillator is largely (but not completely) determined by the mode structure in the cold cavity since the optical guiding of the radiation in the undulator is weak. This is not the case in a RAFEL where the mode is guided through the undulator. As a result, the mode structure that forms as the RAFEL saturates differs substantially from the cold cavity modes, and our choice of a Rayleigh range of 0.5 m serves mainly to determine the radii of curvature of the mirrors. Since the radiation is guided in the undulator, the spot size at the undulator exit may be smaller than it would be in the cold cavity, which means that the Rayleigh range of the radiation as it exits the undulator is smaller than it would be in the cold cavity. This implies, in turn, that the optical mode will expand more rapidly as it propagates to the downstream mirror. Alternatively, decomposing the smaller spot size at the undulator exit in cold cavity modes, necessarily leads to higher order transverse modes in the optical field. After propagating to the outcoupler, the superposition of these modes determine the fraction of the optical field coupled out through the hole, and, similarly, after propagation to the undulator entrance, the superposition sets the field profile at undulator entrance. Small variations in the exponential growth rate, \textit{e.g.}, due to changing coupling of the electrons to the optical field at the undulator entrance, lead to relatively larger effects on the optical guiding of the radiation. This in turn changes the spot size at the undulator exit and, hence, the energy coupled out of the resonator and the spot size at the undulator entrance.

The oscillation in mode size has also been observed in simulation of low-gain oscillators \cite{vanderslot_2009} where the magnitude of the oscillation depends on the amount of optical guiding, which may vary between the FEL gain codes used \cite{vanderslot_2008}. Furthermore, it is found that mirror aberrations, \textit{e.g.}, spherical aberration, also effect the variation in optical mode size observed from pass to pass ~\cite{vanderslot_2008}. 

\begin{figure}[t] 
\centering
\includegraphics[width=8cm]{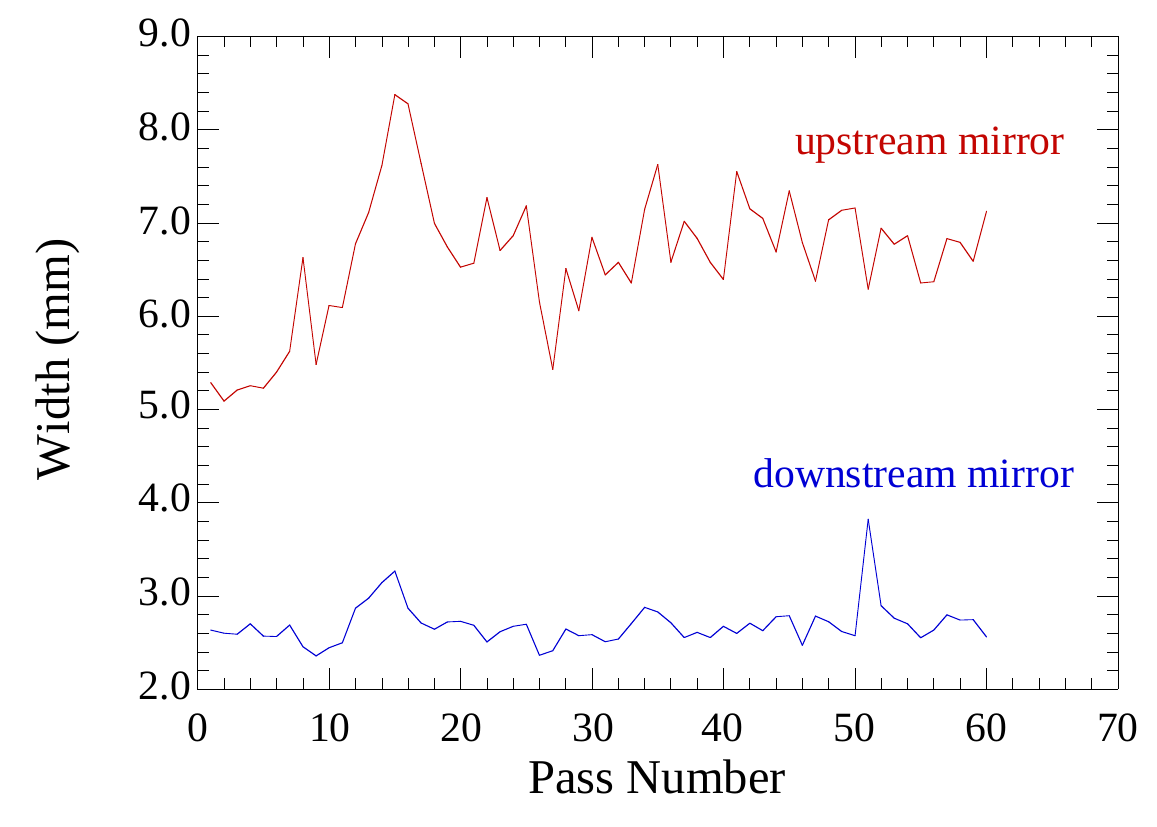}
\caption{\label{fig:ir_rafel_opt_width}Variation in the width (rms diameter) of the optical beam incident on the upstream mirror (red) and downstream mirror (blue) as a function of the number of passes for a cavity detuning of $\Delta L_{\textrm{cav}}=-8\lambda$ and an undulator field strength of $B_u=6.658$~kG ($K_{\textrm{rms}}$ = 1.055). The other parameters are given in Table~\ref{tab:ir_rafel}.}
\end{figure}

This is illustrated in Figure~\ref{fig:ir_rafel_opt_width} where we plot the pass-to-pass variation in the width of the optical mode on the downstream and upstream mirrors for $B_u$ = 6.585 kG ($K_{\textrm{rms}}$ = 1.043) and $\Delta L_{\textrm{cav}}$ = -8$\lambda$. It is clear that both the spot size and the fluctuations of the spot size on the upstream mirror are greater than those on the downstream mirror due to the optical properties of the resonator. At saturation the location of the smallest optical beam size moves over the axis of the undulator and this changes the optical magnification. Consequently, the size of the optical field at both mirrors as well as at the entrance of the undulator changes from pass to pass as can be observed in Figure~\ref{fig:ir_rafel_opt_width}.

\begin{figure}[t] 
\centering
\includegraphics[width=8cm]{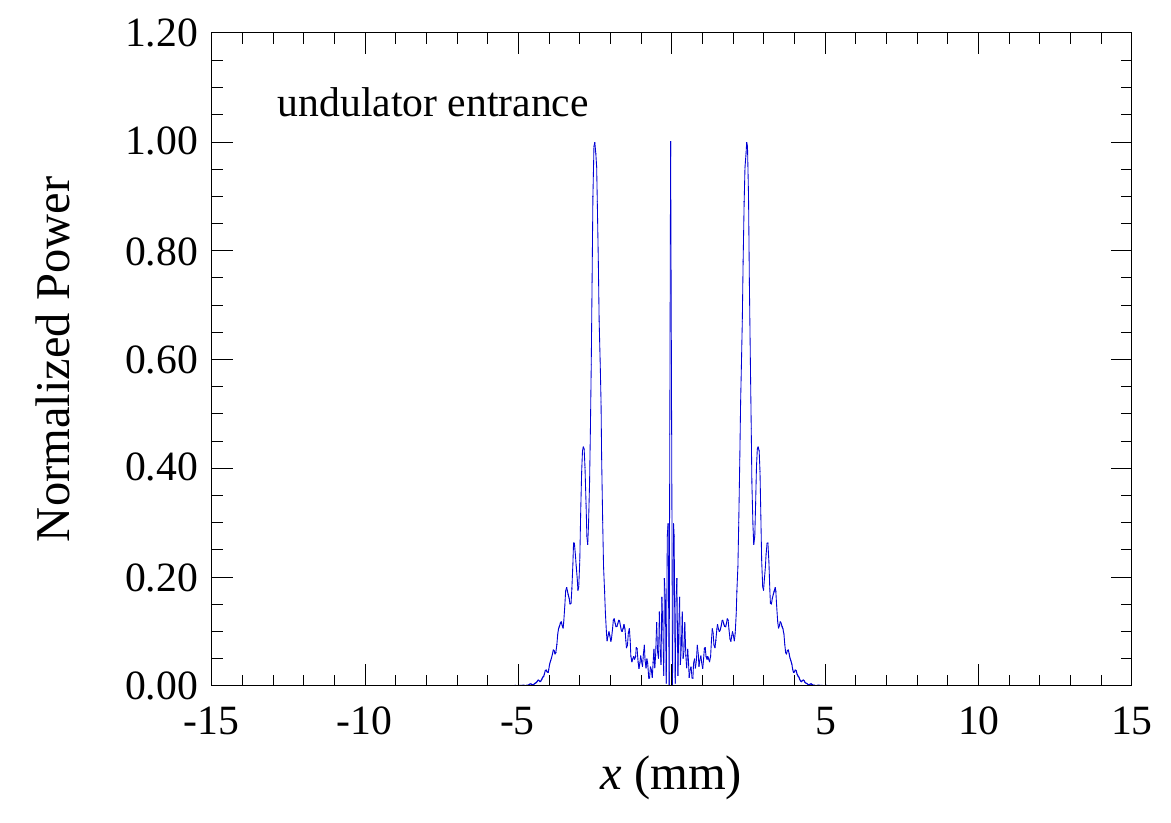}
\caption{\label{fig:ir_rafel_p_x}Cross section of the optical field at the undulator entrance on pass 60 for a cavity detuning $\Delta L_{\textrm{cav}}=-18\lambda$ and an undulator field strength of $B_u=6.658$~kG ($K_{\textrm{rms}}$ = 1.055). The other parameters are given in Table~\ref{tab:ir_rafel}.}
\end{figure}

\begin{figure}
     \centering
     \begin{subfigure}[t]{0.45\textwidth}
         \centering
         \includegraphics[width=\textwidth]{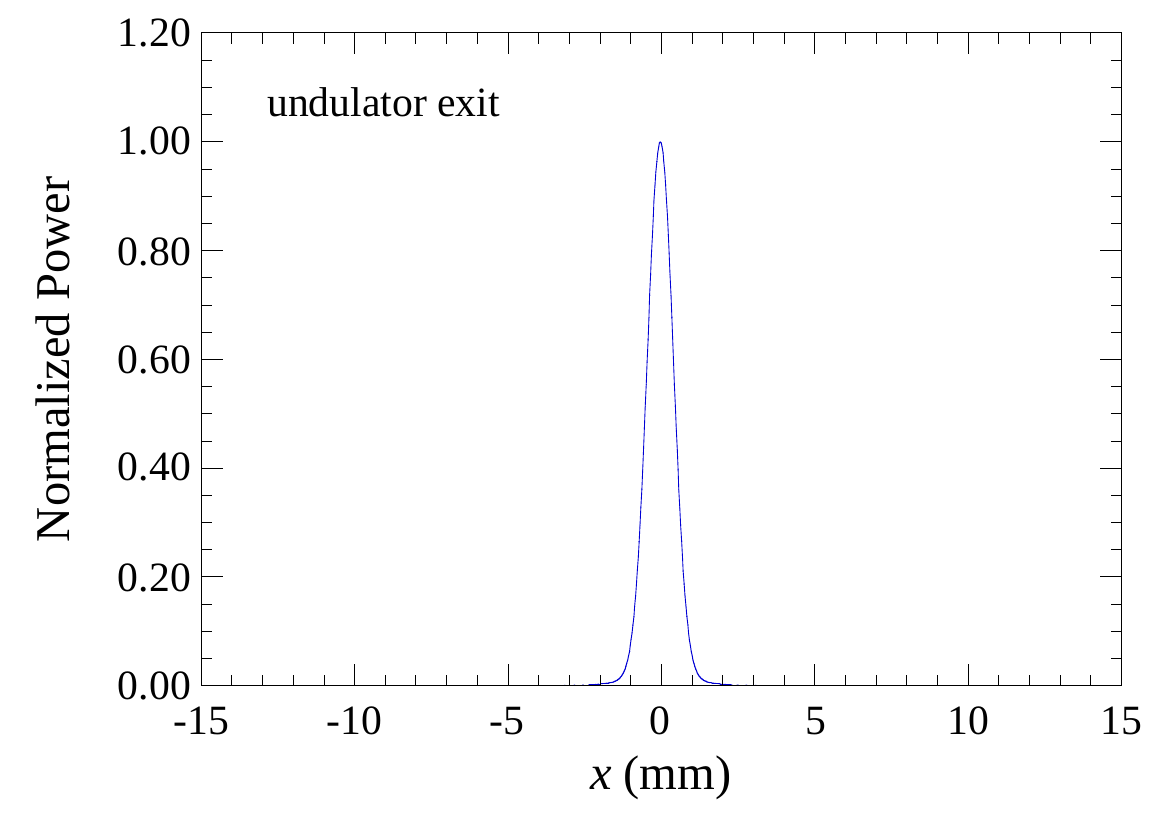}
         \caption{Exit of undulator}
         \label{fig:ir_rafel_P_x_exit}
     \end{subfigure}
     \hfill
     \begin{subfigure}[t]{0.45\textwidth}
         \centering
         \includegraphics[width=\textwidth]{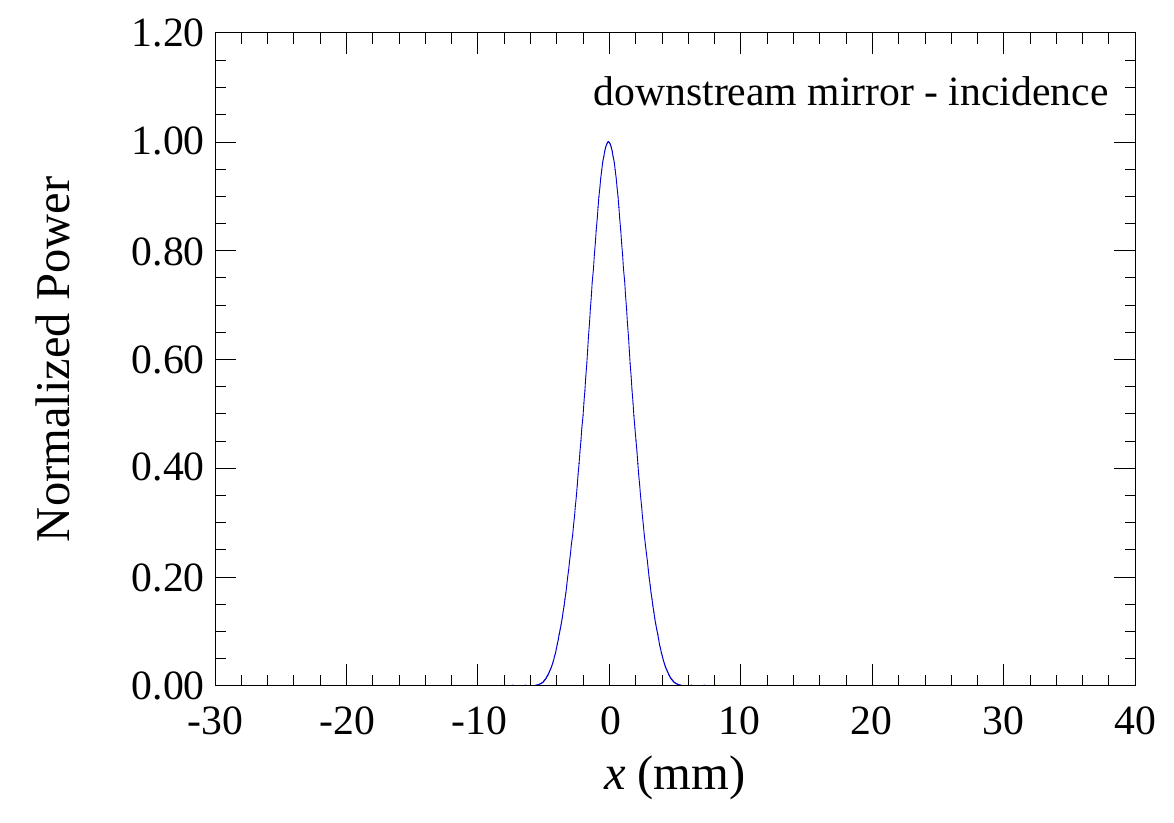}
         \caption{Incident on downstream mirror}
         \label{fig:ir_rafel_P_x_dm}
     \end{subfigure}
     
     \vspace{2mm}
     \begin{subfigure}[t]{0.45\textwidth}
         \centering
         \includegraphics[width=\textwidth]{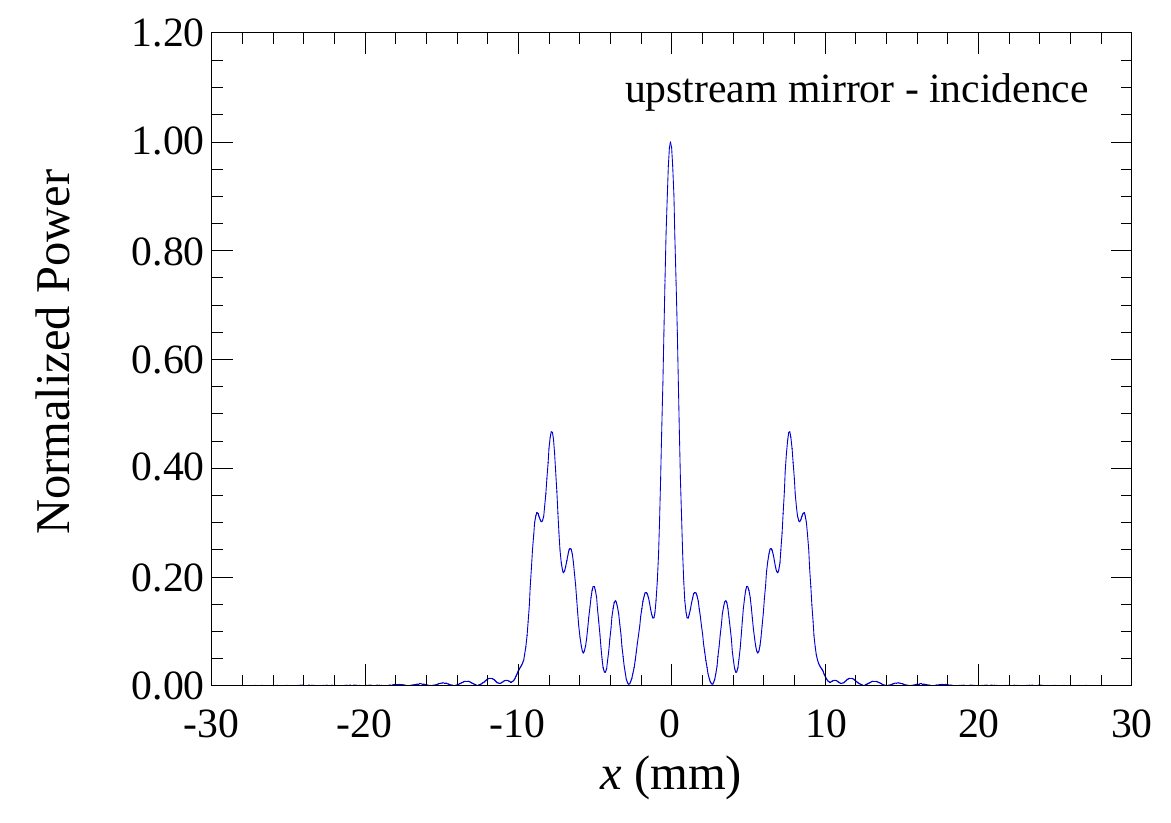}
         \caption{Incident on upstream mirror}
         \label{fig:ir_rafel_P_x_um}
     \end{subfigure}
     \vspace{1mm}
        \caption{Cross section of the optical field at various location on pass 60 for a cavity detuning $\Delta L_{\textrm{cav}}=-18\lambda$ and an undulator field strength of $B_u=6.658$~kG ($K_{\textrm{rms}}$ = 1.055). The other parameters are given in Table~\ref{tab:ir_rafel}.}
        \label{fig:ir_rafel_p_x_n=60}
\end{figure}

The transverse mode structure is not only a result of optical guiding, it is also affected by the hole out-coupling. Consider the case of $B_u$ = 6.658 kG ($K_{\textrm{rms}}$ = 1.055) and $\Delta L_{\textrm{cav}}$ = -18$\lambda$. The cross section of the field delivered to the undulator entrance on pass 60 is shown in Figure~\ref{fig:ir_rafel_p_x} where we plot the normalized power in the $x$-direction (\text{i.e.}, the wiggle plane). Observe that the bulk of the power is at the edge of the optical field, but there is a spike at the center that seeds the subsequent pass through the undulator. Despite the multiple peaks in the cross section at the undulator entrance, the strength of the interaction in the undulator yields a near-Gaussian mode peaked on-axis at the undulator exit, as shown in Figure~\ref{fig:ir_rafel_P_x_exit}. What has happened is that the interaction with the electron beam, which has a diameter of about 0.784 mm, essentially amplifies and guides the central peak shown in Figure~\ref{fig:ir_rafel_p_x} while the power in the wings falls outside the electron beam and is not amplified. This near-Gaussian mode then propagates to the downstream mirror during which it expands by about a factor of three, as shown in Figure~\ref{fig:ir_rafel_P_x_dm}, where the FWHM is about 3.9 mm in width. The FWHM of the modal superposition at the undulator exit is about 1.2 mm.

\begin{figure}
     \centering
     \begin{subfigure}[t]{0.45\textwidth}
         \centering
         \includegraphics[width=\textwidth]{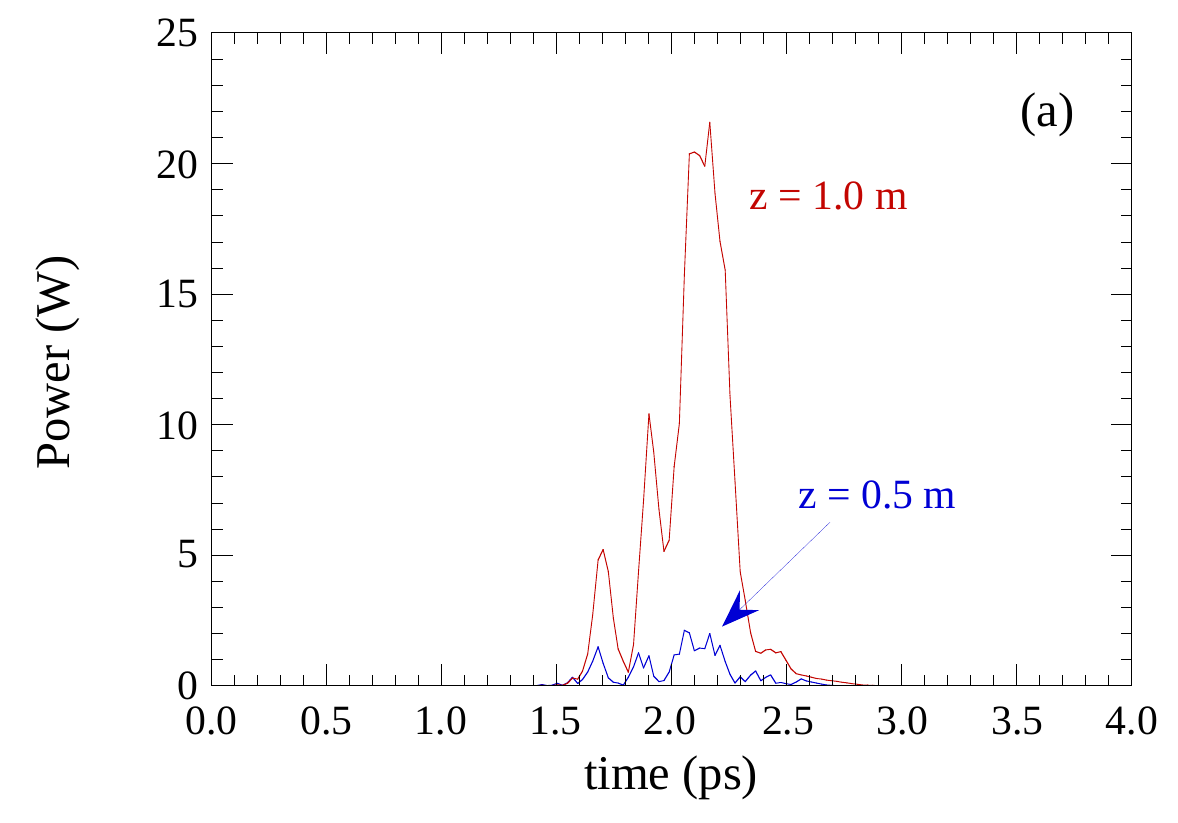}
         \caption{Locations $z=0.5$ (blue) and 1.0 m (red).}
         \label{fig:ir_rafel_P_t_z=0.5_1.0}
     \end{subfigure}
     \hfill
     \begin{subfigure}[t]{0.45\textwidth}
         \centering
         \includegraphics[width=\textwidth]{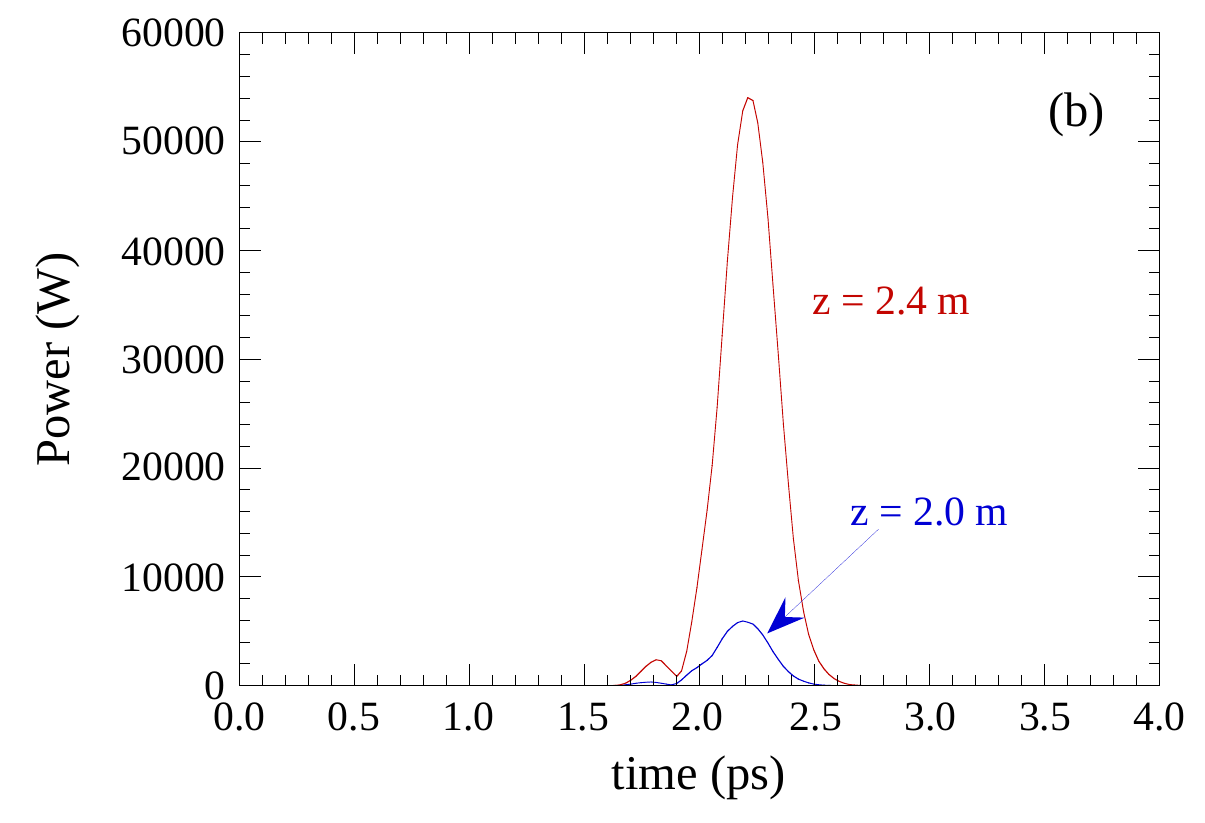}
         \caption{Locations $z=2.0$ (blue) and 2.4 m (red).}
         \label{fig:ir_rafel_P_t_z=2.0_2.4}
     \end{subfigure}
     \vspace{1mm}
        \caption{Temporal pulse shape at various locations along the undulator for the first pass when using a cavity detuning $\Delta L_{\textrm{cav}}=-18\lambda$ and an undulator field strength of $B_u=6.658$~kG ($K_{\textrm{rms}}$ = 1.055). The other parameters are given in Table~\ref{tab:ir_rafel}.}
        \label{fig:ir_rafel_p_t_n=1}
\end{figure}

Ignoring the hole in the out-coupling mirror, the waist (0.59 mm) of the fundamental cold cavity mode is designed to be about $\sqrt{2}$ times the matched electron beam radius in the undulator. The FWHM of the fundamental cold cavity mode at the undulator exit and downstream mirror are 1.84 and 4.82 mm respectively. We thus observe that the optical mode in the RAFEL expands faster from the undulator exit to the downstream mirror than the fundamental cold cavity mode (factor 3.25 and 2.62 respectively). That the RAFEL mode size is still smaller at the downstream mirror is the result of a balance between the faster expansion and smaller spot size of the RAFEL optical mode at the undulator exit compared to the cold cavity mode. The smaller spot size at the undulator exit is due to gain guiding as described above. Both, the smaller spot size at the undulator exit and faster expansion of the RAFEL optical beam again indicate that at the undulator exit, the optical field consists of fundamental and higher order cold-cavity modes. Note, a fundamental Gaussian beam having a waist at the undulator exit with the same size as the RAFEL optical beam would have a Rayleigh range of 1.49 m. Finally, the cross section of the mode incident on the upstream mirror is shown in Figure~\ref{fig:ir_rafel_P_x_um}. After reflection from the upstream mirror and propagation through the resonator to the undulator entrance, this field results in a modal pattern similar to that shown Figure~\ref{fig:ir_rafel_p_x}.

Now consider the formation of temporal coherence. Since the RAFEL starts from shot noise on the beam, the initial growth of the mode starts from spiky noise and, just as in a SASE FEL, develops coherence as it propagates through the undulator. However, the undulator is not long enough to reach saturation on a single pass, so that the evolution to temporal coherence is expected to develop over multiple passes. In Figure~\ref{fig:ir_rafel_p_t_n=1} we plot the temporal pulse shapes found on the first pass (a) at z = 0.5 m (blue) and 1.0 m (red), and (b) at z = 2.0 m (blue) and 2.4 m (red), the latter being at the undulator exit, for an on-axis undulator field amplitude $B_u$ = 6.658 kG ($K_{\textrm{rms}}$ = 1.055). Figure~\ref{fig:ir_rafel_p_t_n=1} shows the full simulation time window, and it should be noted that the electron beam is centered in the time window with a full (parabolic) width of 1.2 ps. The pulse shows many spikes at 0.5 m, but that it has coalesced into about 5 spikes after 1.0 m. The development of temporal coherence continues until at the undulator exit at 2.4 m only two spikes remain.

\begin{figure}[t]
     \centering
     \begin{subfigure}[t]{0.45\textwidth}
         \centering
         \includegraphics[width=\textwidth]{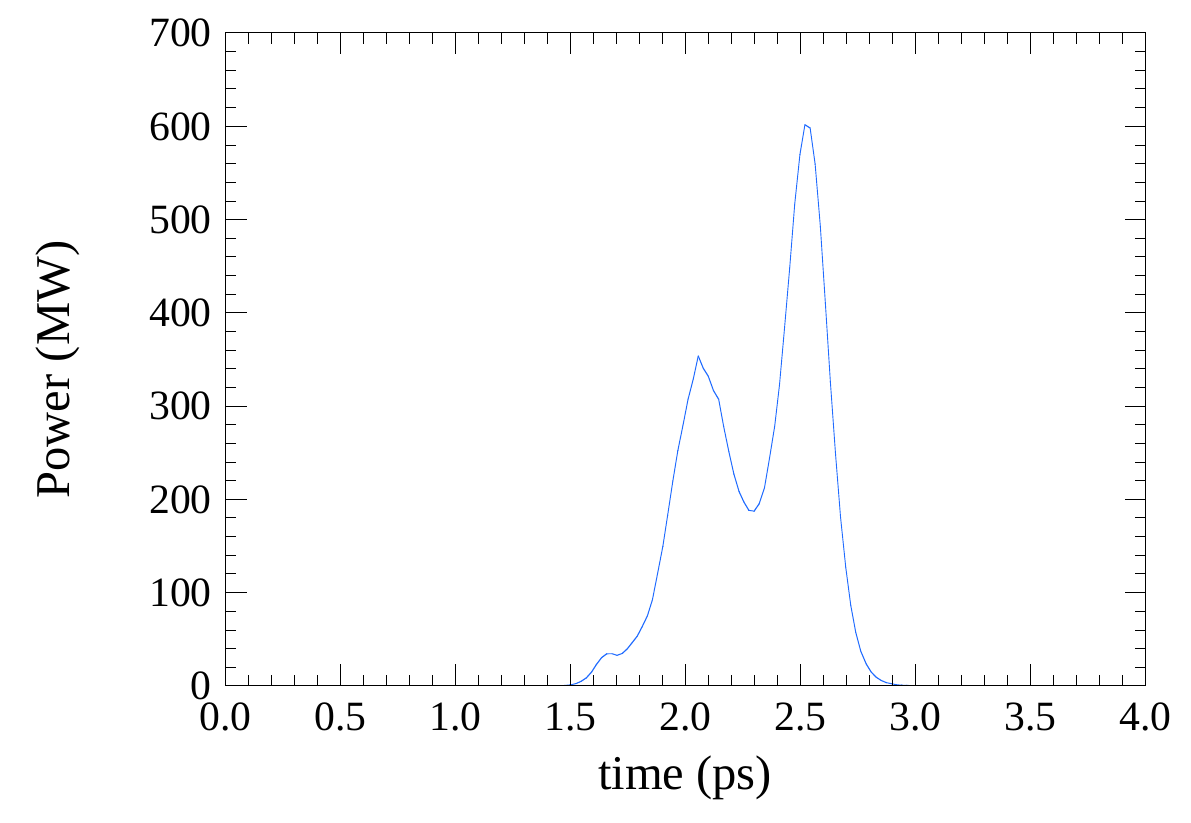}
         \caption{$\Delta L_{\textrm{cav}}=0$}
         \label{fig:ir_rafel_P_t_dLc=0_n=60}
     \end{subfigure}
     \hfill
     \begin{subfigure}[t]{0.45\textwidth}
         \centering
         \includegraphics[width=\textwidth]{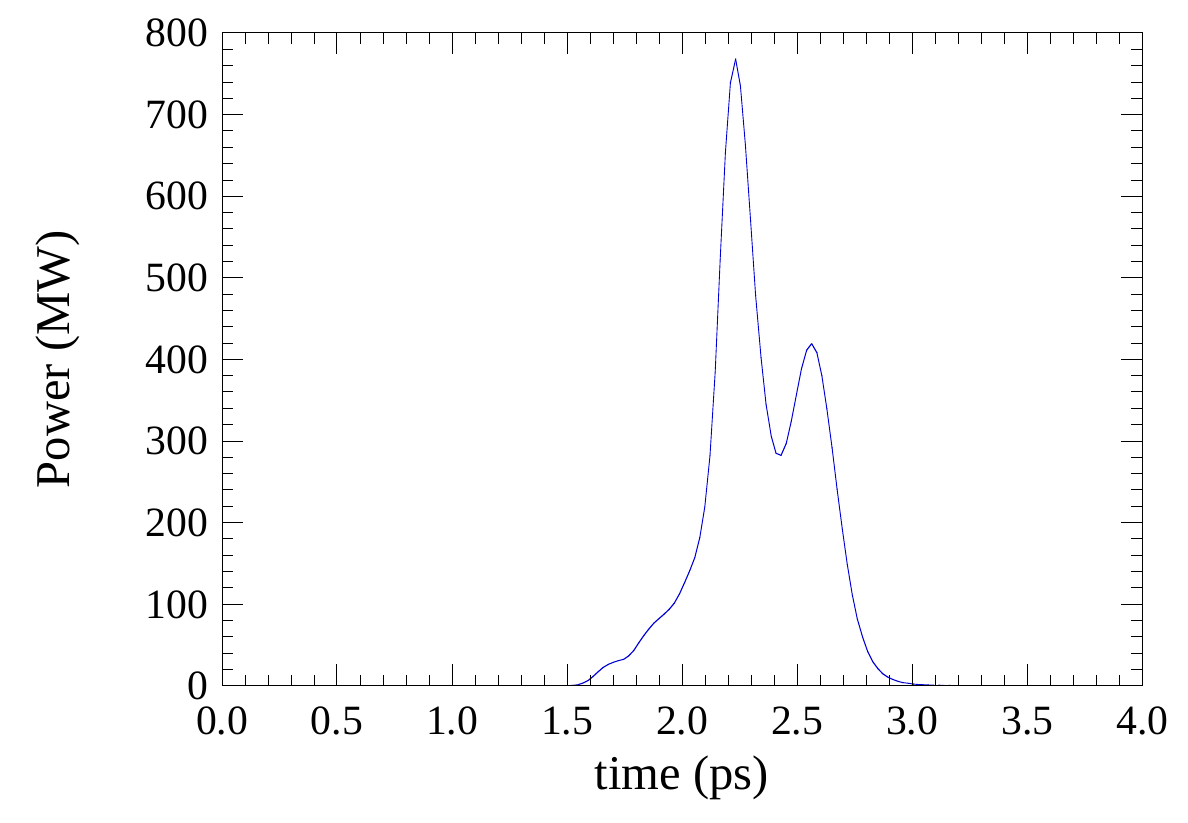}
         \caption{$\Delta L_{\textrm{cav}}=-8\lambda$}
         \label{fig:ir_rafel_P_t_dLc=-8_n=60}
     \end{subfigure}
     
     \vspace{2mm}
     \begin{subfigure}[t]{0.45\textwidth}
         \centering
         \includegraphics[width=\textwidth]{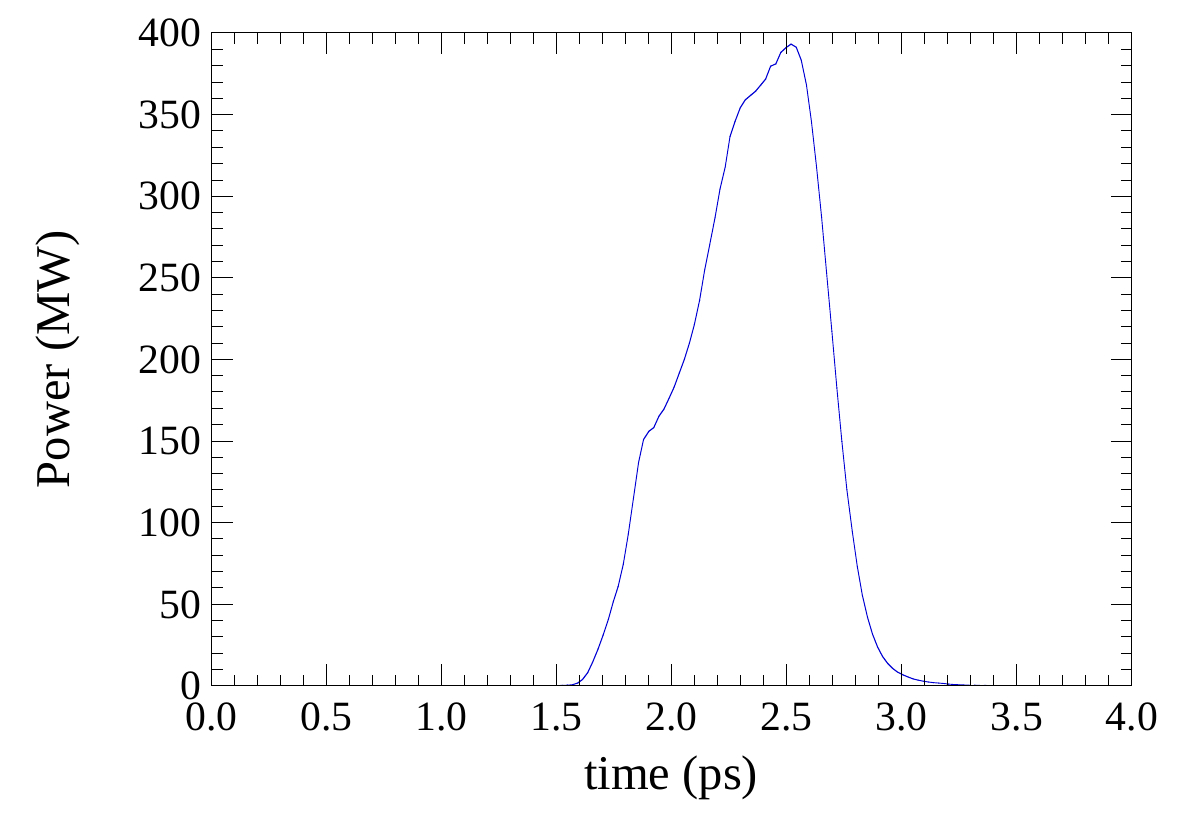}
         \caption{$\Delta L_{\textrm{cav}}=-18\lambda$}
         \label{fig:ir_rafel_P_t_dLc=-18_n=60}
     \end{subfigure}
     \hfill
     \begin{subfigure}[t]{0.45\textwidth}
         \centering
         \includegraphics[width=\textwidth]{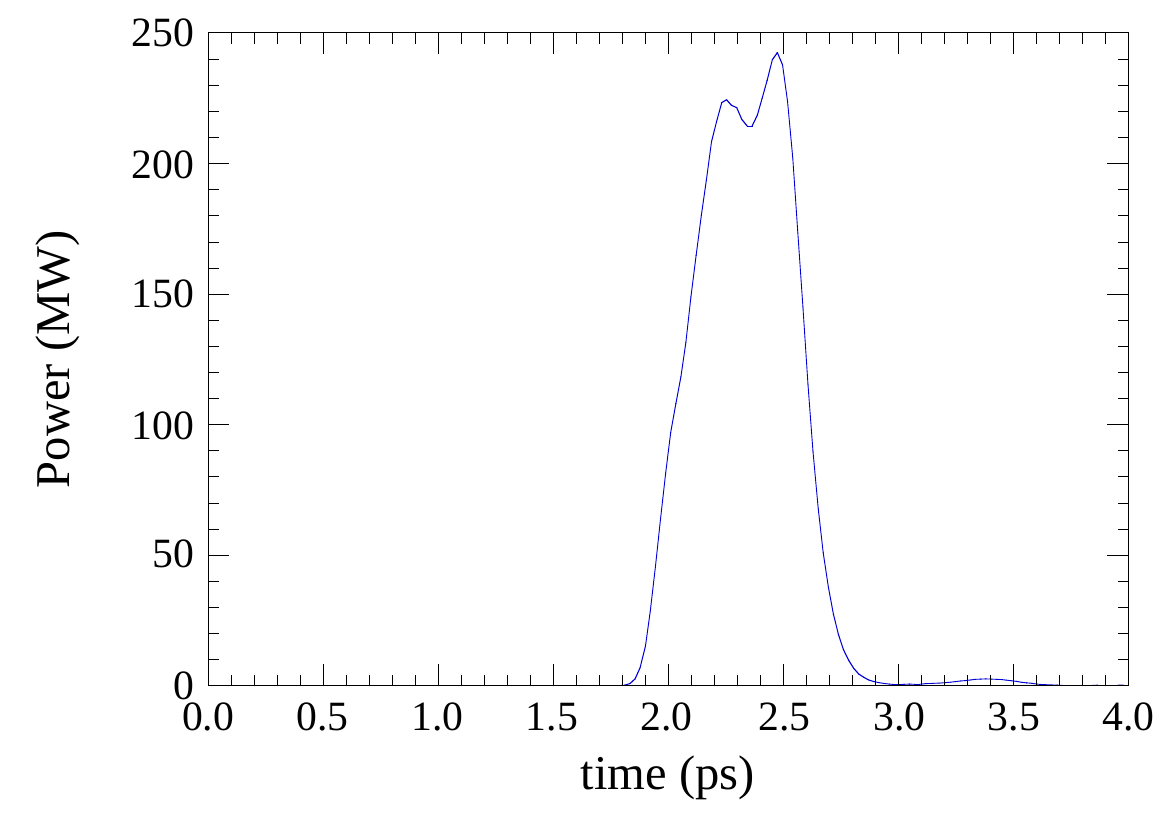}
         \caption{$\Delta L_{\textrm{cav}}=-43\lambda$}
         \label{fig:ir_rafel_P_t_dLc=-43_n=60}
     \end{subfigure}
     \vspace{1mm}
        \caption{Temporal pulse shape at the exit of the undulator after 60 passes and for various cavity detunings. The undulator field strength is $B_u=6.658$~kG ($K_{\textrm{rms}}$ = 1.055). The other parameters are given in Table~\ref{tab:ir_rafel}.}
        \label{fig:ir_rafel_p_t_dLc=x_n=60}
\end{figure}

The multi-pass development of temporal coherence after the first pass depends strongly on the cavity detuning. The temporal pulse shapes on the 60th pass at the undulator exit for $B_u$ = 6.658 kG ($K_{\textrm{rms}}$ = 1.055) and for cavity detunings of $\Delta L_{\textrm{cav}}=  0, -8\lambda, -18\lambda$ and $-43\lambda$  are shown in Figure~\ref{fig:ir_rafel_p_t_dLc=x_n=60}. As demonstrated previously, the synchronized cavity length for a RAFEL is shorter than the synchronized cavity length of a low-gain FEL oscillator. Therefore, as we have used the speed of light \textit{in vacuo} instead of the actual group velocity to define the cavity detuning, we note that $\Delta L_{\textrm{cav}}=0$ actually corresponds to a cavity length that is larger than the synchronized length for the high-gain RAFEL. Consequently, the returning optical pulse will lag behind the center of the electron bunch. The pulse will be amplified as it propagates through the undulator, but as shown in Figure~\ref{fig:ir_rafel_p_t_n=1}, the two spikes formed over the first pass remain. This behavior is also found for small detunings as shown in Figure~\ref{fig:ir_rafel_P_t_dLc=-8_n=60} for $\Delta L_{\textrm{cav}}=-8\lambda$.

If the detuning is closer to the center of the detuning range, then the synchronism between the returning optical pulse and the electrons is a better match, and the multiple spikes are “washed out”. This is shown in Figure~\ref{fig:ir_rafel_P_t_dLc=-18_n=60}, where $\Delta L_{\textrm{cav}}=-18\lambda$  and we  see that a broader pulse has formed. As the cavity length is decreased further, the optical pulse arrives increasingly near the head of the electron bunch. In this case, there is not enough gain to wash out the multi-spike character of the signal. This is shown in Figure~\ref{fig:ir_rafel_P_t_dLc=-43_n=60} for $\Delta L_{\textrm{cav}} = -43\lambda$ where the total power (or pulse energy is much reduced).

\subsection{An X-Ray RAFEL}
The impetus driving research into x-ray free-electron laser oscillators (XFELOs) and RAFELs is the character of SASE emission. Since there are no seed lasers available at x-ray wavelengths, fourth generation FEL light sources \cite{emma_2010, ishikawa_2012, milne_2017, kang_2017, ko_2017, weise_2017} at these wavelengths are based on SASE over a single pass through a long undulator line. While pulse energies of the order of 2 mJ have been achieved at {\AA}ngstrom to sub-{\AA}ngstrom wavelengths, with the potential to reach multi-TW peak powers \cite{tanaka_2013, prat_2015, prat_2015a, emma_2016, freund_2018a, shim_2018}, SASE exhibits shot-to-shot fluctuations in the output spectra and power of about 10 – 20 percent. For many applications, these fluctuations are undesirable, and efforts are underway to find alternatives.

The utility of an XFELO has been under study for a decade \cite{kim_2008, lindberg_2009, kim_2009, Lindberg2011, kim_2012, kim_2016, qin_2017} making use of resonators based upon Bragg scattering from atomic layers within diamond crystals \cite{shvydko_2010, shvydko_2011, shvydko_2013, kolodziej_2016, kolodziej_2018}. The development of these crystals is a major breakthrough in the path toward an XFELO. Estimates indicate that using a superconducting RF linac producing 8 GeV electrons at a 1 MHz repetition rate is capable of producing $10^{10}$ photons per pulse at a 0.86 \AA~wavelength with a FWHM bandwidth of about $2.1 \times 10^{-7}$. This design is consistent with the LCLS-II High Energy Upgrade \cite{lcls-2}. As a consequence, an XFELO on a facility such as the LCLS-II and LCLS-II-HE is expected to result in a decrease in SASE fluctuations in the power and spectrum and to narrow the spectral linewidth.

As with the majority of FELOs to date, the aforementioned XFELOs use low gain/high-$Q$ resonators with transmissive out-coupling through thin diamond crystals \cite{kim_2009}. Potential difficulties with low-gain/high-$Q$ resonators derive from sensitivities to electron beam properties, mirror loading and alignments. In addition, transmissive out-coupling with high intra-cavity power can result in mirror damage. While experiments show that diamond crystals can sustain relatively high thermal and radiation loads \cite{kolodziej_2018}, transmissive out-coupling cannot be easily achieved at the photon energies of interest here. Because of this, x-ray RAFELs \cite{freund_2019, huang_2006, marcus_2020} using a variety of out-coupling schemes are receiving a great deal of attention, and we consider a RAFEL design using a pinhole in one of the mirrors \cite{nguyen_1999, freund_2019, huang_2006} here.

Optics propagation codes such as OPC must treat reflections from the diamond crystal Bragg mirrors where the mirror losses and angles of reflection depend on the crystal orientation/geometry and the x-ray photon energy. X-ray Bragg mirrors typically have a very narrow reflection bandwidth and a narrow angle of acceptance \cite{shvydko_2004}. For the x-ray RAFEL, and for computational efficiency, a temporal Fourier transform is applied at the beginning of the optical path when the optical field is passed from MINERVA to OPC and the propagation is performed in the wavelength domain, \textit{i.e.}, each wavelength is independently propagated through the resonator using the modified Fresnel propagator (Equation~\ref{eq:opc_fresnel_modified}). The inverse Fourier transform is calculated at the end of the optical path, before the field is handed back to MINERVA. As the optical field inside the cavity is typically not collimated, a spatial Fourier transform in the transverse coordinates is calculated for each of the wavelengths when a Bragg mirror is encountered. Each combination of transverse and longitudinal wavenumber corresponds to a certain photon energy and angle of incidence on the Bragg mirror and these parameters are used to calculate the complex reflection and transmission coefficients of the Bragg mirror \cite{shvydko_2004}. After applying the appropriate parameter to the optical field, depending on whether it is reflected or transmitted, the inverse spatial Fourier transform is calculated and the field is propagated to the next optical element along the path until the end is reached.

\begin{figure}[t] 
\centering
\includegraphics[width=12cm]{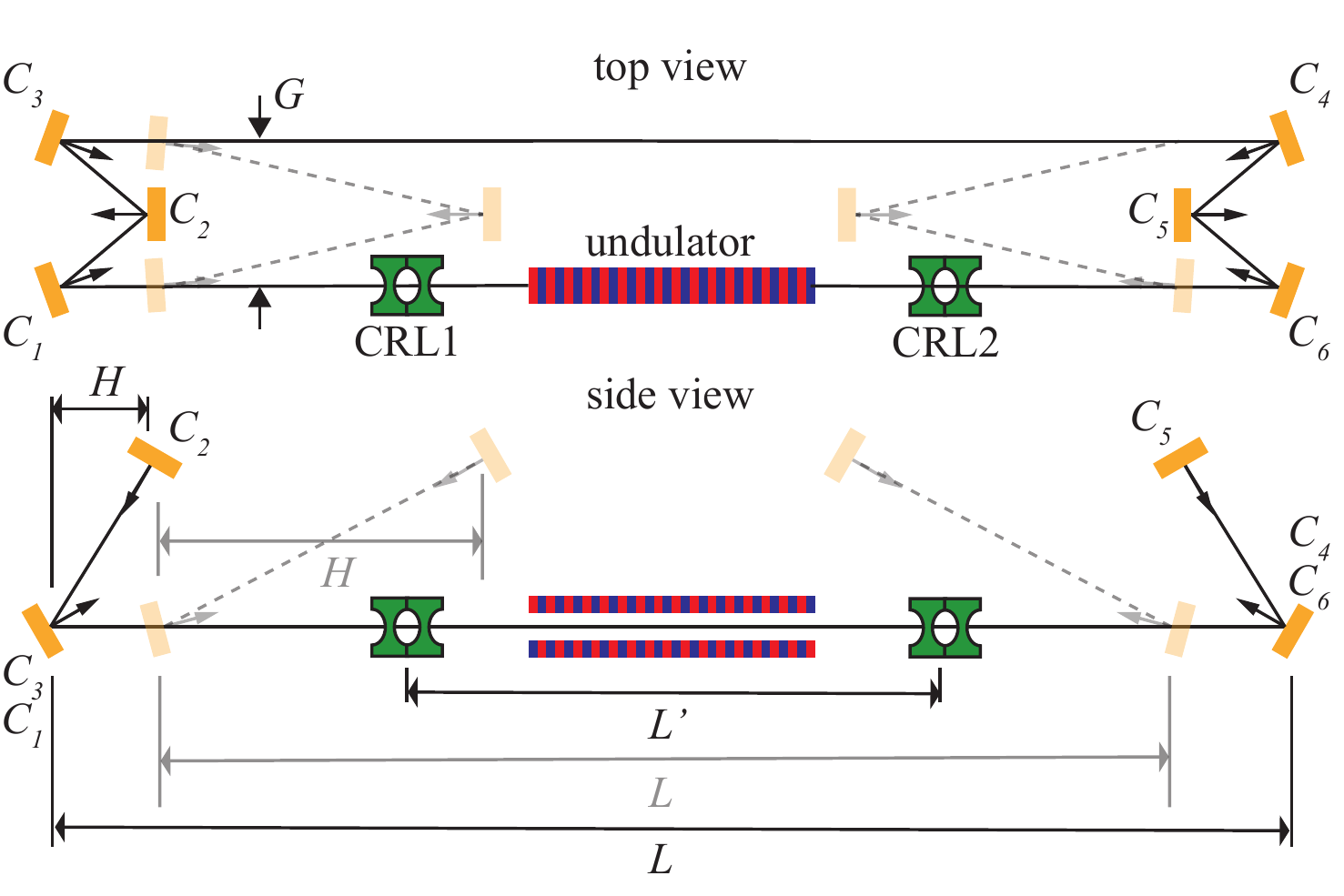}
\caption{\label{fig:xfelo_res}Schematic view of a 6-mirror ring resonator for an x-ray FEL oscillator using plane Bragg mirrors $C_1 \dots C_6$ and compound refractive lenses, CRL1 and CRL2 for focusing. After \cite{shvydko_2013}.}
\end{figure} 

Here, we consider a six-crystal, tunable, compact cavity \cite{shvydko_2013} as illustrated in Figure~\ref{fig:xfelo_res}, which shows both a top and a side view. The crystals are arranged in a non-coplanar (3-D) scattering geometry. There are two backscattering units comprising three crystals ($C_1$, $C_2$, and $C_3$) on one side of the undulator and three crystals ($C_4$, $C_5$, and $C_6$) on the other side. Collimating and focusing elements are shown as CRL1,2, which could be grazing-incidence mirrors but are represented in Figure~\ref{fig:xfelo_res} by another possible alternative – compound refractive lenses \cite{snigirev_1996, lengeler_1999}. In each backscattering unit, three successive Bragg reflections take place from three individual crystals to reverse the direction of the beam from the undulator. Assuming that all the crystals and Bragg reflections are the same, the Bragg angles can be chosen within the range 30º < $\theta$ < 90º; however, Bragg angles close to $\theta$= 45º should be avoided to ensure high reflectivity for both linear polarization components, as the reflection plane orientations for each crystal change. The cavity allows for tuning the photon energy in a large spectral range by synchronously changing all Bragg angles. In addition, to ensure constant time of flight, the distance $L$ (which brackets the undulator), and the distance between crystals as characterized by $H$ have to be changed with $\theta$. The lateral size $G$ is kept constant as the resonator is tuned.

Because the $C_1C_6$ and $C_3C_4$ lines are fixed, intracavity radiation can be out-coupled simultaneously for several users at different places in the cavity, although we only consider out-coupling through $C_6$ at the present time. Out-coupling through crystals $C_3$ and $C_6$ are most favourable, since the direction of the out-coupled beams do not change with photon energy, but out-coupling for more users through crystals $C_1$ and $C_4$ is also possible. Such multi-user capability is in stark contrast with present SASE beamlines which support one user at a time. We consider that the electron beam propagates from left to right through the undulator and the out-coupling is accomplished through a pinhole in the first downstream mirror ($C_6$).

Consider the LCLS-II beamline \cite{lcls-2} with the HXR undulator corresponding to an electron energy of 4.0 GeV, a bunch charge in the range of 10 – 30 pC with an rms bunch duration (length) at the undulator of 2 – 173 fs (0.6 – 52 $\mu$m) and a repetition rate of 1 MHz. The peak current at the undulator is 1000 A with a normalized emittance of 0.2 – 0.7 mm-mrad, and an rms energy spread of about 125 – 1500 keV. The HXR undulator \cite{lcls-2} is a plane-polarized, hybrid permanent magnet undulator with a variable gap, a period of 2.6 cm, and a peak field of 10 kG. Each HXR undulator has 130 periods. In the simulation each undulator is modeled using Equation~\ref{eq:mag_fpf} and we consider that the first and last period describe an entry/exit taper. There is a total of 32 segments that can be installed. The break sections between the undulators are 1.0 m in length and contain steering, focusing and diagnostic elements, although we only consider the focusing quadrupoles in the simulation which we position in the center of the breaks. The quadrupoles are assumed to be 7.4 cm in length with a field gradient of 1.71 kG/cm. 

A fundamental resonance at 3.05 keV (= 4.07 \AA) implies an undulator field of 5.61 kG. We assume that the electron beam has a normalized emittance of 0.45 mm-mrad and a relative energy spread of $1.25 \times 10^{-4}$, corresponding to the nominal design specification for LCLS-II. This yields a Pierce parameter of $\rho = 5.4 \times 10^{-4}$. In order to match this beam into the undulator/FODO line, the initial beam size in the $x(y)$-direction is 37.87(31.99) $\mu$m with Twiss $\alpha_x = 1.205$ ($\alpha_y = -0.8656$). Note that this yields Twiss $\beta_x = 24.9$ m and $\beta_y = 17.80$ m.

The resonator dimensions were fixed by means of estimates of the gain using the Ming Xie parameterization \cite{mingxie_1995}, and MINERVA simulations indicated that about 40 – 60 m of undulator would be required to operate as a RAFEL. As such, the distance, L, between the two mirrors framing the undulator was chosen to be 130 m, which is also the distance separating the two mirrors on the back side of the resonator (elements $C_3$ and $C_4$). In studying the cavity tuning via time-dependent simulations, these two distances are allowed to vary while holding fixed the configurations of the \textit{backscattering} units. The compound refractive lenses, which are modeled as thin lenses by OPC, are placed symmetrically around the undulator and are designed to place the optical focus at the center of the undulator \textit{in vacuo}. In this study, the focal length is approximately 94.5 m.

In order to out-couple the x-rays through a transmissive mirror at the wavelength of interest, the diamond crystal would need to be impractically thin (about 5 $\mu$m); hence, we consider out-coupling through a hole in the first downstream mirror ($C_6$). We consider all the mirrors to be 100 $\mu$m thick. Due to the high computational requirements of time-dependent simulations, we begin with an optimization of the RAFEL with respect to the hole radius and the undulator length using steady-state (\textit{i.e.}, time-independent) simulations.

\begin{figure}[t] 
\centering
\includegraphics[width=8cm]{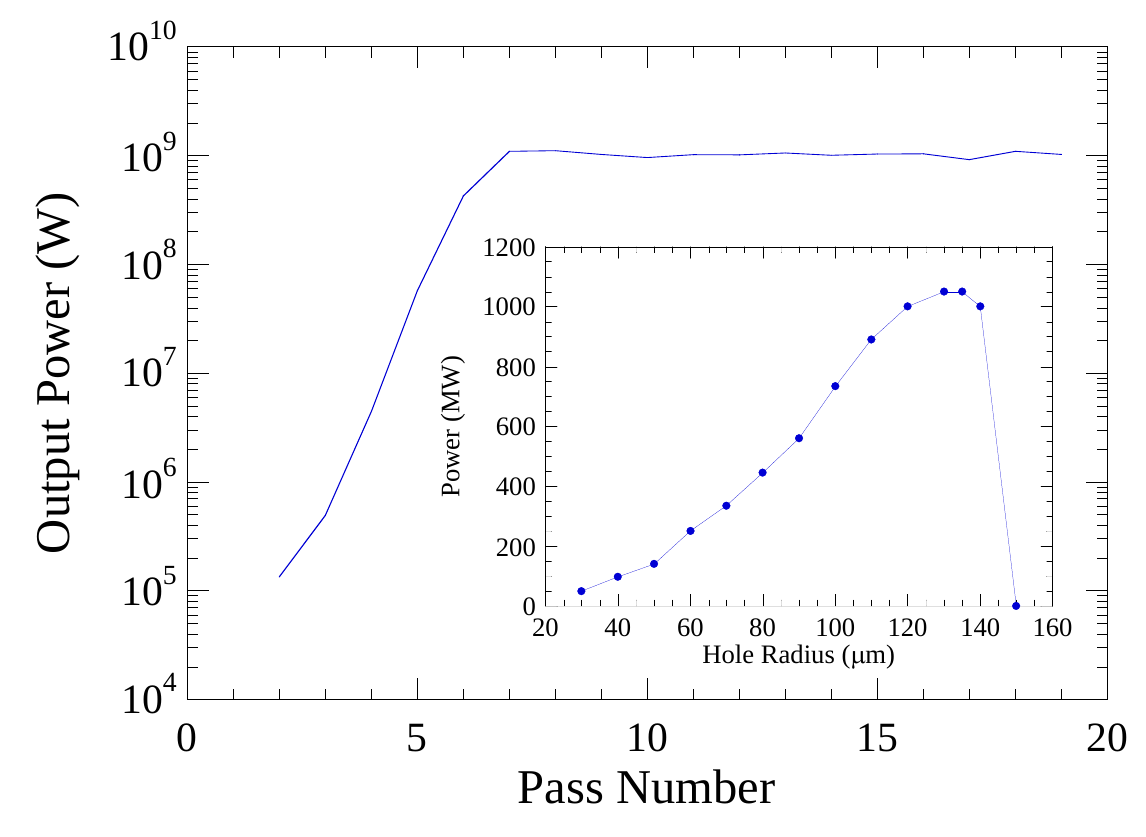}
\caption{\label{fig:x_rafel_p_n}Output power versus number of passes through the resonator for a hole radius of 135~$\mu$m. The inset shows the output power as a function of hole radius. Other simulation parameters as described in the text.}
\end{figure} 

The choice of hole radius is important because if the hole is too small then the bulk of the power remains within the resonator while if the hole is too large then the losses become too great and the RAFEL cannot lase. The results for the optimization of the hole radius indicate that the optimum hole radius is 135 $\mu$m which allows for 90 percent out-coupling, where we fixed the undulator line to consist of 11 HXR undulator segments. This is shown in Figure~\ref{fig:x_rafel_p_n} where we plot the output power as a function of pass number for the optimum hole radius and the variation in the saturated power with the hole radius (inset).

\begin{figure}[t] 
\centering
\includegraphics[width=8cm]{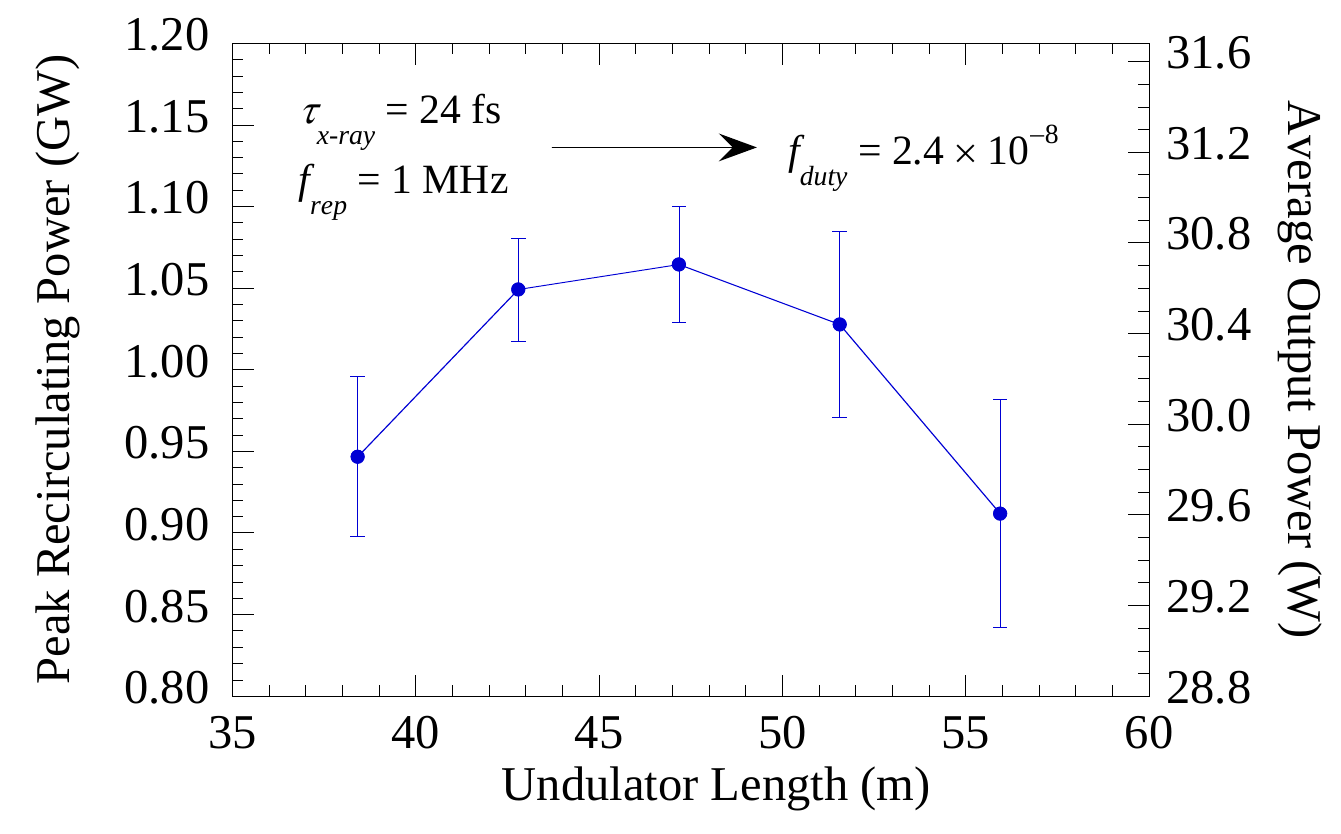}
\caption{\label{fig:x_rafel_pcir_Lw}Recirculating output power versus length of the undulator $L_u$ for a hole radius of 135 $\mu$m. Other simulation parameters as described in the text.}
\end{figure} 

A local optimization on the undulator length for a hole radius of 135 $\mu$m is shown in Figure~\ref{fig:x_rafel_pcir_Lw} where we plot the peak recirculating power (left axis) and the average output power (right axis). The error bars in the figure indicate the level of pass-to-pass fluctuations in the power which is generally smaller than the level of shot-to-shot fluctuations in SASE. Note that while this represents steady-state simulations, the average power is calculated under the assumption of an electron bunch with a flat-top temporal profile having a duration of 24 fs which yields a duty factor of $2.4 \times 10^{-8}$. Each point in the figure refers to a given number of HXR undulators ranging from 9 – 13 segments. It is evident from the figure that the optimum length is 47.18 m corresponding to 11 segments.

\begin{figure}[t] 
\centering
\includegraphics[width=8cm]{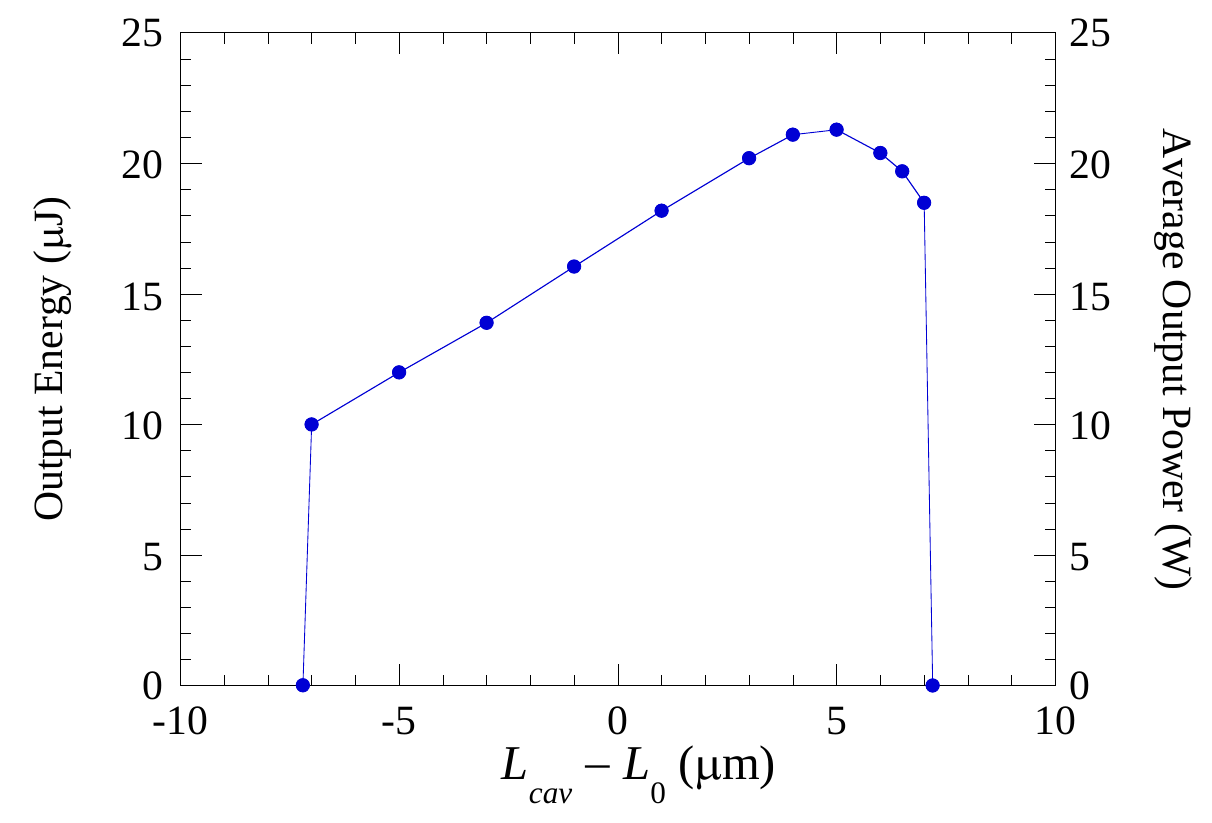}
\caption{\label{fig:x_rafel_P_cav_detune}Output energy as a function of the cavity detuning $\Delta L_{\textrm{cav}}=L_{\textrm{cav}}-L_0$ for a hole radius of 135~$\mu$m. The simulation parameters are as described in the text.}
\end{figure} 

Time-dependent simulations were performed under the assumption of electron bunches with a flat-top temporal profile having a full width duration of 24 fs and a peak current of 1000 A. This corresponds to a bunch charge of 24 pC. The detuning curve defining what cavity lengths are synchronized with the repetition rate of the electrons is shown in Figure~\ref{fig:x_rafel_P_cav_detune}. For simplicity we take the synchronous cavity length $L_0$ as $L_0 = c/f_{\textrm{rep}}$ (\textit{cf.} Equation~\ref{eq:zero_detuning_low_gain}), where a single optical pulse is assumed to be inside the cavity and the effect of gain on the group velocity is ignored. Here $L_0 = 299.7924580$ m. The range of  cavity lengths with overlap between the optical pulse and an electron pulse with a length of about 7.2\,$\mu$m is given by $L_0 - 7.2\:\mu\textrm{m} < L_{\textrm{cav}} < L_0 + 7.2\:\mu$m, where $L_{\textrm{cav}}$ is taken here to be the total roundtrip length of the cavity.

\begin{figure}[t] 
\centering
\includegraphics[width=8cm]{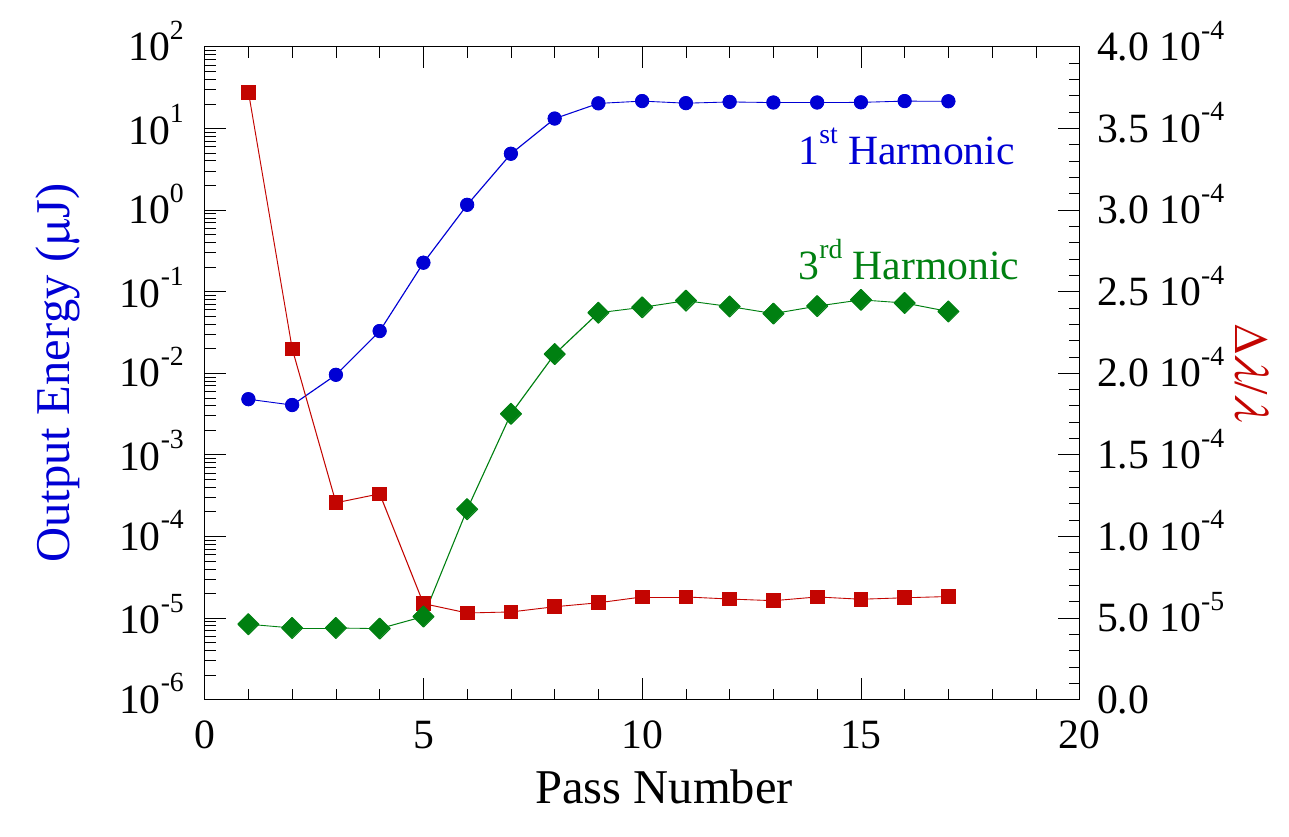}
\caption{\label{fig:x_rafel_E_n}Evolution of the fundamental pulse energy (blue, left) and the 3rd harmonic (green, left), as well as the relative linewidth (red, right) when the cavity detuning is $\Delta L_{\textrm{cav}}=5$ $\mu$m and for an electron beam energy spread $\Delta E_b/E_b = 1.25 \times 10^{-4}$. Other parameters as described in the text.}
\end{figure} 

The evolution of the output energy at the fundamental and the 3rd harmonic, and the spectral linewidth of the fundamental, vs pass are shown in Figure~\ref{fig:x_rafel_E_n} for a detuning of 5 $\mu$m which is close to the peak in the detuning curve (Figure~\ref{fig:x_rafel_P_cav_detune}). While it is not evident in the figure, the rms fluctuation in the energy from pass to pass is of the order of about 3 percent, which is lower than the shot-to-shot fluctuations expected from SASE. At least as important as the output power is that the linewidth contracts substantially during the exponential growth phase and remains constant through saturation. Starting with a linewidth of about $3.7 \times 10^{-4}$ after the first pass corresponding to SASE, the linewidth contracts to about $6.0 \times 10^{-5}$ at saturation. The SASE linewidth after the first pass through the undulator is slightly smaller than the predicted linewidth based on 1-D theory \cite{huang_2006} which is approximately $4.5 \times 10^{-4}$. Hence, the RAFEL is expected to have both high average power and a stable narrow linewidth, which is to a large extend determined by the spectral filtering of the Bragg mirrors.

The 3rd harmonic grows parasitically from high powers/pulse energies at the fundamental in a single pass through the undulator \cite{freund_2000} and has been shown to reach output intensities of 0.1 percent that of the fundamental in a variety of FEL configurations, and this is what we find in the RAFEL simulations. As shown in Figure~\ref{fig:x_rafel_E_n}, the 3rd harmonic intensity remains small until the fundamental pulse energy reaches about 1 $\mu$J after which it grows rapidly and saturates after about 12 passes. This is close to the point at which the fundamental saturates as well. The saturated pulse energies at the 3rd harmonic reach about 0.067 $\mu$J. Given a repetition rate of 1 MHz, this corresponds to a long-term average power of 67 mW.

\begin{figure}[t]
     \centering
     \begin{subfigure}[t]{0.45\textwidth}
         \centering
         \includegraphics[width=\textwidth]{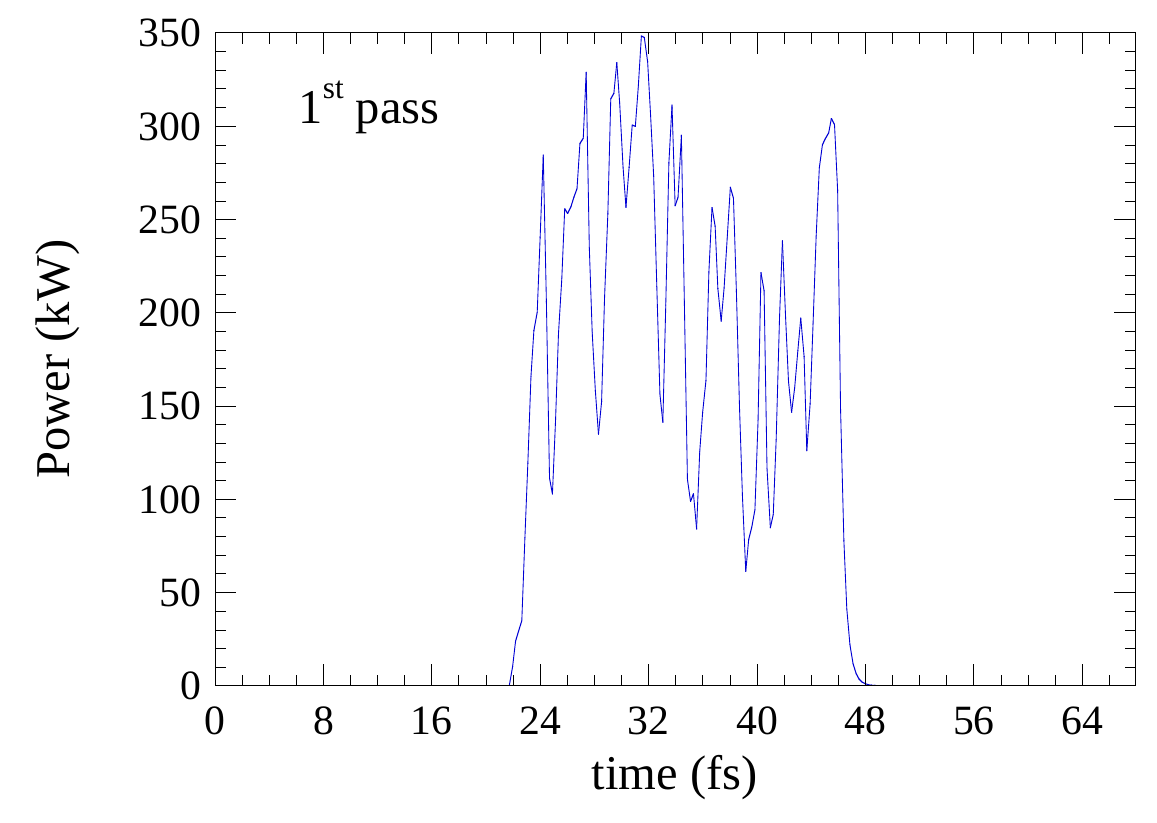}
         \caption{Temporal profile}
         \label{fig:x_rafel_P_t_n=1}
     \end{subfigure}
     \hfill
     \begin{subfigure}[t]{0.45\textwidth}
         \centering
         \includegraphics[width=\textwidth]{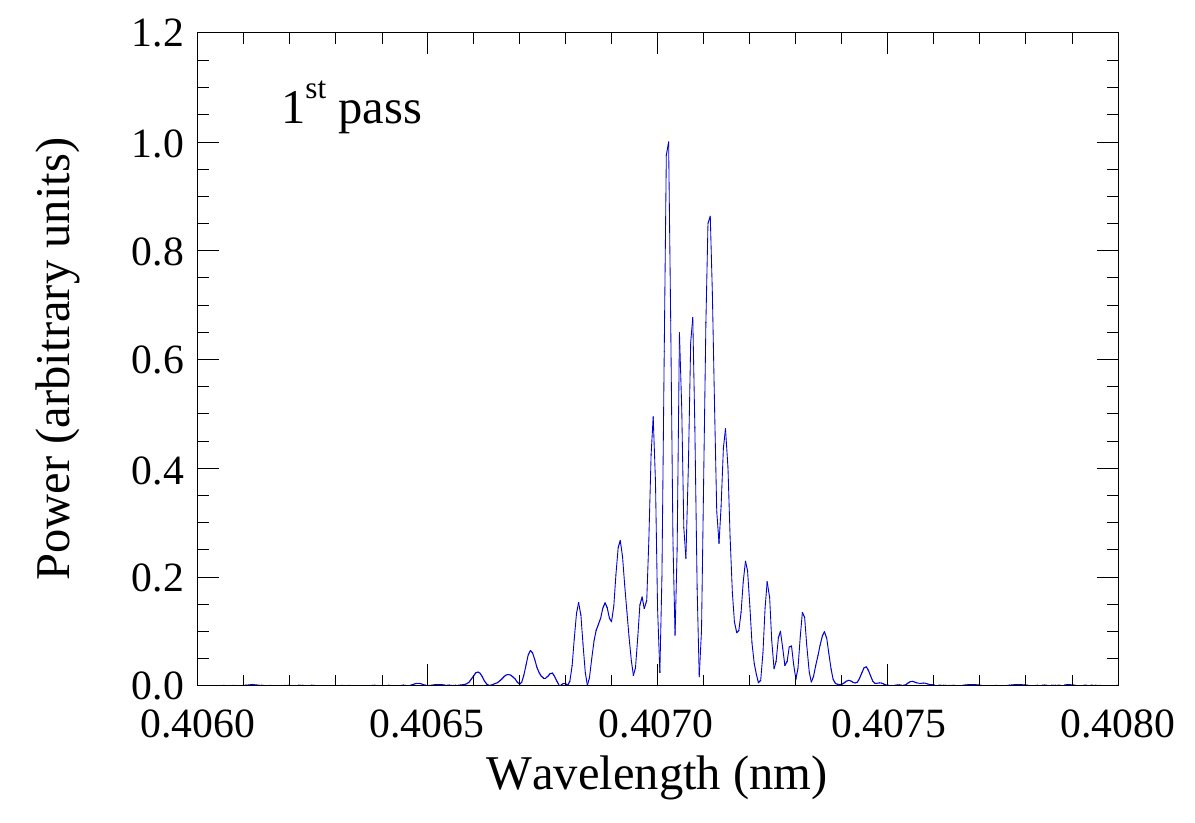}
         \caption{Spectrum}
         \label{fig:x_rafel_P_lambda_n=1}
     \end{subfigure}
     \vspace{1mm}
        \caption{Temporal pulse shape and spectrum of the optical pulse at the exit of the undulator for the first pass. Other parameters as described in the text.}
        \label{fig:x_rafel_p_t_lambda_n=1}
\end{figure}

The reduction in the linewidth after saturation shown in Figure~\ref{fig:x_rafel_E_n} indicates that a substantial level of longitudinal coherence has been achieved in the saturated regime. The RAFEL starts from shot noise on the beam during the first pass through the undulator, and longitudinal coherence develops over the subsequent passes. Hence, we expect that the temporal profile of the optical field will exhibit the typical spiky structure associated with SASE at the undulator exit after the first pass and this is indeed depicted in Figure~\ref{fig:x_rafel_P_t_n=1} which shows the temporal profile at the undulator exit after the first pass. The number of spikes expected, $N_{\textrm{spikes}}$, is given approximately by $N_{\textrm{spikes}}$ = $l_b$/(2$\pi l_c$), where $l_b$ is the rms bunch length and $l_c$ is the coherence length. For the present case, $l_b$ = 7.2 $\mu$m and $l_c$ = 60 nm; hence, we expect that $N_{\textrm{spikes}}$ = 19. We observe about 14 spikes in Figure~\ref{fig:x_rafel_P_t_n=1} which is in reasonable agreement with the expectation. Note that the time axis encompasses the time window used in the simulation. As indicated in Figure~\ref{fig:x_rafel_E_n}, the linewidth after the first pass is of the order of $4.3 \times 10^{-4}$ which is relatively broad and corresponds to the interaction due to SASE. This is reflected in the output spectrum from the undulator after the first pass which is shown in Figure~\ref{fig:x_rafel_P_lambda_n=1}.

\begin{figure}[t]
     \centering
     \begin{subfigure}[t]{0.45\textwidth}
         \centering
         \includegraphics[width=\textwidth]{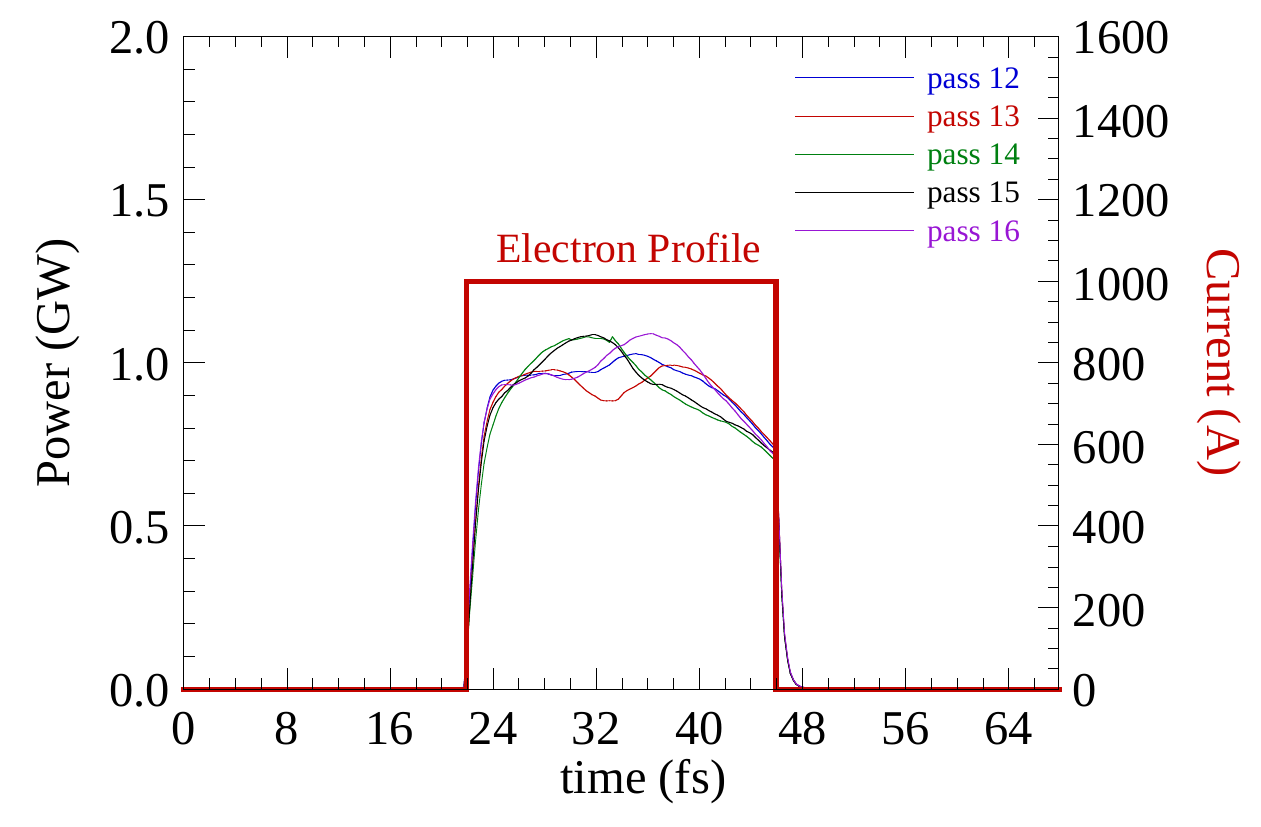}
         \caption{Temporal profile for the optical pulse after different number of passes and current pulse (red).}
         \label{fig:x_rafel_P_t_n=x}
     \end{subfigure}
     \hfill
     \begin{subfigure}[t]{0.45\textwidth}
         \centering
         \includegraphics[width=\textwidth]{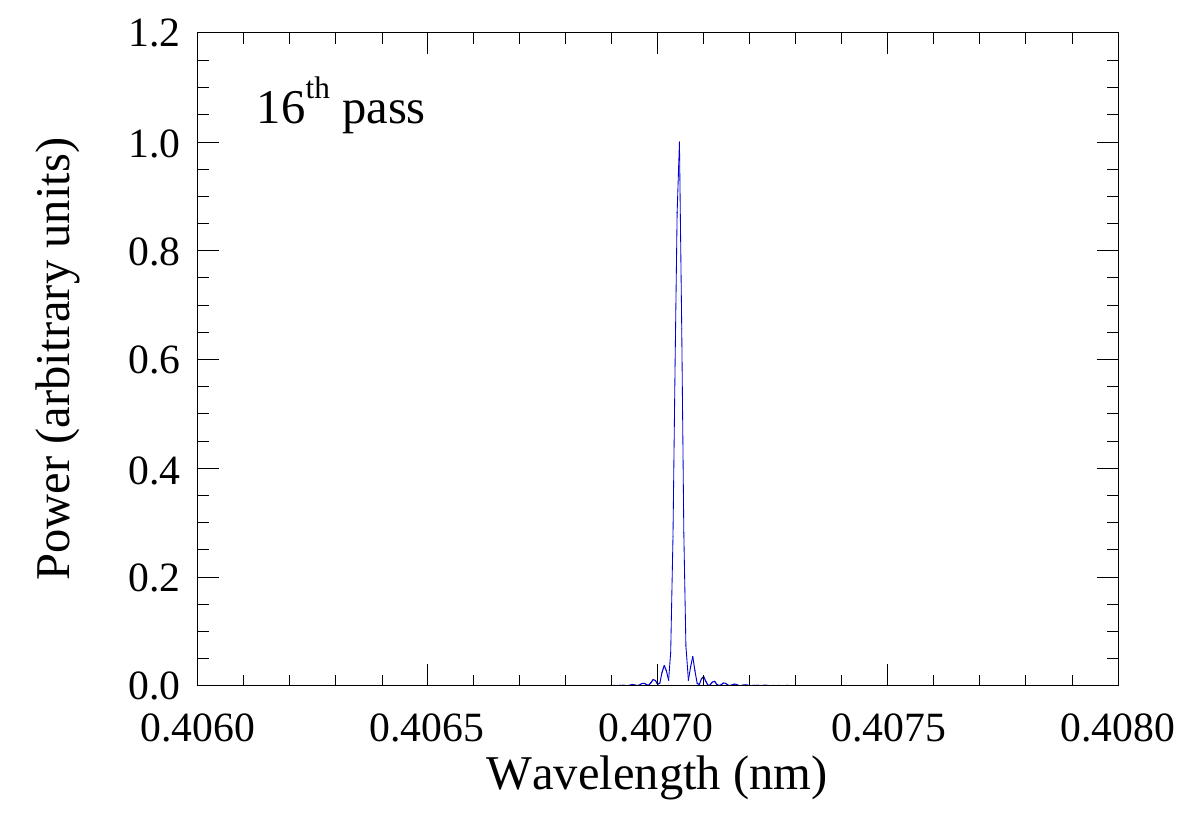}
         \caption{Spectrum at the exit of the undulator after 16 passes}
         \label{fig:x_rafel_P_lambda_n=16}
     \end{subfigure}
     \vspace{1mm}
        \caption{Temporal pulse shape for the electron bunch and optical pulse after different number of passes and the spectrum at pass number 16 of the optical pulse at the exit of the undulator. Other parameters as described in the text.}
        \label{fig:x_rafel_p_t_lambda_n=x}
\end{figure}

The spectral narrowing that is associated with the development of longitudinal coherence as the interaction approaches saturation results in a smoothing of the temporal profile. This is illustrated in Figure~\ref{fig:x_rafel_P_t_n=x}, where we plot the temporal profiles of the optical field at the undulator exit corresponding to passes 12 – 16 which are after saturation has been achieved (left axis). As shown in the figure, the temporal pulse shapes from pass-to-pass are relatively stable and exhibit a smooth plateau with a width of about 23 – 24 fs which corresponds to, and overlaps, the flat-top profile of the electron bunches which is shown on the right axis. Significantly, the smoothness of the profiles corresponds with the narrow linewidth and contrasts sharply with the SASE output after the first pass through the undulator (see Figure~\ref{fig:x_rafel_P_lambda_n=1}). Both the pass-to-pass stability and smoothness of the output pulses contrast markedly with the large shot-to-shot fluctuations and the spikiness expected from the output pulses in pure SASE.

The narrow relative linewidth in this regime of about $7.3 \times 10^{-5}$ at the undulator exit, as shown in Figure~\ref{fig:x_rafel_P_lambda_n=16} after pass 16, as well as the smooth temporal profiles shown in Figure~\ref{fig:x_rafel_P_t_n=x}, are associated with longitudinal coherence after saturation is achieved.

\section{Summary and Conclusion} \label{section:sum_concl}
Development of both low-gain/high-$Q$ and high-gain/low-$Q$ oscillators is an active and ongoing area of research into FEL sources over virtually the entire electromagnetic spectrum \cite{li_2019, sumitomo_2019, lee_2019, xu_2019, zhao_2020, zen_2020, marcus_2020, petrillo_2020, yan_2020, paraskaki_2021}. Both types of FEL oscillators have their own merits. For example, high-gain/low-$Q$ are typically better suited for higher peak power operation due to a reduced mirror loading \cite{vanderslot_2007} and higher extraction efficiency as found in Section~\ref{section:hg_lq}. On the other hand, properties like gain bandwidth, and stability of the output are closer to SASE FELs than low-gain/high-$Q$ FEL oscillators. However, we have not yet explored the full range of feedback and increasing the feedback may shift the performance in this respect closer to that of low-gain/high-$Q$ oscillators.

As we have described in this paper, great progress has been made in our ability to simulate both the interaction through the undulator and in multi-pass configurations through the undulator and the resonator. This was demonstrated in Section~\ref{section:lg_hq} by the excellent agreement found between the simulations and the IR Demo and 10-kW Upgrade experiments at JLab after which the simulations of both an infrared and an x-ray RAFEL were described in Section~\ref{section:hg_lq}. 

The combination of a three-dimensional, time-dependent FEL simulation code with an optics propagation code provides enormous flexibility in being able to study a wide range of undulator and optical resonator configurations. In particular, the use of a general FEL code such as MINERVA allows simulation of variable polarization (Section~\ref{section:minerva}) governed by the choice of undulator geometries (Section~\ref{section:b-models}) and ultra-short electron bunches \cite{campbell_2020} down to the neighborhood of the cooperation length in the undulators. The use of OPC permits the simulation of arbitrary optical resonators including thermal heating and distortion as well as Bragg reflections.

In summary, the numerical algorithms that have been implemented for the simulation of FEL oscillators are now sufficiently mature to serve as reliable design tools for virtually any conceivable undulator and oscillator configuration.

\section*{Funding}
The work described was supported over the course of time by multiple contracts with the U.S. Office of Naval Research, the Department of Defense Joint Technology Office, and the Department of Energy.

\section*{Acknowledgements}
The authors would like to thank numerous staff members at JLab for many helpful discussions about the IR Demo and 10-kW Upgrade experiments. We would also like to thank Heinz-Dieter Nuhn and Paul Emma for providing assistance with understanding the data from the LCLS and thank Luca Giannessi for assistance concerning the SPARC data.
The work described was supported over the course of time by multiple contracts with the U.S. Office of Naval Research, the Department of Defense High Energy Laser Joint Technology Office, and the Department of Energy.

\section*{Conflict of interests}
The authors declare no conflict of interest. Furthermore, the funders had no role in the design of the study; in the collection, analyses, or interpretation of data; in the writing of the manuscript, or in the decision to publish the results.

\section*{Abbreviations}
The following abbreviations are used in this manuscript:\\
\noindent 
\begin{tabular}{@{}ll}
BPM & beam position monitor\\
CRL & compound refractive lens\\
CW & continuous wave\\
FEL & free-electron laser\\
FELO & free-electron laser oscillator\\
FODO & focusing and defocusing\\
FWHM & full width at half maximum\\
HGHG & high gain harmonic generation\\
IR & infrared\\
JLab & Thomas Jefferson National Accelerator Facility\\
LCLS & linac coherent light source\\
linac & linear accelerator\\
RAFEL & regenerative amplifier free-electron laser\\
RF & radio frequency\\
rms & root mean square\\
SASE & self amplified spontaneous emission\\
SDE & source dependent expansion\\
SPARC & sorgente pulsata ed amplificata di radiazione coerente\\
VUV & vacuum ultraviolet\\
XFELO & x-ray free-electron laser
\end{tabular}

\end{document}